\documentclass[preprint2,iop,twocolappendix]{emulateapj}
\usepackage{array}
\usepackage{mathtools}
\usepackage{multirow}

\def\gaia{\textit{Gaia}}
\def\hip{\textit{Hipparcos}}
\def\recvarpi{\frac{1}{\varpi}}
\def\median{_{Md}}
\def\mode{_{Mo}}
\def\true{_{\rm true}}
\def\rlim{r_{\rm lim}}
\def\rmode{r\mode}
\def\rmed{r\median}
\def\rtrue{r\true}
\def\fobs{f_{\rm obs}}
\def\ftrue{f\true}
\def\errVarpi{\sigma_\varpi}
\def\vmini{(V-I_C)}
\def\bias{\overline{x}}
\def\rms{\overline{x^2}^{1/2}}

\def\stddevmod{\sigma_{x,\mode}}
\def\gxp{G_{\rm XP}}
\def\gbp{G_{\rm BP}}
\def\grp{G_{\rm RP}}
\def\posterior{P(\mathbf{\Theta}|\mathbf{X})}
\def\llf{P(\mathbf{X}|\mathbf{\Theta})}
\def\prior{P(\mathbf{\Theta})}

\def\ud{uniform distance}
\def\usd{uniform space density}
\def\expp{exponentially decreasing space density}
\def\mw{Milky Way}
\def\exphrd{exponentially decreasing space density\,+\,HRD\,+\,phot}
\def\mwhrd{Milky Way\,+\,HRD\,+\,phot}

\begin{document}
\title{Estimating distances from parallaxes. II.~Performance of Bayesian distance estimators on a \gaia{}-like catalogue}

\author{Tri L. Astraatmadja and Coryn A. L. Bailer-Jones}
\affil{Max Planck Institute for Astronomy, K\"{o}nigstuhl 17, D-69117, Heidelberg, Germany}

\shorttitle{Estimating distances from parallaxes} 
\shortauthors{Astraatmadja and Bailer-Jones}

\begin{abstract}
Estimating a distance by inverting a parallax is only valid in the absence of noise. As most stars in the \gaia{} catalogue will have non-negligible fractional parallax errors, we must treat distance estimation as a constrained inference problem. Here we investigate the performance of various priors for estimating distances, using a simulated \gaia{} catalogue of one billion stars. We use three minimalist, isotropic priors, as well an anisotropic prior derived from the observability of stars in a Milky Way model. The two priors that assume a uniform distribution of stars--either in distance or in space density---give poor results: The root mean square fractional distance error, $f_{\rm RMS}$, grows far in excess of 100\% once the fractional parallax error, $\ftrue$, is larger than 0.1. A prior assuming an exponentially decreasing space density with increasing distance performs well once its single scale length parameter has been set to an appropriate value: $f_{\rm RMS}$ is roughly equal to $\ftrue$ for $\ftrue<0.4$, yet does not increase further as $\ftrue$ increases up to to 1.0. The \mw{} prior performs well except towards the Galactic centre, due to a mismatch with the (simulated) data. Such mismatches will be inevitable (and remain unknown) in real applications, and can produce large errors. We therefore suggest to adopt the simpler \expp{} prior, which is also less time-consuming to compute. Including \gaia{} photometry improves the distance estimation significantly for both the \mw{} and \expp{} prior, yet doing so requires additional assumptions about the physical nature of stars.
\end{abstract}

\keywords{methods: data analysis -- methods: statistical -- surveys -- parallaxes -- stars: distances -- stars: fundamental parameters}

\section{Introduction}
In a universe devoid of measurement errors, estimating distances from parallaxes would involve simply taking the reciprocal of the measured parallax. However, we do not live in such universe. Measurement errors are always present and estimates must take these into account. This is the main question of this paper: Given a measured parallax $\varpi$ and its measurement uncertainty $\sigma_\varpi$, how can we best estimate the distance $r$ and its uncertainty?
\begin{figure*}
\includegraphics[width=\hsize]{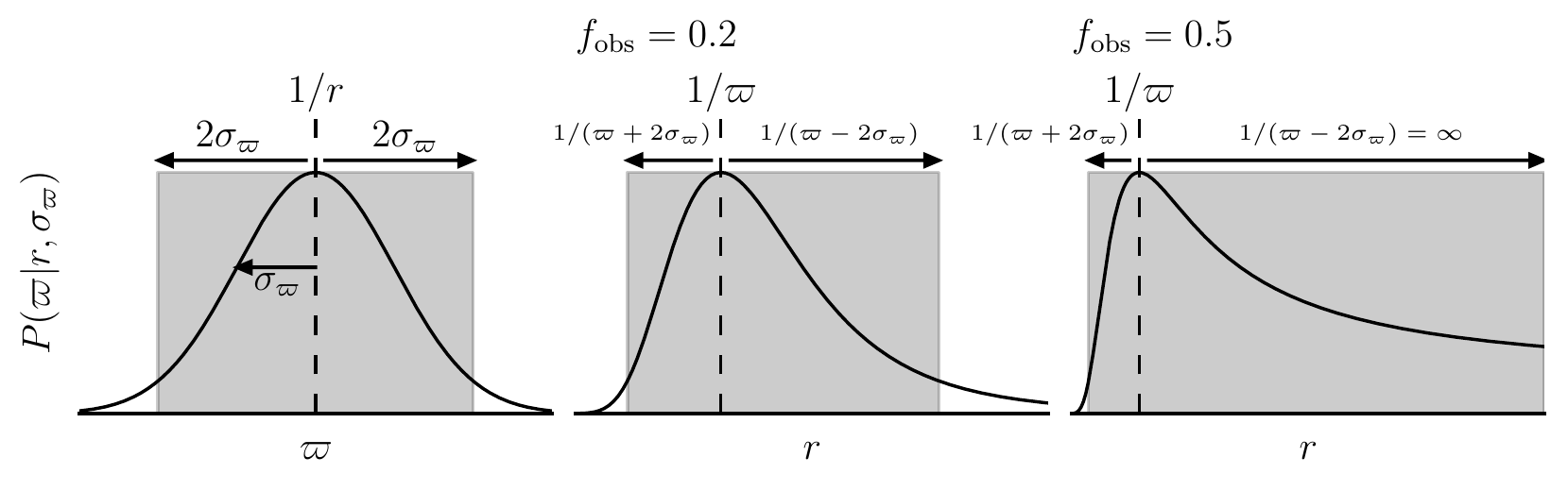}
\caption{The left panel is the probability $P(\varpi|r,\errVarpi)$ as a function of $\varpi$. The middle panel is $P(\varpi|r,\errVarpi)$ as a function of $r$ for an object with $\fobs=0.2$, and the right panel is as the middle panel but for $\fobs=0.5$. The shaded areas indicate the $2\sigma$ credible interval about $1/r$ and the corresponding transformed credible interval.}
\label{fig:rDist}
\end{figure*}
\begin{figure}
\includegraphics[width=\hsize]{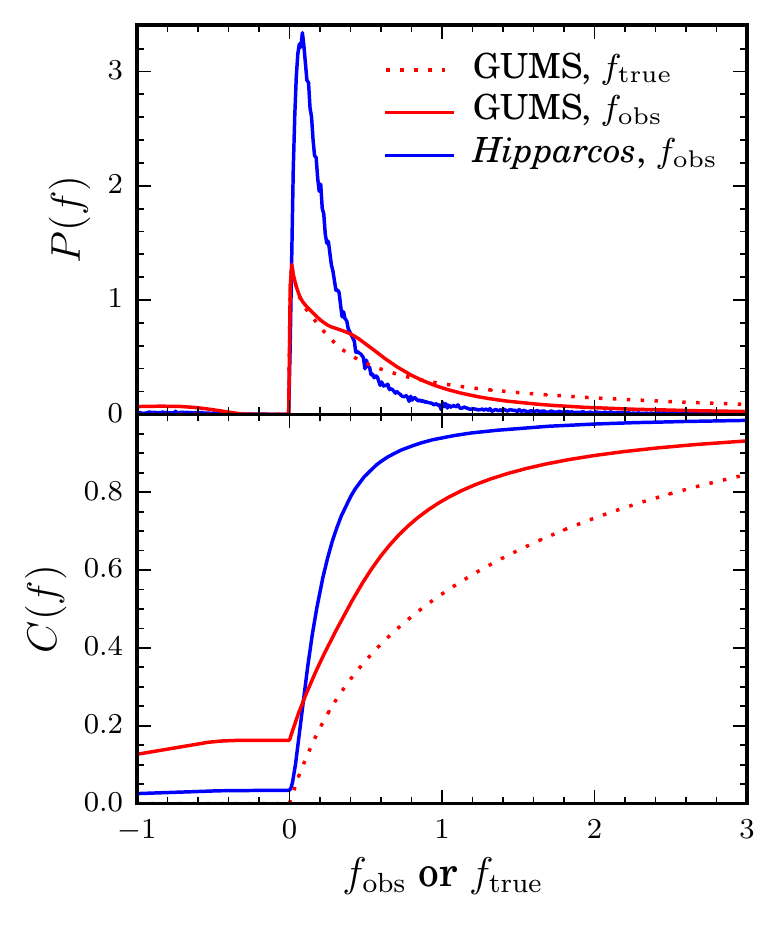}
\caption{The probability density (top row) and cumulative distribution (bottom row) of $\fobs$ (solid lines) and $\ftrue$ (dashed lines) of stars in the \hip{} (blue lines) and the GUMS (red lines) catalogue. $\ftrue$ is strictly positive by definition. Note that the plot only covers a subrange of all possible $f$ values. GUMS \citep{rob12} is a synthetic catalogue that simulates the expected content of the \gaia{} catalogue, and will be discussed further in Sect.~\ref{subsec:gums}.}
\label{fig:fObsDist}
\end{figure}

This question is of pressing importance, because upcoming data releases from the \gaia{} mission will measure parallaxes for all objects down to its magnitude limit of about $G=20$ (approximately $10^9$ objects) with expected accuracies of 0.01\,mas--1\,mas \citep{lin12}. The Large Synoptic Survey Telescope (LSST) hopes to extend this by 4 magnitude with a parallax accuracy of the order of 1\,mas \citep{ive08}. 

A common practice is to report $1/\varpi$ as the distance estimate and give its uncertainty from a first-order Taylor expansion as $\errVarpi/\varpi^2$. But this is problematic because $\varpi$ is noisy and the transformation to distance is highly nonlinear. If we assume a Gaussian noise model for our parallaxes, then the measurement can be considered to have been drawn from a Gaussian with mean $1/r$ and standard deviation $\sigma_\varpi$
\begin{equation}
P(\varpi|r,\errVarpi) = \frac{1}{\sqrt{2\pi}\errVarpi}\exp\left[-\frac{1}{2\errVarpi^2}\left(\varpi-\frac{1}{r}\right)^2\right],\;\sigma_\varpi\geq 0.
\label{eq:likelihood}
\end{equation}
We see in the left panel of Fig.~\ref{fig:rDist} that while this distribution is symmetric in $\varpi$, it is not in $r$. The distribution is skewed even for an object with a small fractional parallax error $\fobs=\errVarpi/\varpi$ of 0.2 (central panel of Fig.~\ref{fig:rDist}). Because the likelihood does not go to zero even at infinite distance, confidence intervals could extend up to infinity even for finite parallax errors, giving meaningless results (see \cite{cbj15} for more discussion). This is illustrated in the right-panel of Fig.~\ref{fig:rDist}. For such skewed distributions it is not obvious that the mode is even a good summary.

Another problem with using $1/\varpi$ as the distance estimator is that it does not work for parallaxes with negative values. 
The measurement model of Eq.~\ref{eq:likelihood} has a nonzero probability of drawing negative parallaxes, and this probability grows with increasing fractional parallax error and increasing distance.
Such measurements occur in reality, and tell us that the source is distant and/or has a large measurement error. 

One way to ameliorate these problems is by discarding negative parallaxes and retaining only stars with high accuracy (say $\fobs\leq 0.2$). However, as we can see from Fig.~\ref{fig:fObsDist} (bottom panel, solid red line), we expect 80\% of objects in the \gaia{} catalogue to have either negative parallaxes or $\fobs>0.2$. This is 
quite different from the \hip{} catalogue \citep{per97}, in which about 55\% of the sources have $\fobs\geq 0.2$ (bottom panel of Fig.~\ref{fig:fObsDist}, solid blue line). Discarding data with large measurement errors is a waste of hard-won data, and will bias subsequent analyses by removing fainter or farther sources \citep{are99, smi06}.

There are of course instances when one is not interested in distances explicitly. Sometimes we want to fit a model and can often do this by using it to predict parallaxes which we then compare directly with the measured parallaxes (e.g. \cite{rom52, jun71, van98}). In this case working in the parallax-domain is preferable as it is much easier to deal with the uncertainties. But here we have no model and are interested specifically in the problem of distance inference for single objects.

In a previous tutorial paper, one of us looked at this problem and showed that one has to take a proper inference approach which cannot avoid the specification of a prior \citep[henceforth Paper I]{cbj15}. This paper examined three priors on $r$: the \ud{} prior, the \usd{} prior, and the \expp{} prior. The properties and performance of distance estimators based on these priors was then assessed using noisy toy data. This showed the problems of using naive priors like the \ud{} prior and suggested instead that priors which converge asymptotically to zero at infinite distance are to be preferred.

Here we extend this analysis in two important ways. First, we do distance inference with a number of priors and estimators on a simulation of the entire \gaia{} catalogue (Sect.~\ref{subsec:gums}--\ref{subsec:simu}). Second, we introduce a new prior based on our knowledge of the distribution of stars in the Galaxy which includes their observability (including accounting for extinction effects) by \gaia{} (Sect.~\ref{subsec:mw}). We assess the performance of the four priors in a manner similar to Paper I (Sect.~\ref{sec:results}), and further investigate the reliability of distance uncertainties based on credible intervals derived from the posterior distributions (Sect.~\ref{subsec:formalErrors}). We also considers the inclusion of photometric measurements and a Hertzsprung-Russell Diagram (HRD) prior together with the \mw{} prior and compare its performance to parallax-only distance inference (Sect.~\ref{sec:mwhrd}).

A summary of the notations used in this paper is shown in Table~\ref{tab:notation}.
\begin{table}
\caption{Notations used in this paper.}
\label{tab:notation}
\begin{center}
\begin{tabular}{ll}
\hline\hline
$\varpi$ & Measured parallax\\
$\errVarpi$ & Parallax measurement uncertainty\\
$\rtrue$ & True distance\\
$\rmed$ & Distance from the median of the posterior\\
$\rmode$ & Distance from the mode of the posterior\\
$\rlim$ & Limiting distance imposed in several priors\\
$\fobs$ & $\errVarpi/\varpi$, observed fractional parallax error\\
$\ftrue$ & $\errVarpi\rtrue$, true fractional parallax error\\
$P(\theta)$ & Probability distribution function (PDF) of $\theta$\\
$P(\theta|x)$ & Posterior distribution of $\theta$ given $x$\\
$P^*(\theta)$ & Unnormalised PDF\\
$C(\theta)$ & Cumulative distribution function (CDF) of $\theta$\\
\hline
\end{tabular}
\end{center}
\end{table}

\section{The methods and the models}
\label{sec:method}
\subsection{Distance inference}
Suppose that a star at true distance $r$ is measured to have a parallax $\varpi$ with measurement error $\sigma_\varpi$. We want to find the posterior probability density (PDF) $P(r|\varpi,\sigma_\varpi)$. Using Bayes' theorem this is expressed in terms of the likelihood of the observation $P(\varpi,\errVarpi)$, and the prior probability $P(r)$ of the distance, as
\begin{equation}
P(r|\varpi,\errVarpi) = \frac{1}{Z}P(\varpi|r,\errVarpi)P(r),
\end{equation}
where
\begin{equation}
Z = \int^\infty_0 dr P(\varpi|r,\errVarpi)P(r)
\end{equation}
is the normalisation constant.

The likelihood is given in Eq.~\ref{eq:likelihood}. This expresses the probability density over the measurement---the parallax---given the true distance, and is the one used in the \gaia{} data processing.

The prior expresses our knowledge and assumptions about the distance distribution, independent of the specific measurement. As was explained in Paper I, we have essential knowledge which should be used if parallaxes are to be interpreted correctly in the general case.

\begin{table*}
\caption{The three isotropic priors used in this paper, summarised from \cite{cbj15}.}
\label{tab:priors}
\begin{center}
\begin{tabular}{>{\raggedright\arraybackslash}p{3.6cm}ll}
\hline\hline
Prior & Form & Mode of the posterior\\
\hline
Uniform distance & 
$P_{\rm ud}(r) = 
\begin{dcases*}
\frac{1}{r_{\rm lim}} & for $0<r\leq r_{\rm lim}$,\\
0 & otherwise.
\end{dcases*}$
&
$\rmode = 
\begin{dcases*}
\recvarpi & for $0<\recvarpi\leq\rlim$,\\
\rlim & for $\recvarpi>\rlim$ or $\varpi\leq 0$.\\
\end{dcases*}$
\\
\hline
Uniform space density & 
$P_{\rm usd}(r) = 
\begin{dcases*}
\frac{3}{r_{\rm lim}^3}r^{2} & for $0<r\leq r_{\rm lim}$,\\
0 & otherwise.
\end{dcases*}$ &
$\begin{aligned}
r_{\rm mode} &= \frac{1}{4\fobs^2\varpi}\left(1-\sqrt{1-8\fobs^2}\right)\\
\rmode &=
\begin{dcases*}
r_{\rm mode} & for $\varpi>0$, $\fobs<1/\sqrt{8}$, $r_{\rm mode}\leq\rlim$,\\
\rlim & otherwise.
\end{dcases*}
\end{aligned}$
\\
\hline
Exponentially decreasing space density &
$P_{\rm exp}(r) =
\begin{dcases*}
\frac{1}{2L^3}r^2\exp\left(-\frac{r}{L}\right) & for $r>0$,\\
0 & otherwise.
\end{dcases*}$
&
Solve $\frac{r^3}{L}-2r^2+\frac{\varpi}{\errVarpi^2}r-\frac{1}{\errVarpi^2} = 0$ for $r$.
\\
\hline
\end{tabular}
\end{center}
\end{table*}
\begin{figure*}
\includegraphics[width=\hsize]{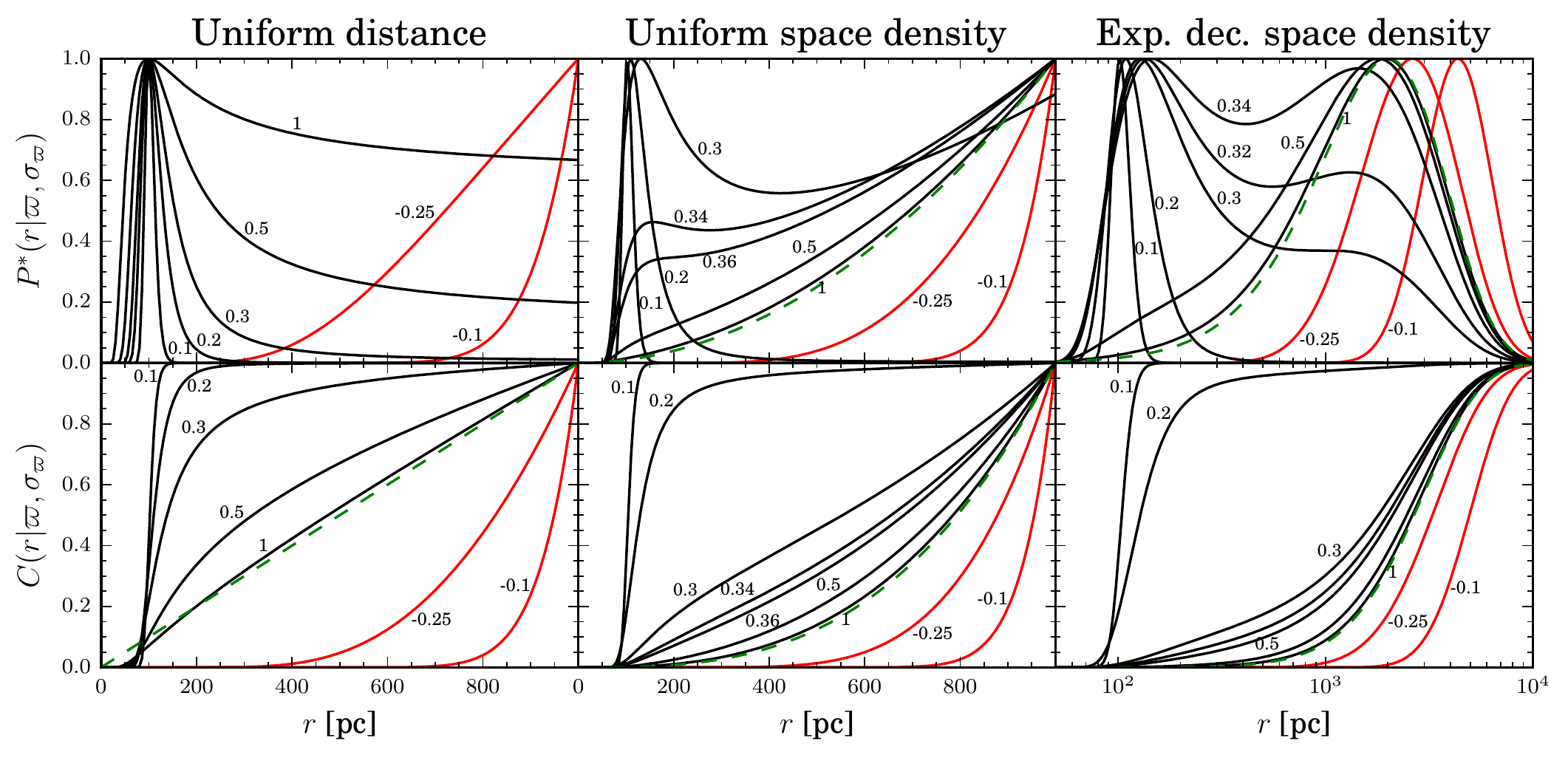}
\caption{The unnormalised posterior PDF (top row) and the corresponding CDF (bottom row) for the \ud{} prior (left column), \usd{} prior (middle column), and the \expp{} prior (right column). The posteriors are shown for various $\fobs$ as indicated by the value near the corresponding line. Positive $\fobs$ are shown in black lines, while negative $\fobs$ are shown in red lines. Green dashed lines indicate the corresponding priors, but $P^*_{\rm ud}(r)$ is not shown because it is simply a uniform function in $r$. The magnitude of the observed parallax is $|\varpi| = 10$\,mas in all cases. For the \ud{} and the \usd{} priors, the distance cutoff is $\rlim=1$\,kpc and for the \expp{} prior the scale length is $L=1$\,kpc. Note that the PDF and CDF of the \expp{} prior are shown on a logarithmic scale in $r$.}
\label{fig:posteriors}
\end{figure*}
\subsection{The three isotropic priors}
\label{subsec:3priors}
The first three priors considered in this paper are the three isotropic priors discussed in Paper I. These are the \ud{} prior, the \usd{} prior, and the \expp{} prior. We will introduce the anisotropic \mw{} prior in Sect.~\ref{subsec:mw}.

The form of the priors and expression for their mode is given is Table~\ref{tab:priors}, and plots for various values of $\fobs$ are shown in Fig.~\ref{fig:posteriors}. Since these priors have been discussed extensively in Paper I, readers are referred to that paper for details. We just note here some particular behaviours.

Using the \ud{} prior, the mode of the posterior is always $\rmode = 1/\varpi$, except for negative parallaxes or when $1/\varpi>\rlim$, in which case $\rmode=\rlim$. This is markedly different from the other priors, in which $\rmode$ is also influenced by the parallax uncertainty. We can see this in the top rows of Fig.~\ref{fig:posteriors}. The mode of the posteriors with the \usd{} and the \expp{} priors shift to larger distances as $\fobs$ increases, while the mode of the posteriors with the \ud{} priors are independent of $\fobs$.

The median distance of all the posteriors, on the other hand, shifts to larger distances as $\fobs$ increases. The median distance $\rmed$ of the posterior with the \usd{} prior shifts faster than the posterior using the \ud{} because the \usd{} prior assumes that there are more objects located at large distances.

In place of the sharp cut-off at $r=\rlim$, the \expp{} prior drops asymptotically to zero as $r\rightarrow\infty$. While more desirable, this can produce a bimodality over a small range of $\fobs$ as a transition occurs from ``data-dominated'' posterior to a ``prior-dominated'' posterior. In the right column of Fig.~\ref{fig:posteriors}, this occurs at $\fobs\sim 0.3$. When $\fobs$ gets large enough the data are less informative, so the bimodality disappears and the posterior is dominated by the prior, as can be seen for the case of $\fobs=0.5$ and $\fobs=1$. For $f\gtrsim 1$ the posterior becomes indistinguishable from the prior (shown in the green line). This is just the behaviour we expect and want.

A negative parallax is an indicator that the object is located at a large but highly uncertain distance. Thus from the assumed prior we can still obtain information on the distance, and our confidence in this distance estimate depends on $\fobs$. In Fig.~\ref{fig:posteriors} (red lines) we show a negative parallax of $\varpi=-10$\,mas with $\fobs=-0.1$ and $\fobs=-0.25$.
As the negative $\fobs$ shifts toward zero, the mode of the posterior shift toward larger distances. This is because as $|\fobs|$ gets smaller, we are more confident that the true parallax is close to zero and the object is very far away. As $|\fobs|$ gets larger, the mode shifts to smaller distances and converges on the prior.

\begin{figure*}
\includegraphics[width=\hsize]{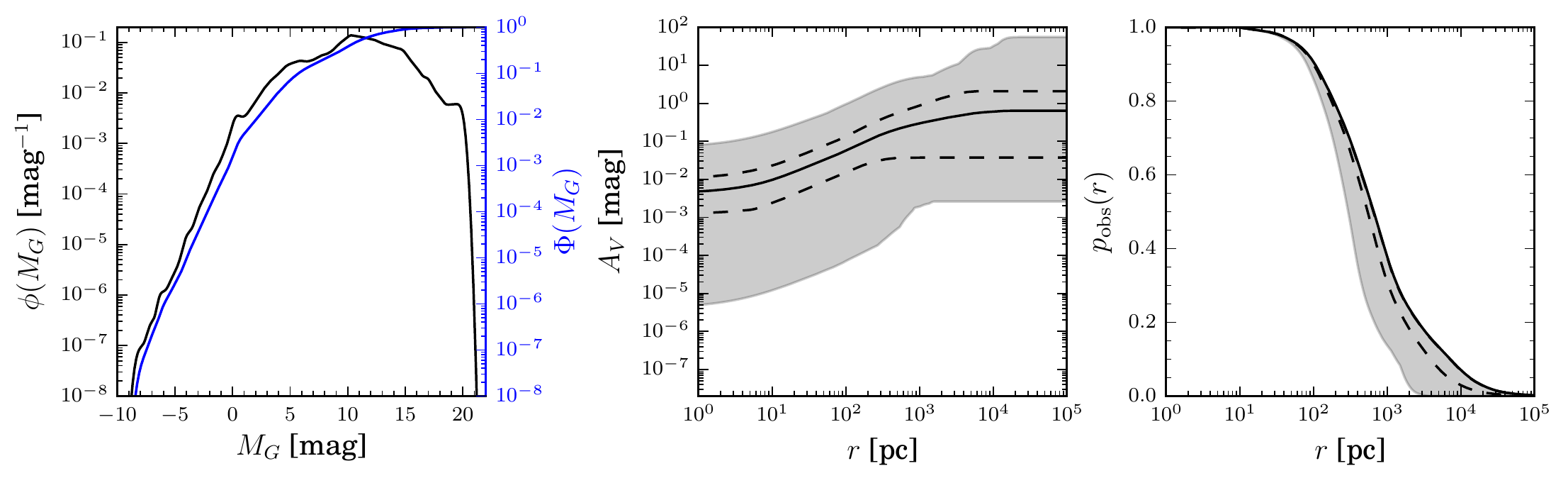}
\caption{Left: The smoothed intrinsic luminosity function $\phi(M_G)$ (black line, left scale) used in the \mw{} prior, and its cumulative distribution function $\Phi(M_G)$ (blue line, right scale). The luminosity function is smoothed using kernel density estimation. Middle: The average line-of-sight $V$-band extinction as a function of distance $r$, shown as the solid black line. The shaded area indicates the full range of extinctions and the dashed lines show the 5\% and 95\% quantiles of the distribution. Right: The line-of-sight fraction of observable stars $p_{\rm obs}(r)$ as a function of distance. The solid black line indicates $p_{\rm obs}(r)$ when extinction effects are neglected, while the shaded area indicates all the possible shape of $p_{\rm obs}(r)$ for all possible line-of-sights. The dashed line show the lower 5\% of $p_{\rm obs}(r)$ (i.e.\ 95\% of all stars at a given $r$ are above the line).}
\label{fig:pObs}
\end{figure*}
\begin{figure*}
\includegraphics[width=\hsize]{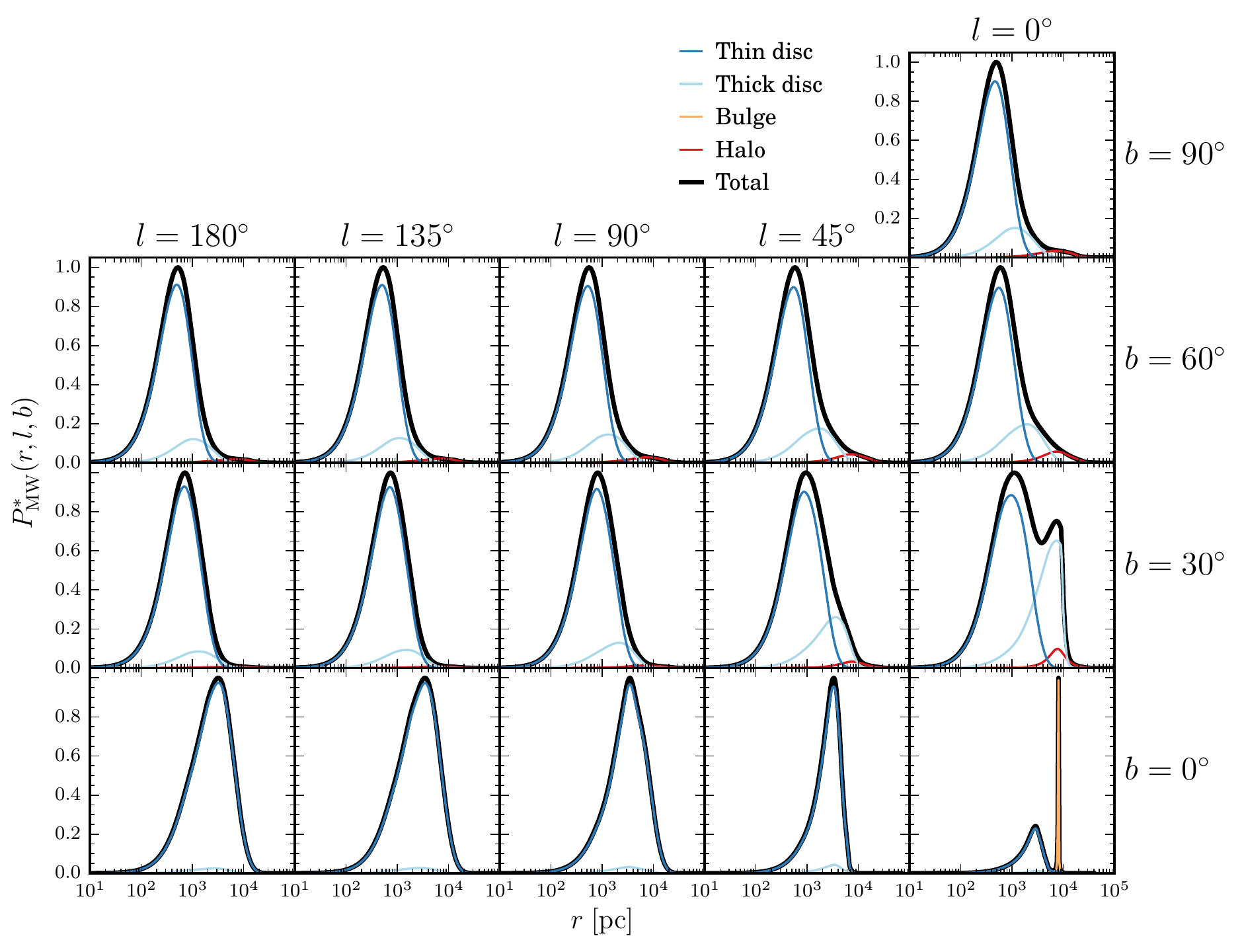}
\caption{The \mw{} prior $P_{\rm MW}^*(r,l,b)$ along the LOS towards the $(l,b)$ directions written on the top and side of the grid, here as a function of heliocentric distance $r$. Each individual components are drawn in different colors, while the black line indicates the total density. The prior is scaled so that the peak distribution is unity.}
\label{fig:prior_mw}
\end{figure*}
\begin{figure*}
\includegraphics[width=\hsize]{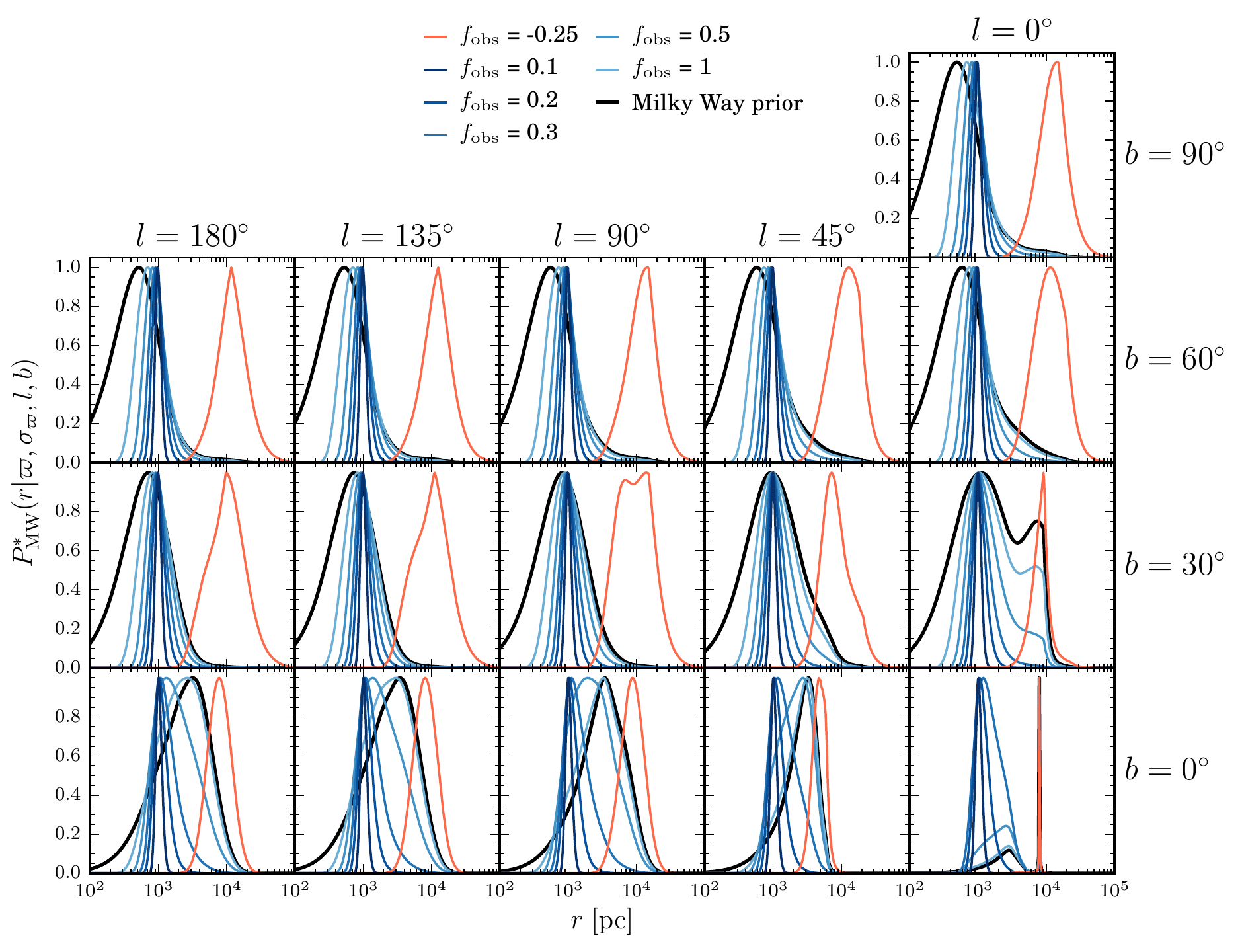}
\caption{The unnormalised posterior PDF $P_{\rm MW}^*(r|\varpi,\errVarpi,l,b)$ for various $\fobs$ as indicated by the legend, and various LOS along the $(l,b)$ direction written on the top and side of the grid. Posteriors with positive $\fobs$ are drawn in various shades of blue, while negative $\fobs$ are drawn in red. Here the observed parallaxes are all kept the same at $|\varpi|=1$\,mas. The black line shows the prior $P_{\rm MW}(r,l,b)$ previously shown in Fig.~\ref{fig:prior_mw}. The PDF are scaled such that they peak at unity. We can see here that the transition of the posterior to the prior for large $\fobs$ also happens, but the shape of the posterior depends also on the LOS direction. Note that the narrow peak at $(l,b) = (0\degr,0\degr)$ contains multiple lines.}
\label{fig:post_mw}
\end{figure*}
\begin{figure*}
\includegraphics[width=\hsize]{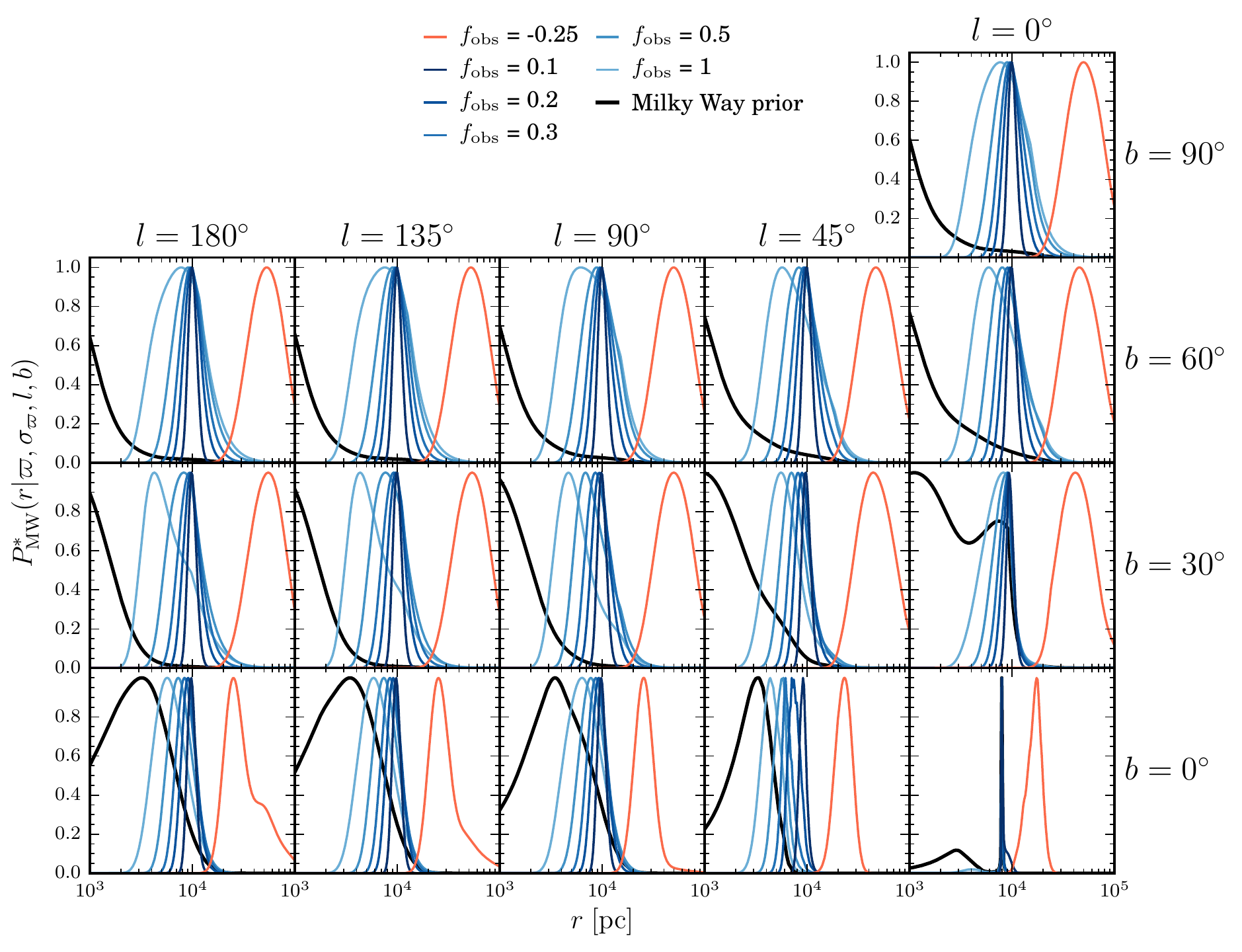}
\caption{As Fig.~\ref{fig:post_mw}, but for $|\varpi|=0.1$\,mas. Note that the range of the horizontal axes is different than those in Fig.~\ref{fig:post_mw}.}
\label{fig:post_mw2}
\end{figure*}

\subsection{The \mw{} prior}\label{subsec:mw}
The above priors have the advantage of being simple, but the disadvantage of being very different from what we already know about the distribution of stars in the Milky Way.
We introduce here a prior based on a three-dimensional density model $\rho_{\rm MW}(\mathbf{r})$ of the Milky Way which also takes into account selection effects of a magnitude-limited survey such as \gaia{}. To model these selection effects we adopt a universal luminosity function $\phi(M_G)$ and an extinction map. 
Since now we include directional information in addition to the measured parallax, we can hope that this will improve our distance estimation.

In constructing the \mw{} prior, we assume that the Milky Way has a stellar number density (the number of stars in a volume element $dV$) described by $\rho_{\rm MW}(r,l,b)$, and a universal luminosity function (the fraction of stars with absolute magnitude between $M_G$ and $M_G+dM_G$) described by $\phi(M_G)$. The probability density of observable stars in a volume element $dV$ is then
\begin{equation}
P^*_{\rm MW}(V) = \rho_{\rm MW}(r,l,b) p_{\rm obs}(r,l,b),
\end{equation}
where
\begin{equation}
p_{\rm obs}(r,l,b) = \int^{M_{G,{\rm faint}}(r,l,b)}_{M_{G,{\rm bright}}} dM_G\phi(M_G)
\label{eq:pObs}
\end{equation}
describes the fraction of observable stars along the line of sight (LOS) direction $(l,b)$ as a function of distance $r$.
Here $M_{G,{\rm bright}}$ is the bright end of the luminosity function, and
\begin{equation}
M_{G,{\rm faint}}(r,l,b) = m_{G,{\rm lim}} - 5\log_{10}r + 5 - A_G(r,l,b)
\label{eq:faintlimit}
\end{equation}
is the faint observable limit of the luminosity function, which is a function of distance $r$ and direction $(l,b)$, and depends on both the limiting magnitude $m_{G,{\rm lim}}$ of \gaia{} (which we set to 20) and the $G$-band extinction $A_G$. 
We could neglect extinction effects by setting $A_G = 0$, thus making $M_{\rm lim}$ the faintest possible limit. But as we shall see later, this overestimates the limit at low galactic latitudes, especially in regions close to the galactic centre. Neglecting extinction effects here would erroneously increase our visibility towards the galactic centre and consequently risk providing an unrealistic distance inference for a large number of stars. We therefore include extinction effects in our prior. The full form of the \mw{} prior is then
\begin{equation}
P^*_{\rm MW}(r, l, b) = r^2\rho_{\rm MW}(r,l,b)p_{\rm obs}(r,l,b).
\label{eq:prior_mw}
\end{equation}
The stellar number density $\rho_{\rm MW}(r,l,b)$ can then be expressed further as the sum of three components: the bulge $\rho_b$, the disc $\rho_d$, and the halo $\rho_h$. These are detailed in Appendix~\ref{app:mwmodel}.

Earlier we defined $\rho_{\rm MW}(r,l,b)$ as the stellar number density of the Milky Way. However, the model that we use actually describes the stellar mass density. We are therefore assuming that the stellar mass traces the stellar counts in the same way everywhere. We are not assuming that all stars have the same mass.

To construct the fraction of observable stars $p_{\rm obs}(r,l,b)$, we use the the stellar colour-magnitude diagram (CMD)\footnote{We only use the absolute magnitude distribution in the GUMS catalogue for the construction of our \mw{} prior. We do not use the spatial distribution} from \cite{rob12}, which is modelled using the initial mass function (IMF) and star formation rate (SFR) observed in the solar neighborhood. The CMD for the thin disc assumes multiple formation epochs with its own IMF and SFR history, while the bulge, thick disc, and halo population assumes a single burst of star formation (the GUMS catalogue and its construction will be described in more detail in Sect.~\ref{subsec:gums}). In this paper, however, we combine all these diagrams into a single luminosity function $\phi(M_G)$.

We transform the $V$-band absolute magnitude $M_V$ into the $G$-band absolute magnitude $M_G$ using the transformation polynomial from \cite{jor10}:
\begin{multline}
M_G = M_V - 0.0257 - 0.0924\vmini\\- 0.1623\vmini^2 + 0.0090\vmini^3 \ .
\end{multline}
To construct a smooth luminosity function from the values extracted from the binned CMD, 
we use kernel density estimation weighted by the number of stars per bin.
We use a Gaussian kernel with kernel width $\sigma = 0.25$ magnitudes. The resulting luminosity function $\phi(M_G)$ and its CDF $\Phi(M_G)$ is shown in the left panel of Fig.~\ref{fig:pObs}. The black line shows the smoothed function, while the blue line shows the CDF of the luminosity function. We can then use this CDF with Eq.~\ref{eq:pObs} to calculate the fraction of observable stars $p_{\rm obs}(r,l,b)$, but in order to calculate the faint end of the integration, in Eq.~\ref{eq:faintlimit} we have to take extinction effects into account. For this we use the extinction map of \cite{dri03}, examples of which are shown in Fig.~\ref{fig:extMap}.

The central panel of Fig.~\ref{fig:pObs} shows the LOS extinction as a function of distance $r$, averaged over all directions. Since the variation in $A_V$ can cover approximately 4 orders of magnitude, it is thus important to take extinction into account. Using the extinction map we can determine $M_{G,{\rm faint}}(r,l,b)$ for any position in the Galaxy. The $V$-band extinction $A_V$ is converted to the $G$-band extinction $A_G$ using the ratio $A_G/A_V=0.695$, a factor we computed from simulations of stellar spectra redenned using the \cite{fit99} extinction curve.

We now perform the integral in Eq.~\ref{eq:pObs} to get $p_{\rm obs}(r,l,b)$. This is shown in 
the right panel of Fig.~\ref{fig:pObs} as a function of $r$. The shaded area shows the full range of observabilities across all $(l,b)$. At 1\,kpc, for example, we can see between about 15\% and 50\% of all stars in the GUMS simulation, depending on direction.

Combining $\rho_{\rm MW}(r,l,b)$ and $p_{\rm obs}(r,l,b)$ we can then calculate the \mw{} prior $P^*_{\rm MW}(r,l,b)$ using Eq.~\ref{eq:prior_mw}. This is shown in Fig.~\ref{fig:prior_mw} for 16 LOS directions in galactic coordinates $(l,b)$. Due to symmetry in the model of the mass distribution, we only show the LOS directions in the northeastern quadrant of the Milky Way. The variations in the other quadrants will be similar, barring some difference in details due to the extinction. The prior shows a direction dependence, but it is clear that the (thin) disc is the most dominant component in almost any direction, being the most massive among the others. At higher latitudes and distances, the halo starts to be discernible, as does the thick disc at intermediate distances. The bottom right panel shows the prior towards the galactic centre. We see a very large and narrow peak around 8\,kpc, which is the galactic centre. The peak is mostly dominated by the bulge. Had we not included extinction in our prior, this peak would be even larger due to the enormous density of stars in that direction, which would led to erroneous features in the posteriors. 

Example posteriors resulting from these priors are shown in Fig.~\ref{fig:post_mw}, for the same LOS directions as Fig.~\ref{fig:prior_mw}. The measured parallax is $|\varpi|=1$\,mas in all cases. The different lines in each panel show different values of $\fobs$.
Here the posterior is almost always unimodal, but nevertheless bimodalities can still appear, for example at $(l,b) = (0\degr, 30\degr)$. The bimodalities are a consequence of both the prior and the likelihood (we saw bimodalities with the \expp{} which was unimodal).
In general bimodalities only appear when $\fobs\gtrsim 0.5$, and only in the directions where there are comparable contributions from multiple components.
If we look at the previously mentioned LOS direction $(l,b) = (0\degr,30\degr)$ again, at $\fobs=1$ a second peak in the posterior appears, which indicates that---given this very large measurement error---the posterior considers the possibility that the source is a thick disc object.

Fig.~\ref{fig:post_mw2} shows the same plots but with a different measured parallax, $\varpi = 0.1$\,mas. Here the posteriors are always unimodal.

\begin{figure}
\includegraphics[width=\hsize]{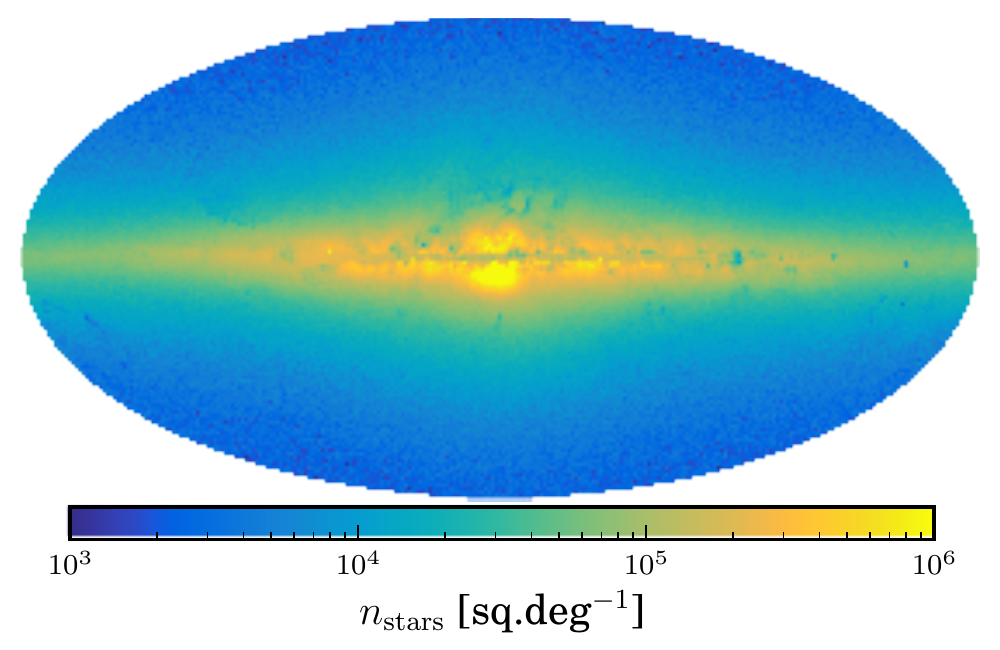}
\caption{The Mollweide projection of the spatial distribution of all stellar objects in the GUMS catalogue with $G\leq 20$, in the galactic coordinate system.}
\label{fig:gums}
\end{figure}

\begin{figure}
\includegraphics[width=\hsize]{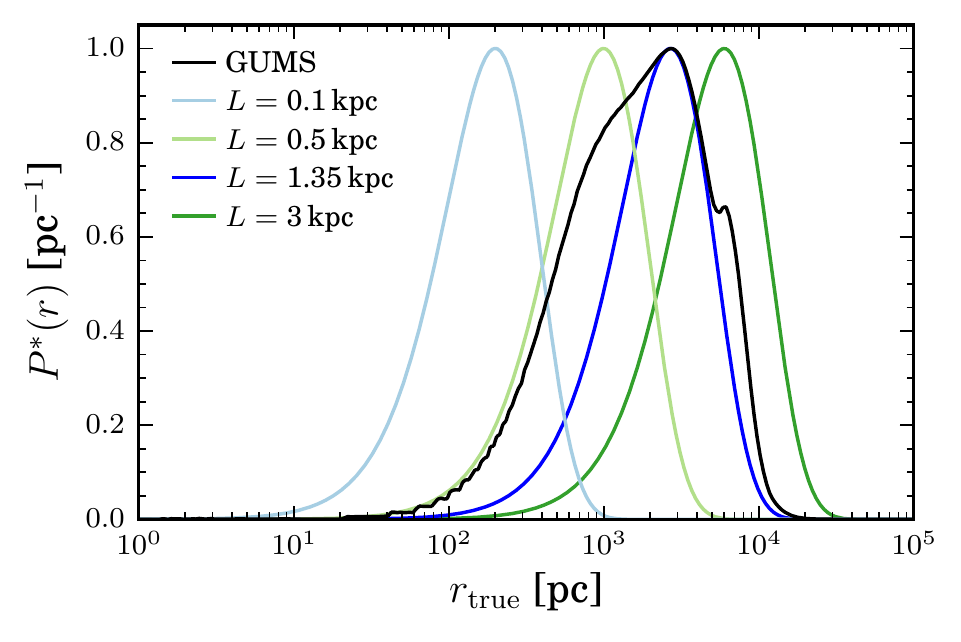}
\caption{The distribution of the true distances $\rtrue$ of all stellar objects in the GUMS catalogue (black line). The other colored lines indicates the \expp{} prior for different scale lengths $L$. The PDFs are scaled such that they peak at unity.}
\label{fig:gums_rdist}
\end{figure}
\begin{figure*}
\includegraphics[width=\hsize]{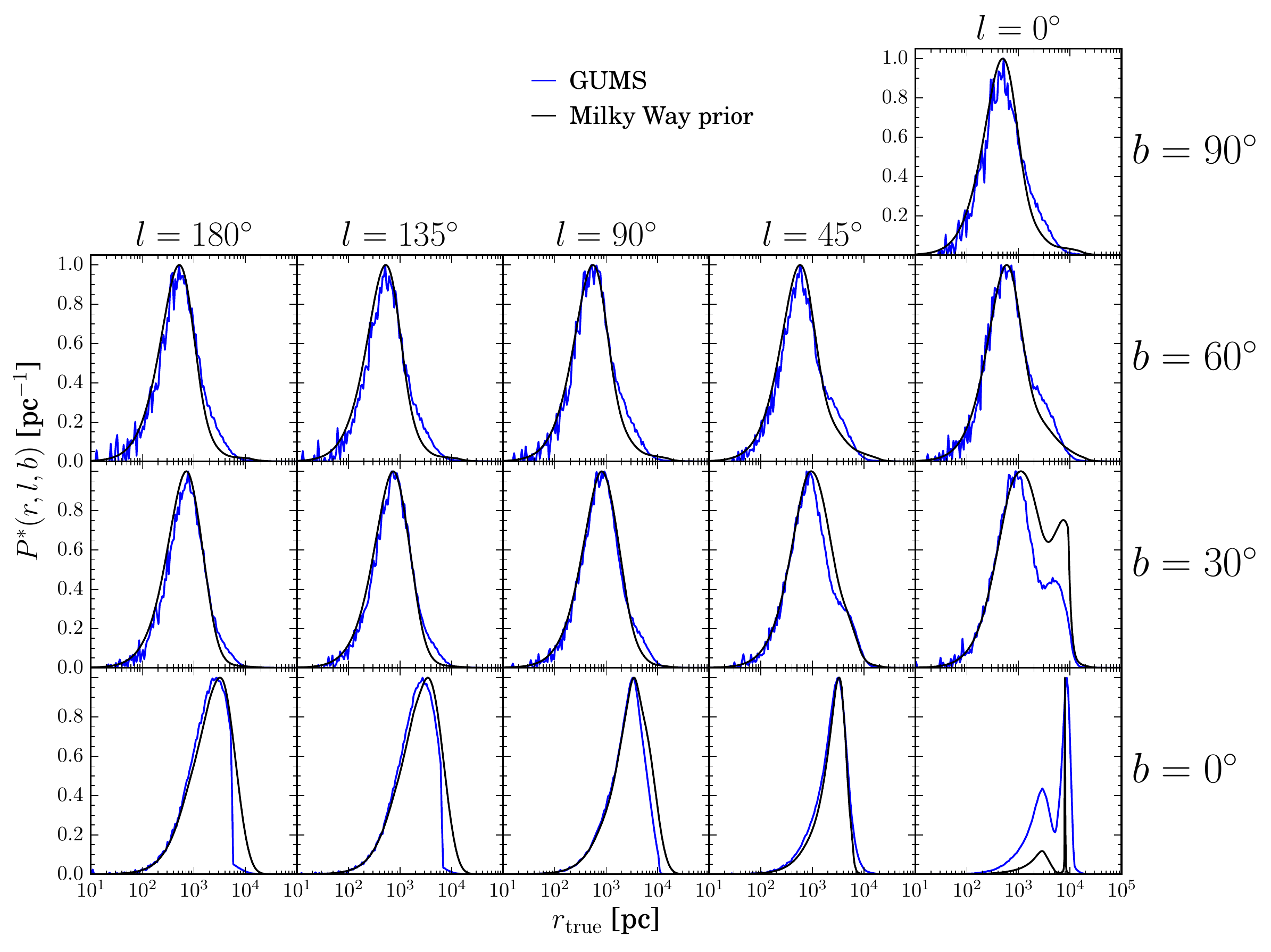}
\caption{The distribution of the true distances $\rtrue$ of all stellar objects in the GUMS catalogue, for selected LOS directions (as used in Figs.~\ref{fig:prior_mw}--\ref{fig:post_mw2}. Each LOS for $b>0$ comprises stars in an area of 13.43 square degrees (all stars at \texttt{healpix} \citep{gor05} level $n_{\rm side} = 16$ centered on that direction). At $b=0$ we use $n_{\rm side} = 128$ which corresponds to an area of 0.21 square degrees. The black lines show the \mw{} prior for comparison. Both distributions are scaled such that they peak at unity in each panel. The sudden drop in the GUMS distribution at $b=0^\circ$ and $l=90^\circ,\ 135^\circ$, and 180$^\circ$ is an artefact of the GUMS catalogue, which terminates the disc at a galactocentric distance of 14\,kpc \citep{rob12}.}
\label{fig:gums_rdist_directional}
\end{figure*}

\subsection{The simulated data: The GUMS catalogue}
\label{subsec:gums}
We want to investigate the performance of the priors described in Sect.~\ref{subsec:3priors}--\ref{subsec:mw} on objects in a \gaia{}-like catalogue. For this we can use the \gaia{} Universe Model Snapshot (GUMS) catalogue \citep{rob12}, which simulates what \gaia{} was expected to observe (prior to launch) down to its limiting magnitude.

Stars are generated in GUMS using the Besan\c{c}on galaxy model (BGM, \citealt{rob03}). It is based on a four component model of the Galaxy using an evolutionary model, with a specified initial mass function and star formation history. Interstellar extinction is calculated from the map of \cite{dri03}. For each star GUMS reports the apparent magnitudes, colours, positions and distance, as well as other quantities not required in our present work. GUMS only simulates objects that can be reasonably well-observed by \gaia{}, so it omits objects fainter than $G=20$ or more than 100\,kpc from the Sun. The final catalogue contains $\sim$$1.6\times 10^9$ stellar objects, the spatial distribution of which are shown in Fig.~\ref{fig:gums}.

Fig.~\ref{fig:gums_rdist} show the distribution of the true distances $\rtrue$ of all stellar objects in the GUMS catalogue. The distribution peaks at $\sim$2.7\,kpc, which corresponds to the scale length of the thin disc adopted by \cite{rob12}. A small peak is also present at $\sim$6\,kpc, which corresponds to the maximum distribution of observable stars towards the galactic centre (the galactic centre itself is invisible due to extinction).
We overplot the \expp{} prior for 4 different scale length $L$, $L=\{0.1, 0.5, 1.35, 5\}$\,kpc. None of these distributions closely resembles the GUMS distribution, but the mode of the prior for $L=1.35$\,kpc peaks at the same point as those of the GUMS distribution. 

Fig.~\ref{fig:gums_rdist_directional} shows the distance distribution in the GUMS model for different directions (the same as previously used in Figs.~\ref{fig:prior_mw}--\ref{fig:post_mw2}), which we also compare to our \mw{} prior. Being a prior, of course we do not expect it to agree exactly with (our simulation of) reality! Yet it is realistic to believe that our current knowledge can produces a distance distribution which is in broad agreement with what \gaia{} will observe. Overall we see a good match between GUMS and our model, but nonetheless some deviations can be seen. The most significant deviations occur at low latitudes in directions toward the galactic plane at intermediate and large distances, but in other directions we see nevertheless small devations at large distances. This is due to the different adopted model for the thick disc and the halo (and in the case of $(l,b)=(0\degr,0\degr)$, the difference is due to the different bulge-to-disc mass ratio), and the fact that we are using a single universal HRD instead of an HRD for each component of the Milky Way. Such prior mismatches are bound to occur in reality, so using GUMS will be a meaningful test of the performance of this prior. 

\begin{figure}
\includegraphics[width=\hsize]{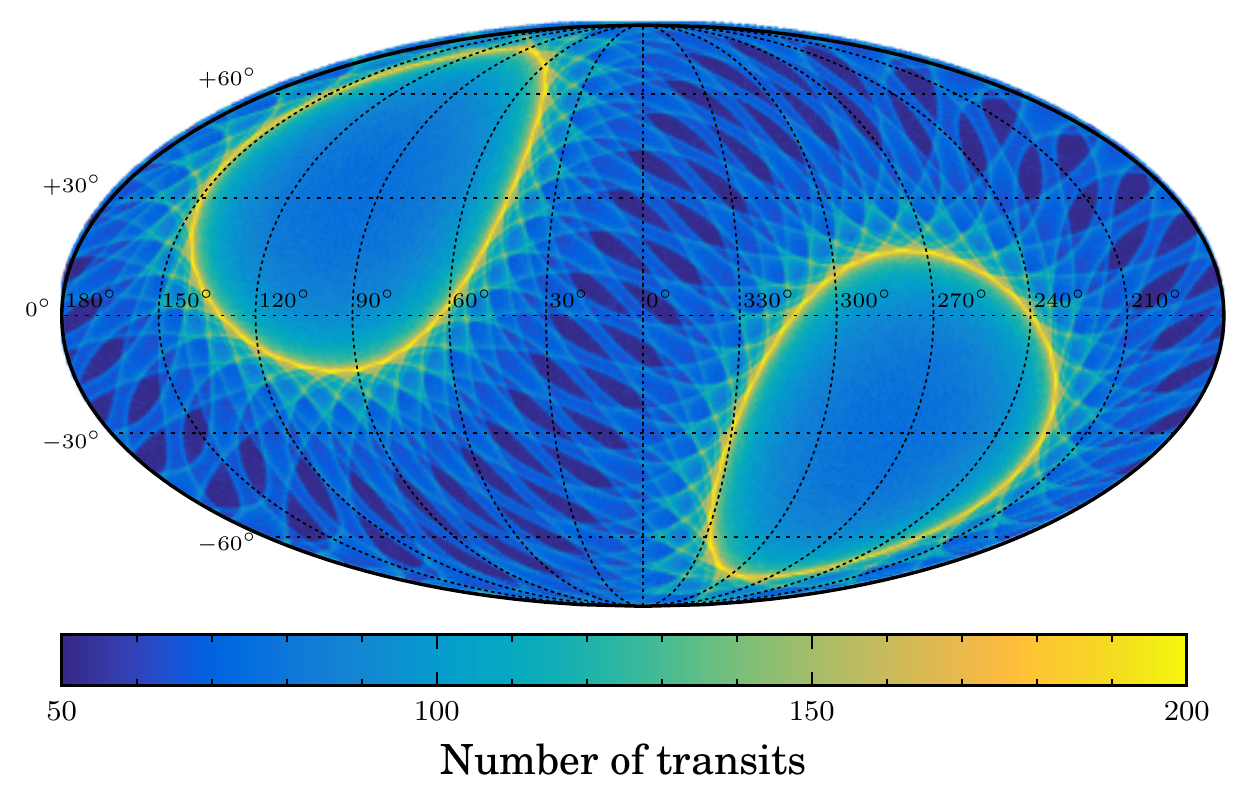}
\caption{The number of transits as a function of $(l,b)$ assuming a nominal \gaia{} scanning law for five years. Data kindly provided by Berry Holl.}
\label{fig:ntr}
\end{figure}

\begin{figure}
\includegraphics[width=\hsize]{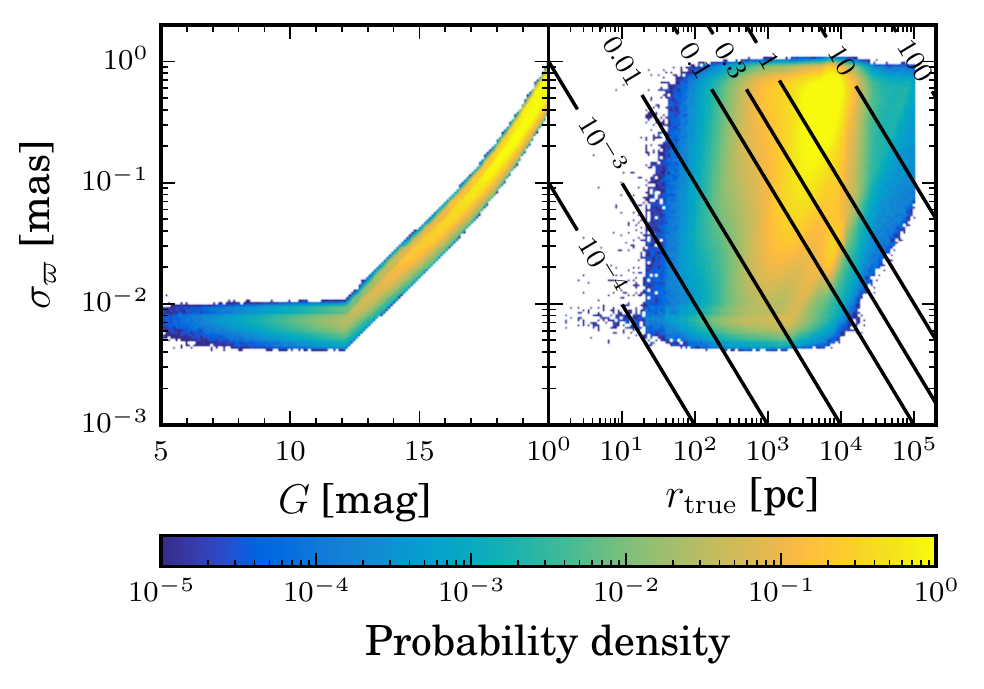}
\caption{The distribution of the parallax measurement errors $\errVarpi$ for simulated \gaia{} observations of the GUMS catalogue. 
The colour shows the probability density per unit log parallax error and per unit magnitude (left) or per unit distance (right).
The diagonal lines in the right panel show the locii of constant $\ftrue = \errVarpi\rtrue$ as indicated by the labels.}
\label{fig:errVarpi_GR}
\end{figure}

\begin{figure}
\includegraphics[width=\hsize]{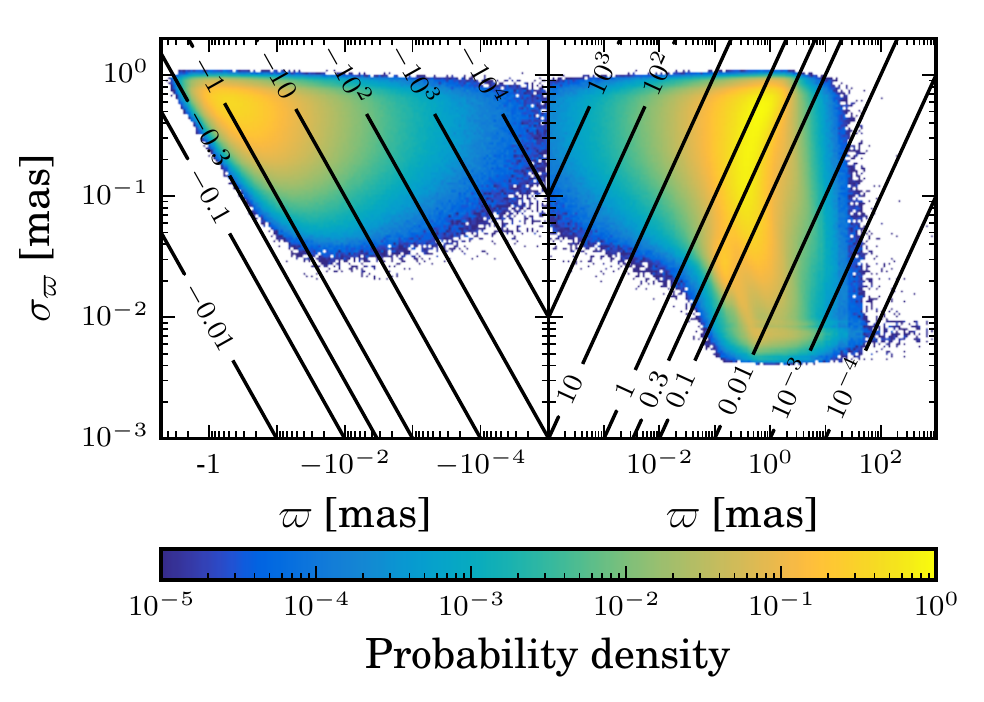}
\caption{The distribution of $\errVarpi$ as shown in Fig.~\ref{fig:errVarpi_GR},
but now as a function of the simulated measured parallaxes, $\varpi$.
The colour shows the probability density per unit log parallax error and per unit log parallax.
The left panel is for negative parallaxes while the right panel is for positive parallaxes. 
The diagonal lines show the locii of constant $\fobs = \errVarpi/\varpi$ as indicated by the labels.
For negative parallaxes, no stars have measurement errors less than 0.01\,mas.}
\label{fig:varpiSigmaVarpiDist}
\end{figure}

\subsection{The parallax noise model}
GUMS provides noise-free measurements. For our simulation we need to adopt a noise model for \gaia{}. We take into account the increased scattered light levels found after launch, which slightly reduce the astrometric accuracy at the faint end compared to pre-launch expectation.
Using the post-launch noise model from \cite{deb14}, the expected Gaussian one sigma uncertainty
in the parallax ($\sigma_\varpi$ in Eq.~\ref{eq:likelihood})
for a source of magnitude $G$ and colour $\vmini$, observed in 
$n_{\rm tr}$ transits of the \gaia{} focal plane is
\begin{multline}
\sigma_\varpi\;[\mu{\rm as}] = \left(\frac{\langle n_{\rm tr}\rangle}{p_{{\rm det},G}(G)n_{\rm tr}}\right)^{1/2}\times\\
 \left(-1.631 + 680.766z + 32.732z^2\right)^{1/2}\times\\ 
 \left[0.986 + (1 - 0.986)\vmini\right].
\label{eq:sigmaVarpi}
\end{multline}
where
\begin{equation}
z = \max\left(10^{0.4(12.09 - 15)}, 10^{0.4(G-15)}\right),
\label{eq:zDeb14}
\end{equation}
and $p_{{\rm det},G}(G)$ is the detection probability per transit (\gaia{} does real time detection independently on every transit). For a five year mission, $\langle n_{\rm tr}\rangle$ (the sky-averaged number of transits) is 70, which takes into account mission gaps and dead time. $n_{\rm tr}$ depends on the \gaia{} scanning law, and is shown as a function of $(l,b)$ for a five year mission in Fig.~\ref{fig:ntr}. $p_{{\rm det},G}(G)$ drops with fainter magnitudes, but is still 95\% at $G=20$ (Table~10 in \cite{jor10}).

Using the above equation, we calculate $\sigma_\varpi$ for every source in GUMS.
Fig.~\ref{fig:errVarpi_GR} shows the distribution as a function of magnitude and distance.
The spread in $\errVarpi$ for objects with the same $G$ is due primarily to the variable number of transits, 
and to a lesser extent due to colour variations across sources.

At the bright end the curve is largely independent of $G$. 
In reality there is still a dependence due to the activation of Time Delayed Integration (TDI) gates used to avoid saturation. This complication is ignored here and a constant parallax noise floor of $\errVarpi\sim 7$\,$\mu$as is used for sources brighter than $G\leq 12.09$.

In the right panel of Fig.~\ref{fig:errVarpi_GR} we see that almost all stars with $\rtrue\lesssim 100$\,pc will have $\ftrue=\errVarpi\rtrue$ below 0.1. The number of stars with $\ftrue\leq 0.1$ decreases quickly as we move to larger distances.
Only a very small fraction of stars with $\rtrue\gtrsim 10$\,kpc have $\ftrue<0.1$. Comparing the
left and right panels, we see that these stars are near the turn-off at $G\sim 12$: they are distant but bright (mostly giants), so relatively accurate parallaxes can still be obtained.

\subsection{Simulated observed catalogue}
\label{subsec:simu}
We now use the data in the GUMS catalogue $(l,b,\rtrue,G,V-I_C)$ together with the measurement model 
(the likelihood in Eq.~\ref{eq:likelihood}) and the specification of $\sigma_\varpi$ in
Eq.~\ref{eq:sigmaVarpi} and \ref{eq:zDeb14} to simulate the parallax for every star in the GUMS catalogue, by drawing a parallax at random from the likelihood.
Fig.~\ref{fig:varpiSigmaVarpiDist} shows the distribution of the parallax uncertainties again, but now as a function of these measured parallaxes. 
The right panel is for positive parallaxes. There is, of course, a very broad distribution in parallax errors for a given parallax, due to the wide range of actual distances and magnitudes of the stars. Stars with small measured parallaxes tend to be distant and faint, and so generally have larger parallax errors and likewise larger fractional parallax errors. The left panel shows negative parallaxes, which is approximately 15\% of all stars.
We can obtain negative parallaxes when $1/\rtrue$ is very close to zero and the parallax measurement error $\errVarpi$ is large. We see in the left panel that the minimum $\fobs$ for negative parallaxes is around 0.2.

Given the simulated parallaxes and corresponding uncertainties, we compute the posterior PDF over distance for each star, using the four priors described in Sect.~\ref{subsec:3priors} and \ref{subsec:mw}.
For the \ud{} and \usd{} priors, we set $\rlim = 100$\,kpc as the limiting distance. This is a distance large enough that covers all stellar objects within the GUMS catalogue. For the scale length $L$ of the \expp{}, we take four different values: $L = \{0.1, 0.5, 1.35, 5\}$\,kpc. We already saw in Fig.~\ref{fig:gums_rdist} that for $L=1.35$\,kpc the posterior closely matches the distribution of $\rtrue$ in the GUMS catalogue, but nevertheless in the case of a mismatch in the scale length we are interested in seeing how it affects the distance determination.

From the posteriors we estimate distances using the mode and median. To calculate the median, as well as a 90\% percent credible interval, the posterior is numerically integrated by sampling on a dense regular grid in $\log r$. The modes for the isotropic priors are given in Table~\ref{tab:priors}, while the mode of the \mw{} prior is found using a golden section search on the aforementioned grid, initialised on the point closest to the mode. We do not use the mean, as some posteriors are very skewed, for which it is unrepresentative.

\begin{figure*}
\includegraphics[width=\hsize]{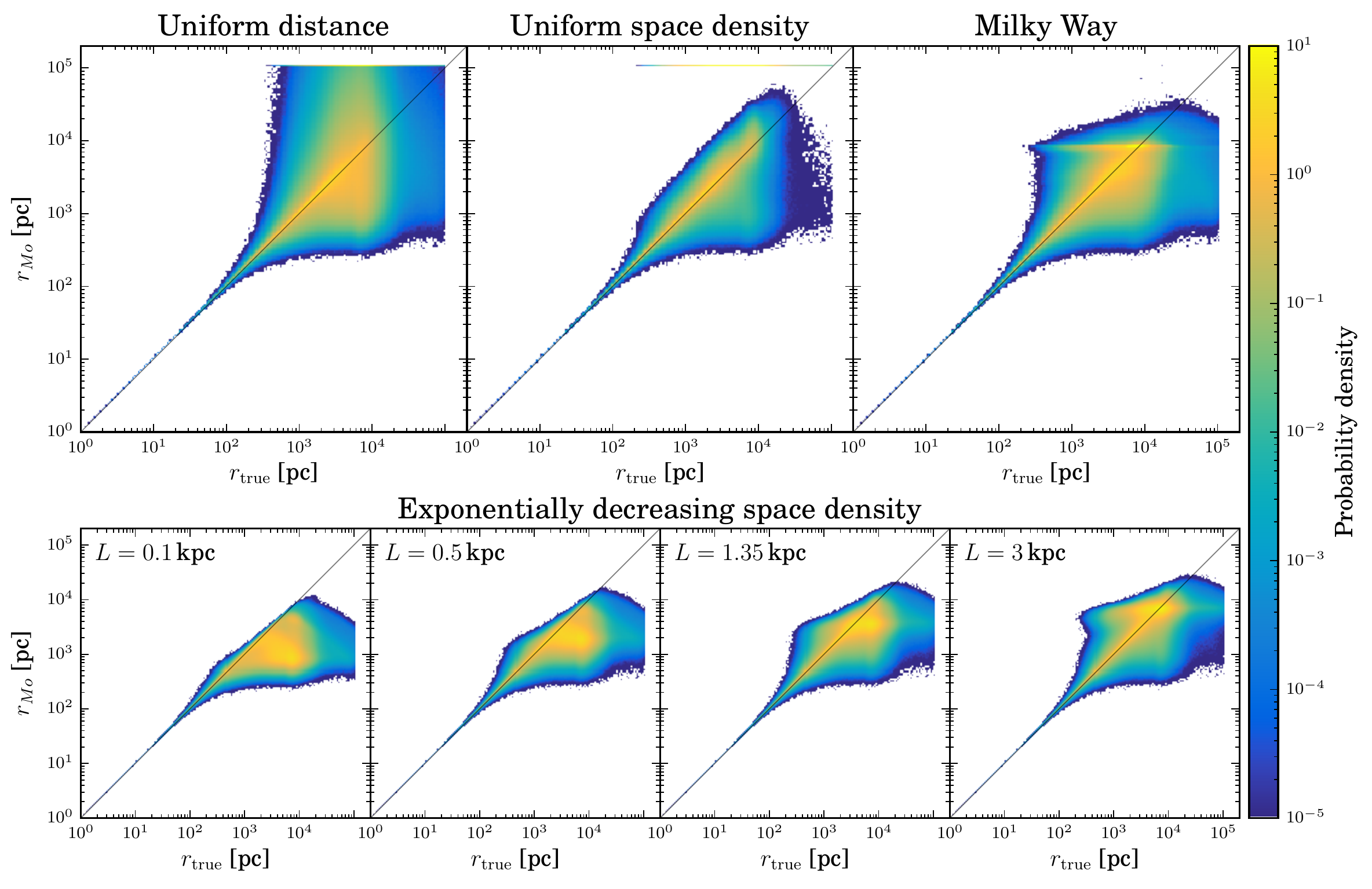}
\caption{Comparison of the median distance estimate $\rmed$ for all stars in the GUMS catalogue to their true distances $\rtrue$, using the \ud{} prior (top left), the \usd{} prior (top middle), the \mw{} prior (top right), and the \expp{} prior (four panels at the bottom) with four different values of scale length $L$ as indicated by the legend. The black diagonal line indicates a perfect match between the estimated and true values. The colour shows the probability density per unit log distance squared.}
\label{fig:comparisons_mo}
\end{figure*}
\begin{figure*}
\includegraphics[width=\hsize]{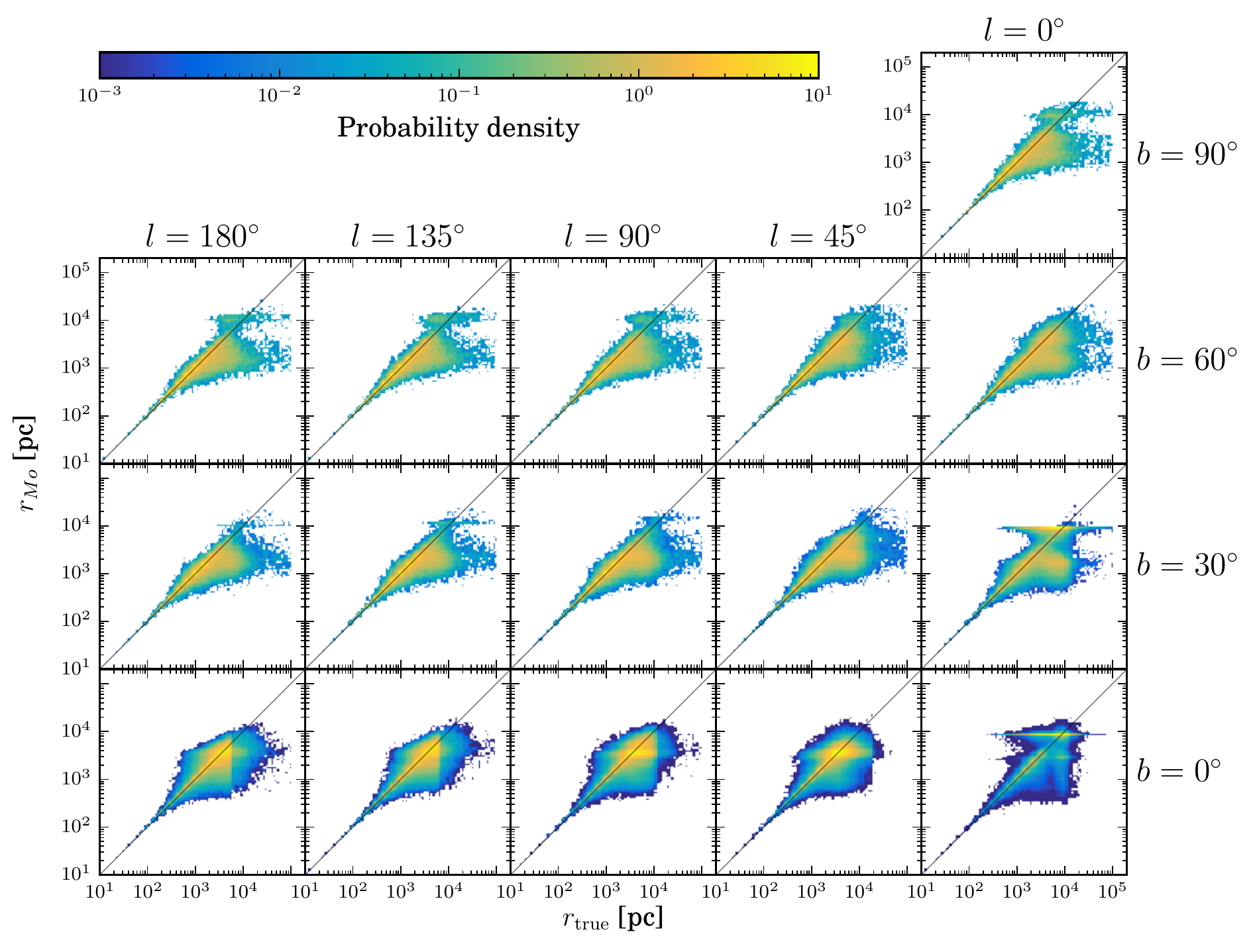}
\caption{A comparison between the mode distance $\rmode$ for all stars in the GUMS catalogue with their true distances $\rtrue$, using the \mw{} prior. Here the comparison is drawn for 16 selected LOS directions as previously used. The black lines indicate a perfect match between the estimated and true value. The colour shows the probability density per unit log distance squared. The sudden drop in density at $b=0^\circ$ and $l=90^\circ,\ 135^\circ$, and 180$^\circ$ is due to the termination of the disc in the GUMS catalogue, which is also observed in Fig.~\ref{fig:gums_rdist_directional} (q.v.).}
\label{fig:comparison_mo_mw_dir}
\end{figure*}
\begin{figure}
\includegraphics[width=\hsize]{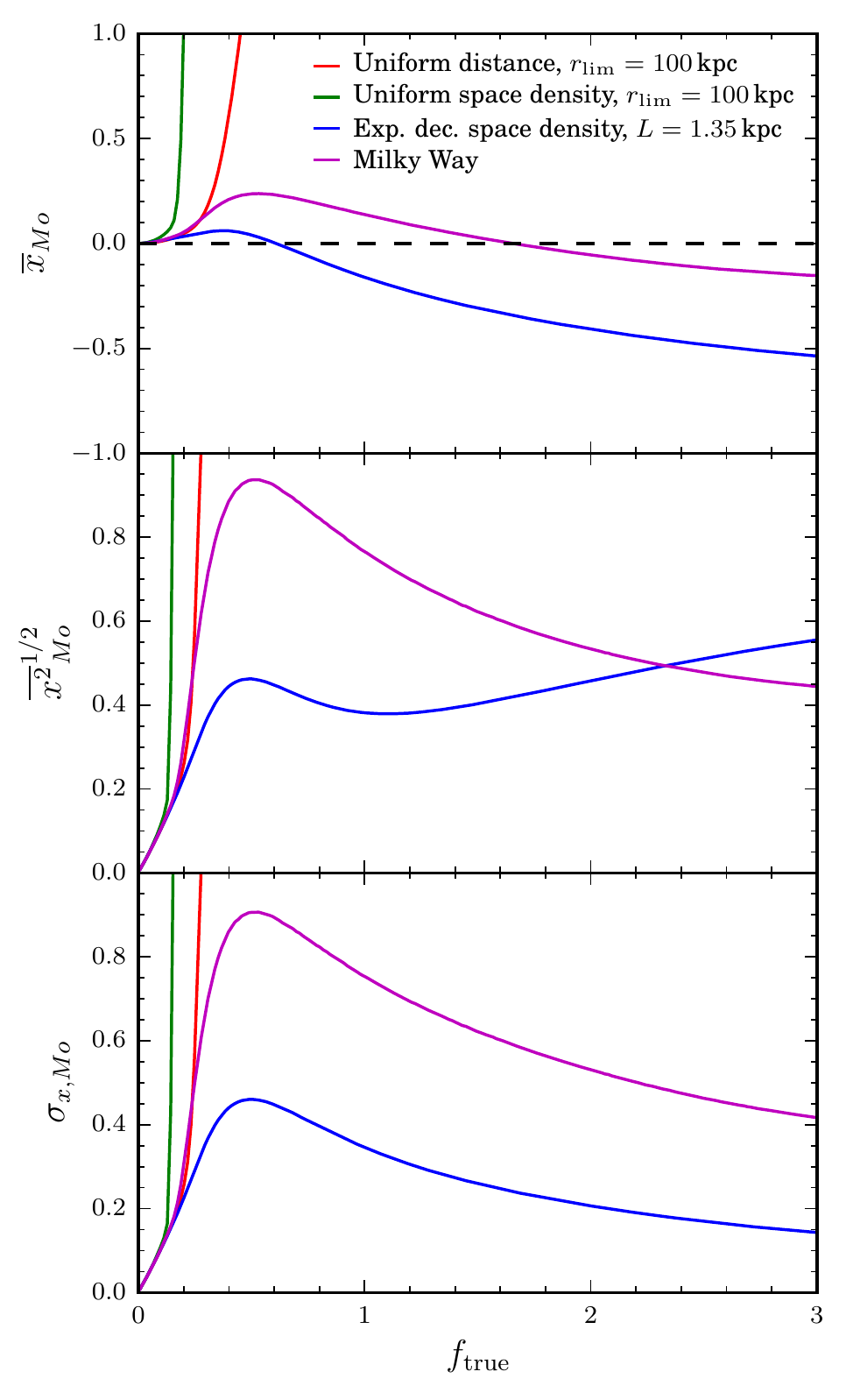}
\caption{Performance of the mode distance $\rmode$ for all priors (coloured lines) as a function of $\ftrue$. The three panels show the variation of the bias $\bias\mode$ (top), the RMS $\rms\mode$ (middle), and the standard deviation $\stddevmod$ (bottom).}
\label{fig:summary_performance_ftrue_mode}
\end{figure}
\begin{figure*}
\includegraphics[width=\hsize]{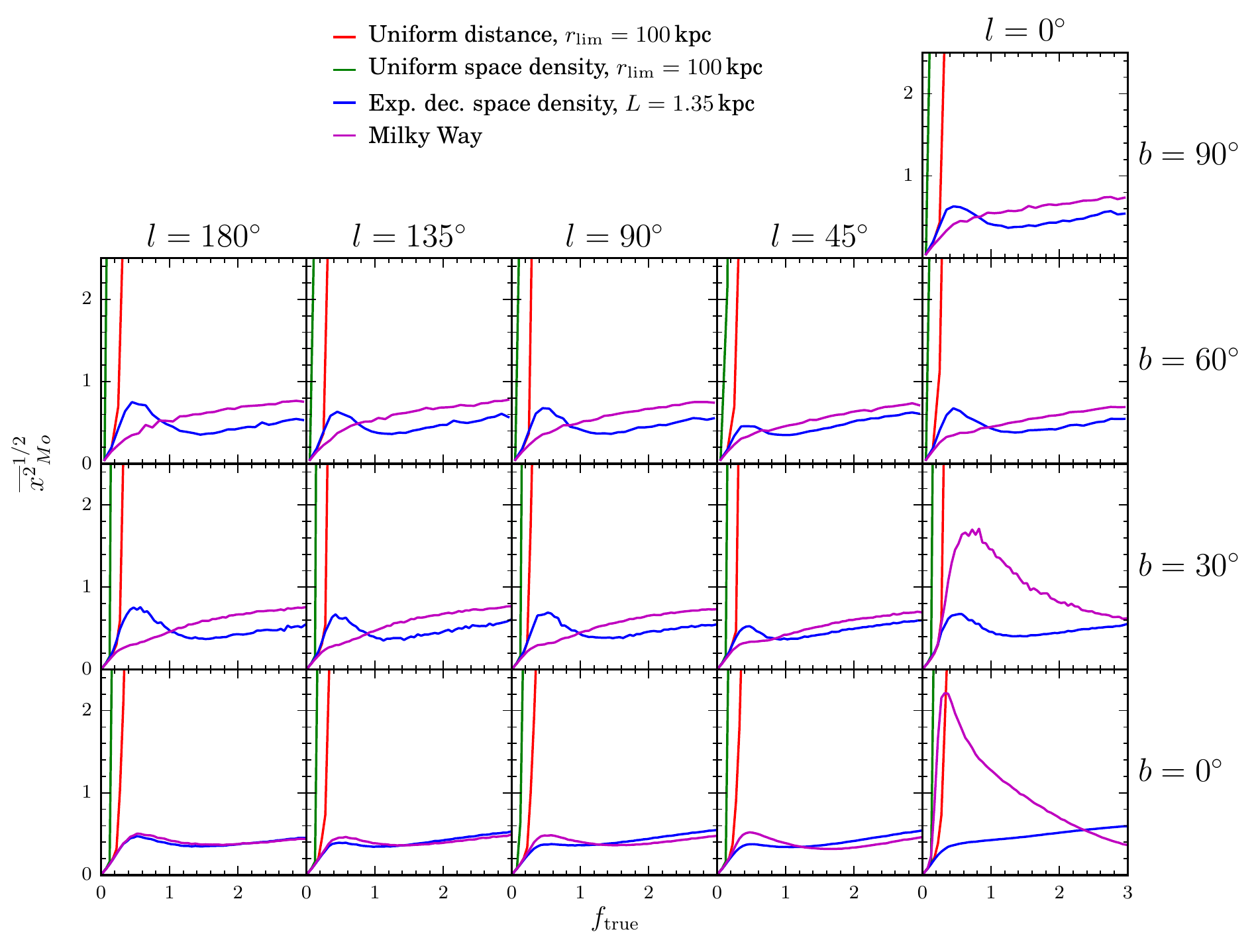}
\caption{The RMS of the scaled residuals using the mode, $\rms\mode$, for the 16 selected LOS directions for all four priors.}
\label{fig:summary_performance_ftrue_mode_dir}
\end{figure*}
\begin{figure}
\includegraphics[width=\hsize]{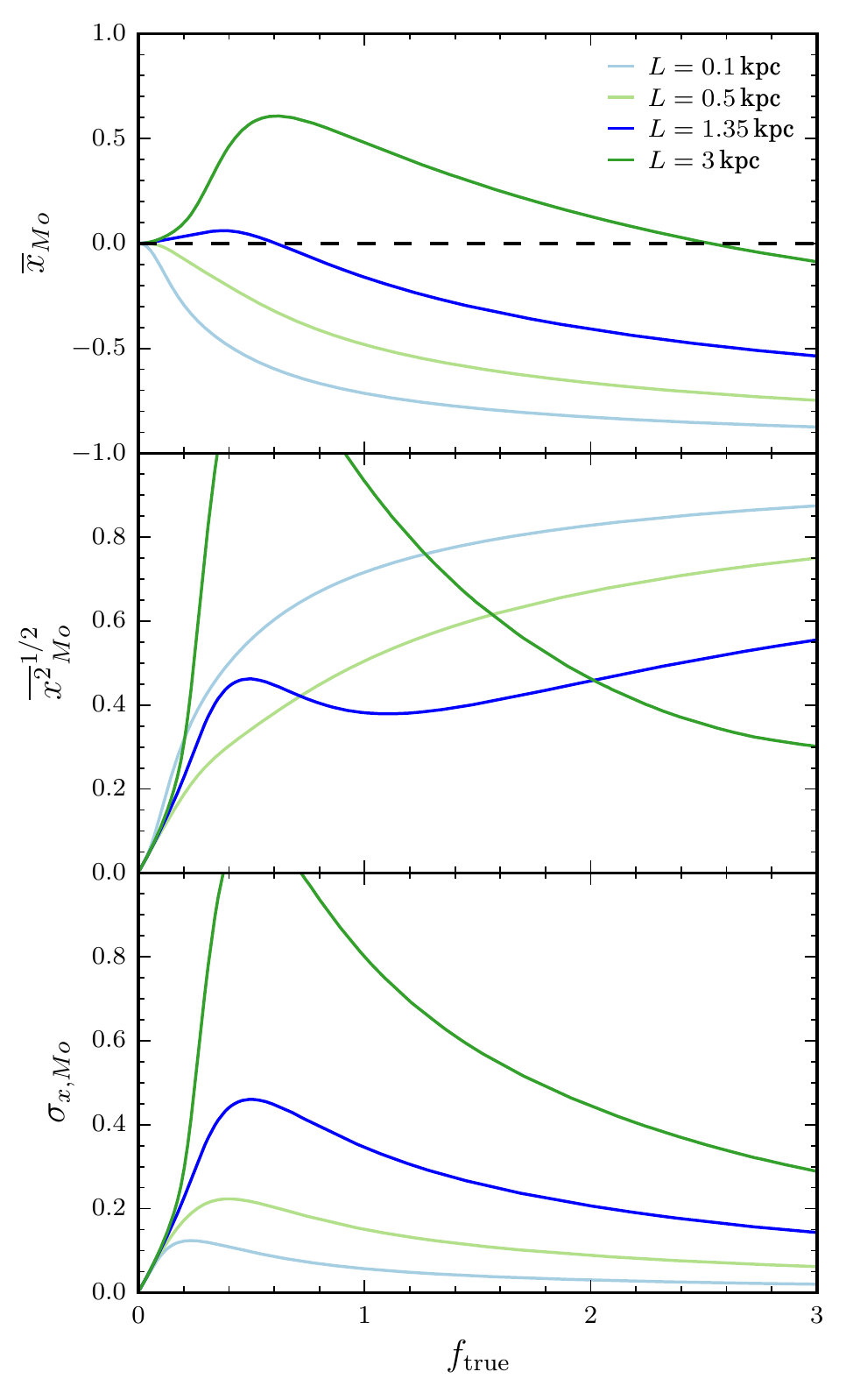}
\caption{As Fig.~\ref{fig:summary_performance_ftrue_mode}, but for the \expp{} prior with various scale length $L$. The dark blue line corresponding to $L=1.35$\,kpc is the same one as those in Fig.~\ref{fig:summary_performance_ftrue_mode}. Scale lengths other than $L=1.35$\,kpc generally give poorer performance, as these are a poorer match to the scale lengths in the Galaxy.}
\label{fig:summary_performance_ftrue_mode_exp}
\end{figure}
\section{Test results and discussion}
\label{sec:results}
In this section we analyse the performance of the priors. In the interest of brevity, here we only look at the mode since we found that overall the mode performs better than the median. The results from using the median are presented and discussed in Appendix~\ref{app:comparisons_md}.

\subsection{Comparison with the true values}
Fig.~\ref{fig:comparisons_mo} compares the mode distance estimate $\rmode$ with the true values $\rtrue$ for all stars, using each of the four prior. As we already saw in Fig.~\ref{fig:errVarpi_GR} (right panel), all stars within 100\,pc will have their distances measured extremely accurately for all priors. Furthermore, there is a high density around the diagonal line (note that the density is on a log scale), indicating that also at large distances a lot of stars achieve accurate distances with all priors. Beyond 10\,kpc, however, this high density around the diagonal line diminishes for all priors, as the number of stars with accurately measured parallaxes get rarer.

The spread in $\rmode$ arises because of the large range of values of $\ftrue$ and $\varpi$ present at a given value (or rather a small range) of $\rtrue$. The variation of this spread with $\rtrue$ is therefore complicated and cannot be easily explained, as it depends on a large number of factors, including the shape of the posterior and the complexity of distribution of stars in the GUMS models.

The distribution of the \ud{} prior is of particular interest, because the mode of the posterior PDF with the \ud{} prior is the inverse of the observed parallax $\varpi$ (provided it is not too small or negative). Although hard to see in the plot, there are many stars with $\rmode=100$\,kpc, which is the maximum distance $\rlim$ imposed by the prior. This is $\rmode$ for negative parallaxes and parallaxes with $1/\varpi>\rlim$. We also see the same cluster in the results using the \usd{} prior. Here there are more stars with $\rmode=100$\,kpc because not only stars with negative parallaxes have $\rmode=100$\,kpc, but also all stars with $\fobs>1/\sqrt{8}$. These aside (which are easily identified in the results), the \usd{} prior looks superior to the \ud{} prior, because the former has a smaller spread above the $\rmode=\rtrue$ line.

In the distribution of the \mw{} prior we see a horizontal elongation across $\rtrue$ at around $\rmode\sim 8$\,kpc. These are stars in the direction towards the galactic centre, with very high observed fractional parallax errors $\fobs$. For such stars, the posterior is practically the same as the prior (seen in Fig.~\ref{fig:posteriors} and \ref{fig:post_mw}), so the inferred distance is close to the mode of the prior, which in this case is around 8\,kpc (the distance to the galactic centre in this prior). We see some other elongations corresponding to the modes of the prior in other directions, but these are much weaker.

Similar elongation is also observed in the distribution of the \expp{} prior---shown in the bottom row of Fig.~\ref{fig:comparisons_mo}---at the mode of this prior, $\rmode=2L$.

Since the \mw{} prior has different shapes depending on the LOS direction, we show in Fig.~\ref{fig:comparison_mo_mw_dir} the $\rmode$ vs.\ $\rtrue$ comparison for the same 16 LOS directions used previously (cf. Fig.~\ref{fig:gums_rdist_directional}). Looking at the distribution, it is more clear that there are elongations of stars with poorly measured parallaxes around a preferred mode. At directions away from the galactic center, however, the elongations are closer to the true distance of most stars, indicating that the prior can help in estimating distances of stars with poor parallax measurements, provided that there are reasonably good match between the prior and the underlying true distribution. We have seen that we have a good match between the prior for the thin disc and its corresponding ``true'' distribution in GUMS, and this is advantageous for the prior as a whole since most stars are thin disc objects. In directions around the galactic centre, however, at $(l,b) = (0\degr,0\degr)$ and $(l,b) = (0\degr,30\degr)$, the elongations are relatively more pronounced, which is due to the prior mismatch in this direction. Note the high density around the galactic centre.

\subsection{Statistical performance as a function of $\ftrue$}
We evaluate the performance of the mode distance estimator $\rmode$ as a function of $\ftrue$. We divide the data into bins of $\ftrue$, then for every object, $i$, in the bin we calculate the scaled residual 
\begin{equation}
x_{Mo,i} = \frac{r_{Mo,i} - r_{{\rm true},i}}{r_{{\rm true},i}} \ .
\label{eq:x}
\end{equation}
For each bin we then calculate the bias and standard deviation as explained in Paper I, as well as the root mean square (RMS) of the scaled residual.
The standard deviation measures the scatter about the bias, whereas the RMS measures the entire error including the bias. The variation of these metrics with $\ftrue$ for the different priors is shown in Fig.~\ref{fig:summary_performance_ftrue_mode}. We can immediately conclude from this plot that both the \ud{} and \usd{} priors performs badly. The latter, for example, becomes heavily biased even at a low values of $\ftrue$ ($\sim$0.2), and the errors (in terms of RMS and standard deviation $\stddevmod{}$) rise rapidly too. Contrary to our expectations, however, the \mw{} prior does not show the best overall performance; rather it is the \expp{} prior with $L=1.35$\,kpc. We see that for large $\ftrue$, the standard deviation of the scaled residual $\sigma_{x,Mo}$ can be as low as $\sim$0.14, and the peak deviation is $\sim$0.46 while for the \mw{} prior its peak deviation is $\sim$0.90, and that occurs at a relatively low $\ftrue$ of $\sim$0.47.

We should be careful in interpreting these results, however. For $\ftrue\gtrsim 1$, the \expp{} prior gives a negative bias which increases in magnitude with $\ftrue$, while the bias in the \mw{} prior decreases towards zero. The large negative bias of the former prior arises because of its isotropy and chosen scale length $L$ of 1.35\,kpc, which produces distance estimates for poorly measured parallaxes all at around $2L=2.7$\,kpc. For a majority of stars with large $\ftrue$, this is far from their true distance. This is the unavoidable consequence of choosing a simple prior. The RMS includes the effect of this bias, and we see in the central panel of Fig.~\ref{fig:summary_performance_ftrue_mode} the RMS of the \expp{} prior increases with $\ftrue$ at larger values of $\ftrue$, whereas the \mw{} prior does not.

One must also appreciate that Fig.~\ref{fig:summary_performance_ftrue_mode} is the performance averaged over all directions in the Galaxy, whereas the \mw{} prior has a directional dependence. In Fig.~\ref{fig:summary_performance_ftrue_mode_dir} we show the RMS as a function of direction for all priors. For directions away from the galactic centre, the \mw{} prior performs better than the \expp{} for $\ftrue\lesssim 1$. In contrast, the \mw{} prior has worse performance in directions toward the galactic centre. As expected, this is the effect of a mismatch between the bulge model as a prior and the ``true'' bulge used in the simulated observations from GUMS (cf. Fig~\ref{fig:gums_rdist_directional}). We can see that away from the galactic centre, the \mw{} prior actually performs best. However, due to the high number of stars in the directions around the galactic centre, the poor performance of the \mw{} prior in Fig.~\ref{fig:summary_performance_ftrue_mode} is dominated by stars in these directions. This should be taken into consideration when using the \mw{} prior.

In Fig.~\ref{fig:summary_performance_ftrue_mode_dir}, we can also see that the \expp{} prior with $L=1.35$\,kpc performs best in directions at the galactic plane, as the chosen scale length is a good match with the ``true'' thin disc used in GUMS. Variations in the behaviour of the \expp{} prior between the panels are due entirely to the different distribution of stars in GUMS, as the prior itself is isotropic.

In terms of standard deviation, the \expp{} prior performs better than the \mw{} prior. In terms of bias the relative performance depends on $\ftrue$. Overall the \expp{} prior with $L=1.35$\,kpc is the best performing prior for $\ftrue\lesssim 2.5$ (which is already a very large error). We show the results of this prior only for $L=1.35$\,kpc because it is the overall best performing (both in averaged and directional performance) among the four scale lengths $L$ investigated, as expected (cf. Fig.~\ref{fig:gums_rdist_directional}. Other scale lengths, however, performs better in different regime of $\ftrue$. This can be seen in Fig.~\ref{fig:summary_performance_ftrue_mode_exp}. For $\ftrue\lesssim 0.6$, those with $L=0.5$\,kpc performs best, while for $\ftrue\gtrsim 2$ it is $L=3$\,kpc.

Note that all of these plots tell us only the performance at a given $\ftrue$, but not how many stars actually have that value of $\ftrue$. This can be seen in Fig.~\ref{fig:fObsDist}, and should be taken into mind when comparing performances.

\begin{figure}
\includegraphics[width=\hsize]{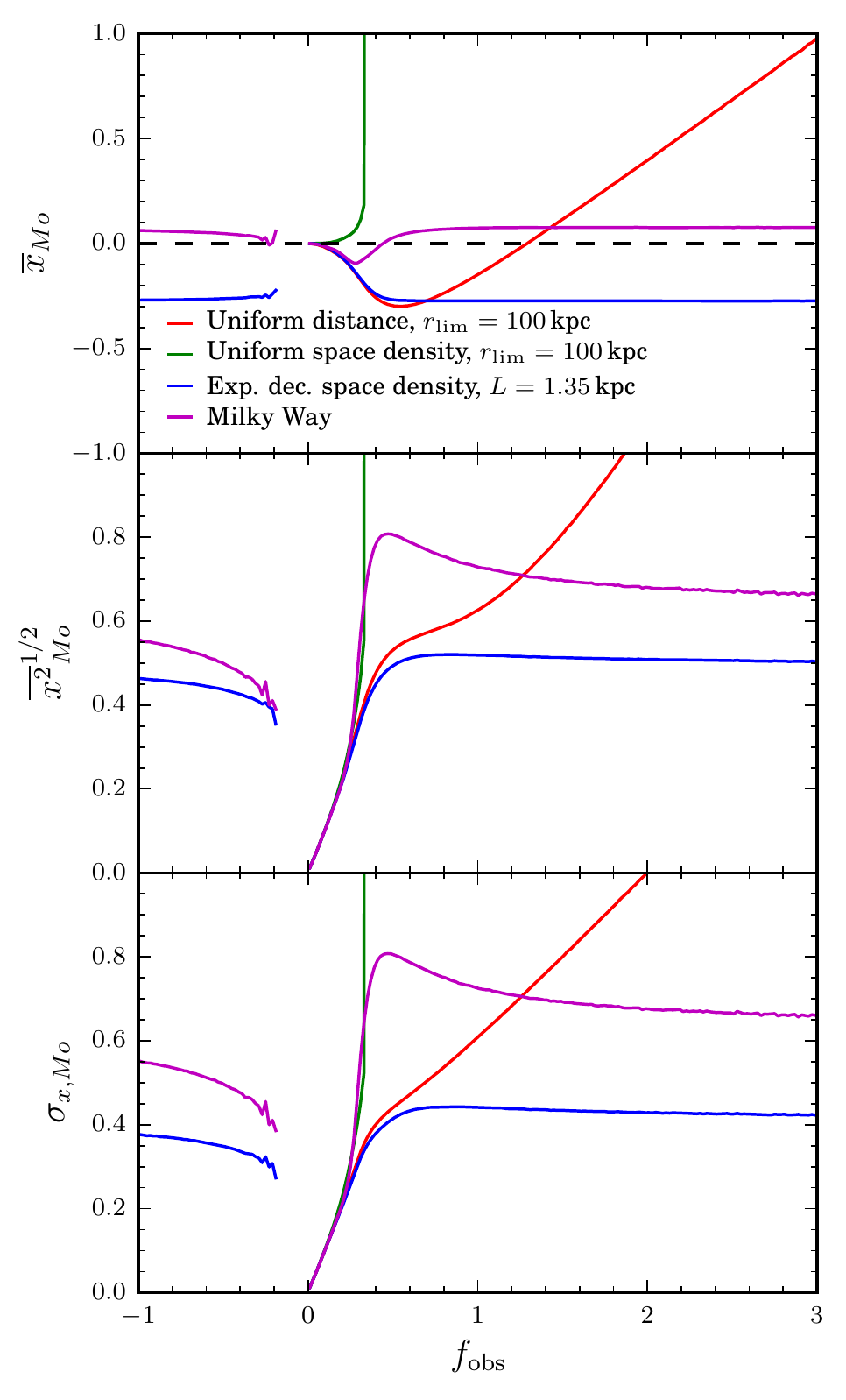}
\caption{As Fig.~\ref{fig:summary_performance_ftrue_mode}, but now plotted as a function of $\fobs=\sigma_\varpi/\varpi$. Note that now the horizontal axis extends to negative values.}
\label{fig:summary_performance_fobs_mode}
\end{figure}
\begin{figure}
\includegraphics[width=\hsize]{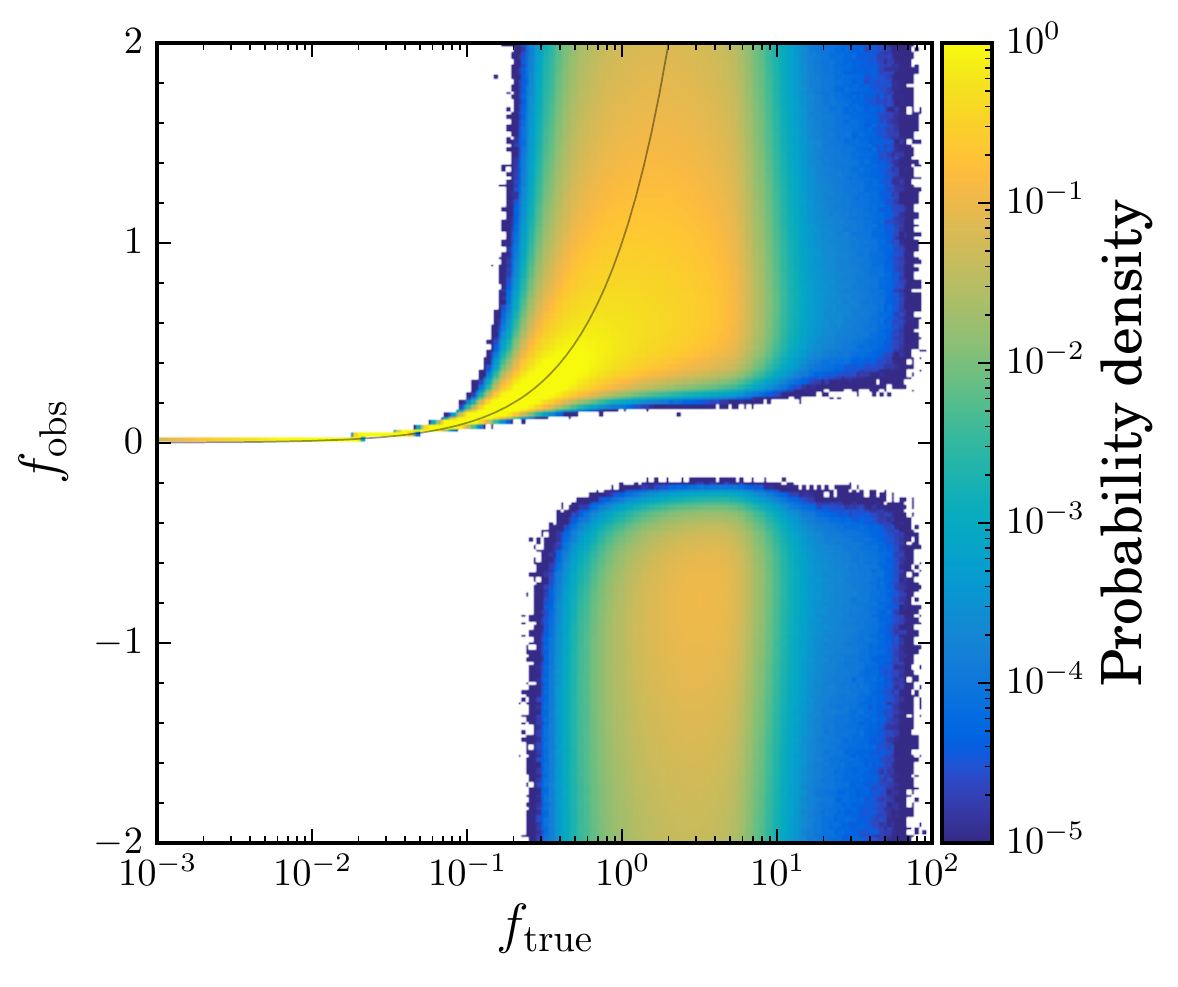}
\caption{The distribution of $\fobs$ and its corresponding $\ftrue$. Note that $\ftrue$ is plotted in log scale, while $\fobs$ is in linear scale and extend to negative values. The black line indicates equal values of $\fobs$ and $\ftrue$. The colour shows the probability density per log $\ftrue$ per $\fobs$.}
\label{fig:fObsTrueDist}
\end{figure}
\begin{figure*}
\includegraphics[width=\hsize]{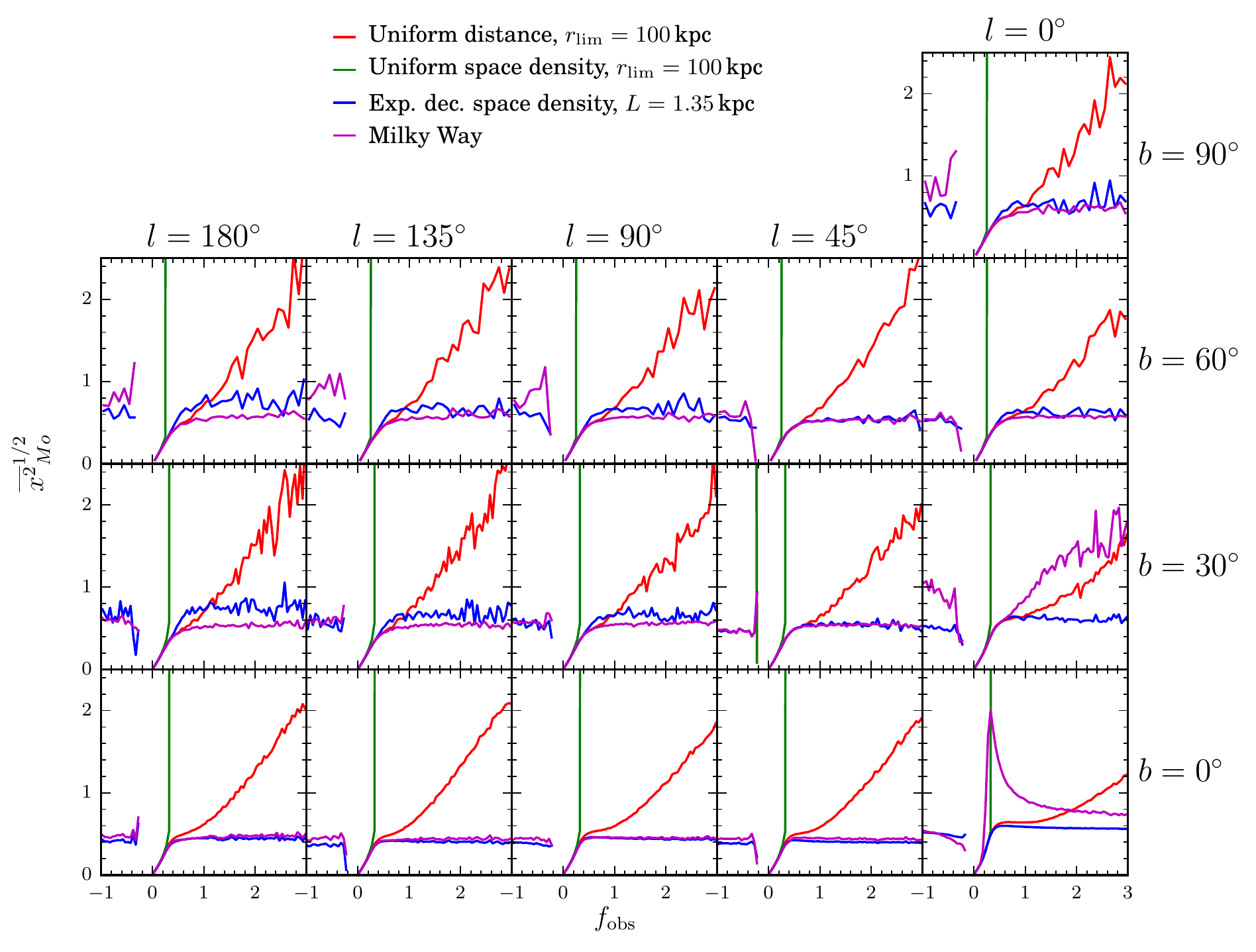}
\caption{As Fig.~\ref{fig:summary_performance_ftrue_mode_dir}, but now plotted against the observed fractional parallax error $\fobs=\sigma_\varpi/\varpi$.}
\label{fig:summary_performance_fobs_mode_dir}
\end{figure*}
\begin{figure}
\includegraphics[width=\hsize]{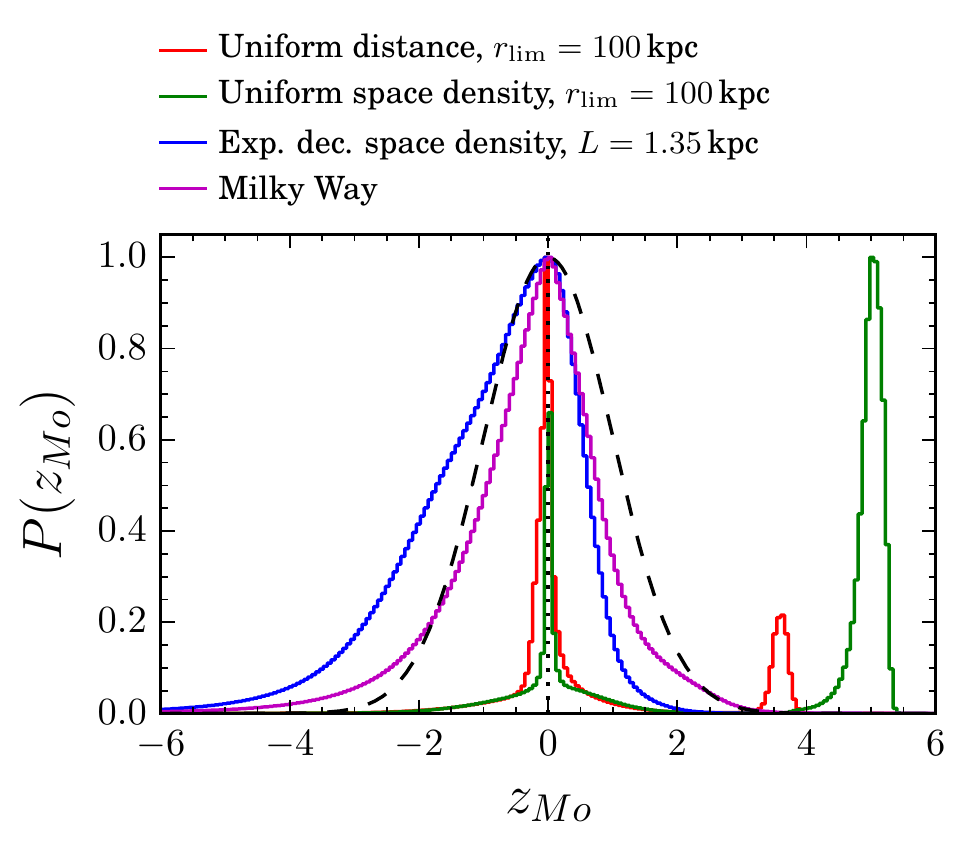}
\caption{The distribution of the standardised residual $z_{\mode}$ (defined in Eq.~\ref{eqn:standardized_residual}), for all priors. The black dashed line indicates the standard normal distribution.} All distributions are scaled to have their maxima at one.
\label{fig:zhist}
\end{figure}
\subsection{Statistical performance as a function of $\fobs$}
In the previous section we showed the performance of the distance estimators as a function of $\ftrue$. These plots are useful for predicting the expected peformance in terms of the {\em expected} fractional parallax errors. But as the true distances are not known in a real application, we will also want to know, for a given {\em measured} fractional parallax error, $\fobs=\sigma_\varpi/\varpi$, what the expected performance is (We would still not know the true performance, of course). Note that $\sigma_\varpi$ is not measured as such---we do not know the true uncertainty---but rather is estimated from a noise model.
Fig.~\ref{fig:summary_performance_fobs_mode} shows the performance of the mode now as a function of $\fobs$, averaged over all directions. This figure includes negative values of $\fobs$ because observed parallaxes can be negative. This is quite a different plot, because the horizontal axis now includes the noisy measurement.

Comparing this to the similar plot against $\ftrue$ (Fig.~\ref{fig:summary_performance_ftrue_mode}) we instantly see a significant difference: When plotted against $\fobs$, the \ud{} prior performs much better for positive parallaxes, and in fact as well as or even better than the \mw{} prior for $\fobs\lesssim 1.2$. The \ud{} prior (and \usd{} prior) remains useless for negative parallaxes, however, because the mode is then always $\rlim=100$\,kpc.

The reason for this change in behaviour for the \ud{} prior is two-fold. First, some stars which previously had smaller values of $\ftrue$ will, due to the noise, achieve larger (absolute) values of $\fobs$. Hence the good performance we previously saw at low $\ftrue$ is partly distributed to larger values of $\fobs$
(Noise will also give some stars even lower values of $\fobs$, but we already saw good performance at low $\ftrue$).
This can also happen for the other priors, of course, but these did not previously have such terrible performance at larger values of $\ftrue$, so the effect is less noticeable. That this does not occur for the \usd{} prior is because for $\fobs$ larger than about 0.3, the mode is always at $\rlim$, which is a very poor estimate for most stars. The \ud{} prior, in contrast, at least has a variable mode in those cases ($1/\varpi$), even if it is often a poor and biased estimate (given that $\varpi$ is noisy). 
When we were plotting the performance against $\ftrue$, the bin can contain stars with negative $\fobs$. In Fig.~\ref{fig:fObsTrueDist} we show the distribution of $\fobs$ and the corresponding $\ftrue$. We see that negative $\fobs$ can even be drawn already from $\ftrue\sim 0.2$. This is the second reason for the \ud{} prior giving better results when plotted against $\fobs$. Stars with negative parallaxes no longer appear in this plot: the line is off to the top left of all the panels in Fig.~\ref{fig:summary_performance_fobs_mode}. 

At negative parallaxes (negative $\fobs$), we see that the \mw{} prior is less biased than the \expp{}, but has a larger standard deviation. This is what we might expect, given that in this regime the prior dominates.

As mentioned before, these summary plots do not tell us how many stars there are at each value of $\fobs$. This can be seen in Fig.~\ref{fig:fObsDist}.

Fig.~\ref{fig:summary_performance_fobs_mode_dir} shows the directional dependence of the mode estimator now as a function of $\fobs$ (cf. Fig.~\ref{fig:summary_performance_ftrue_mode_dir}). The plots are noisy at large $\fobs$ for at high latitudes because of the paucity of stars. We now see that the \mw{} performs better than any other prior, except for directions toward the galactic centre. This is the same behaviour as in Fig.~\ref{fig:summary_performance_ftrue_mode_dir}.

Overall the \expp{} prior performs best, but for directions away from the galactic centre the \mw{} prior performs best. In general we cannot recommend using the \ud{} or \usd{} priors.

\subsection{The formal errors}
\label{subsec:formalErrors}
Just as important as achieving accurate distance estimates is estimating accurate uncertainties in those estimates.
An estimate without an error bar is essentially useless. Here we use the 90\% credible interval $[r_{5},r_{95}]$ of the posterior PDF computed symmetrically about the median (i.e.\ the difference between the 95\% and 5\% quantiles). As many people are still used to think about Gaussian-like one sigma errors, we convert this interval into what would be a one sigma standard deviation if the PDF were Gaussian, by scaling the computed 90\% interval to the 68.3\% interval (this also capture the shape of the tailing distributions). We calculate
\begin{equation}
\sigma_r = \frac{\left(r_{95}-r_{5}\right)}{2s}
\label{eqn:standardized_residual}
\end{equation}
where $s = 1.645$ is the ratio of the 90\% to 68.3\% credible interval in a Gaussian. We then calculate the standardised residuals
\begin{equation}
z\mode = \frac{\rmode-\rtrue}{\sigma_r} \ .
\end{equation}
{\em If} the residuals were Gaussian, we would expect this quantity to have a distribution which is a standardised Gaussian. In any case, for given residuals, values of $z$ much larger than one imply that the uncertainties---as estimated by our credible interval---are underestimated, and values of $z$ much less than one imply that they are overestimated.

Fig.~\ref{fig:zhist} shows the distribution of $z$ for all the priors. The distribution of $z$ for the \ud{} and the \usd{}
have a component which is centered around zero and very narrow, the latter suggesting that the uncertainties are overestimated. But they are both bimodal with significant peaks at large positive values of $z$. These correspond to the stars with large fractional parallax errors or negative parallaxes, when the mode is pushed to $\rlim$. 
We can now see much better how many stars are concentrated into those narrow lines at $\rmode=\rlim$
in Fig.~\ref{fig:summary_performance_ftrue_mode_dir} for these two priors.
Thus these priors only achieve unbiased results for a limited fraction of all objects, and even a posteriori we do not know which of the objects with larger $\fobs$ these are.

The huge secondary peak of the \usd{} prior indicates that the mode is a poor estimator for the \usd{} prior.

The distribution for the \expp{} prior has its mode at $z=0$, but is skewed towards negative $z$.
This is just the negative bias we discussed before. But as the half-width at half-maximum (HWHM) of the distribution is 1.17, which is
close to 1.2 (the value for a Gaussian). At least the uncertainty estimates are reasonable.

The distribution for the \mw{} prior is in many ways much closer to what we want for our estimators: centered on zero and symmetric. Its HWHM is 0.84,
which however suggests that the uncertainties might be overestimated. But we must remember that they have been calculated from a 90\% credible interval, so we do not expect the standardised residuals to follow a standardised Gaussian.

Overall, we conclude that although when averaged over all directions the \mw{} prior does not perform as well as the \expp{} prior in terms of distance estimation, it does performs better in terms of uncertainty estimates. This we see from the distribution of the residuals, which centered on zero and is symmetric (see Fig.~\ref{fig:zhist}). The \ud{} and \usd{} uncertainty estimates are basically useless.

\section{Adding photometric measurements to improve the distance inference}
\label{sec:mwhrd}

Once the fractional parallax error increases above about 100\%, distance estimation with any prior will be inaccurate.
For some types of stars we can get better distance estimates from their colours or spectra, together with photometry, using what are sometimes called photometric distances. This relies on prior information from stellar structure and evolution models, which predict absolute magnitudes based on the colours or spectra. Combined with the measured photometry, and correcting for extinction, a distance estimate can be obtained. For some types of stars, such as red-clumps and giants, distance accuracies of up to respectively 5\%--10\% \citep{bov14} and 20\% \citep{san16} are achieved.

In addition to the $G$-band magnitude, \gaia{} will also observe all stars using low resolution spectrophotometry obtained in the blue and red. These will be used to estimate stellar parameters \citep{cbj13} which in turn can give us photometric distances. \gaia{} will also provide the integrated fluxes from these to produce the $\gbp$ and $\grp$ magnitudes, which are essentally very broad blue and red filters \citep{jor10}. For simplicity we will just examine here how the use of the three magnitudes $G$, $\gbp$, and $\grp$, together with a universal HRD (the prior information on stellar models) can be used to improve distance estimates. Using the full spectra should ultimately improve this further.

Let $\mathbf{X} = (\varpi, G, G-\gbp, G-\grp, \gbp-\grp)$ denote the set of measurements we have.
From these we want to determine the distance, but the measurements are also affected by the absolute magnitude, $M_V$, the intrinsic colour, $\vmini$, as well as the interstellar extinction, $A_V$, so we will need to marginalise over these. We denote these parameters as $\mathbf{\Theta} = (r, A_V, M_V, \vmini)$. 
Bayes' theorem tells us that the posterior PDF over these parameters is
\begin{equation}
\posterior = \frac{1}{Z}\llf\prior,
\label{eq:post_mwhrd}
\end{equation}
where $Z$ is the normalisation constant, $\llf$ is the likelihood to observe the data given the parameters, and $\prior$ is the prior probability of the parameters. 
The marginalised posterior over a single parameter of interest $\Theta_i$ is
\begin{equation}
P(\Theta_i|\mathbf{X}) = \int \posterior \, d\Theta_1\ldots d\Theta_{i-1}d\Theta_{i+1}\ldots d\Theta_n \ .
\end{equation}
For simplicity, we consider the priors to be independent
\begin{equation}
P(\mathbf{\Theta}) = P(r)P(A_V)P(\vmini,M_V) \ .
\end{equation}
$P(\vmini,M_V)$ is the HRD. One could of course adopt more complex priors and additional parameters.

\subsection{The likelihood and forward model}

We will adopt a covariant Gaussian likelihood with mean $\mathbf{X}$ and covariance $\mathbf{\Sigma}$, i.e.\ \begin{equation}
\llf \propto \exp\left(-\frac{1}{2}\left[\mathbf{X}-\mathbf{f}(\mathbf{\Theta})\right]^T\mathbf{\Sigma}^{-1}\left[\mathbf{X}-\mathbf{f}(\mathbf{\Theta})\right]\right) \ .
\end{equation}
$\mathbf{f}(\mathbf{\Theta})$ is the forward model predicting the observations given the parameters:
\begin{equation}
\mathbf{f}(\mathbf{\Theta}) = 
\left(
\begin{array}{c}
1/r\\[0.5em]
M_V + 5\log_{10}r - 5 + A_V + p_0\left[\vmini_{\rm obs}\right]\\[0.5em]
p_1\left[\vmini_{\rm obs}\right]\\[0.5em]
p_2\left[\vmini_{\rm obs}\right]\\[0.5em]
p_3\left[\vmini_{\rm obs}\right]
\end{array}
\right) \ .
\end{equation}
$\vmini_{\rm obs}$ is the observed colour, which is related to the intrinsic $\vmini$ colour (parameter) by
\begin{equation}
\vmini_{\rm obs} = \vmini + A_V\left(1-\frac{A_I}{A_V}\right) \ ,
\end{equation}
where the ratio $A_I/A_V = 0.565$ was calculated using the \cite{fit99} extinction curve and is assumed to be constant throughout. The $p_i$ functions are polynomials which transform the $\vmini$ colours into colours in the \gaia{} photometric system, and are of the form
\begin{equation}
p_i[\vmini] = c_0 + c_1\vmini + c_2\vmini^2 + c_3\vmini^3 \ .
\end{equation}
The values of the coefficients are taken from \cite{jor10} and are tabulated in Table~\ref{tab:poly}.
\begin{table}
\caption{The coefficients of the transformation polynomials $p_i[\vmini]$, from \cite{jor10}.}
\label{tab:poly}
\begin{center}
\begin{tabular}{clrrrr}
\hline\hline
\multicolumn{1}{c}{$i$} & \multicolumn{1}{c}{$p_i[\vmini]$} & \multicolumn{1}{c}{$c_0$} &\multicolumn{1}{c}{$c_1$} &\multicolumn{1}{c}{$c_2$} &\multicolumn{1}{c}{$c_3$}\\
\hline
0 & $G-V$ & -0.0257 & -0.0924 & -0.1623 & 0.0090\\
1 & $G-\gbp$ & 0.0387 & -0.4191 & -0.0736 & 0.0040\\
2 & $G-\grp$ & -0.0274 & 0.7870 & -0.1350 & 0.0082\\
3 & $\gbp-\grp$ & -0.0660 & 1.2061 & -0.0614 & 0.0041\\
\hline
\end{tabular}
\end{center}
\end{table}

The covariance matrix $\mathbf{\Sigma}$ is constructed assuming that the parallax and photometric measurements are independent. The three photometric measurements are independent, but the colours formed from them are not. Thus the covariance matrix is
\begin{equation}
\mathbf{\Sigma} = 
\left(
\begin{array}{ccccc}
\sigma_\varpi^2 & 0 & 0 & 0 & 0\\[0.5em]
0 & \sigma_G^2 & \sigma_G^2 & \sigma_G^2 & 0\\[0.5em]
0 & \sigma_G^2 & \sigma_G^2+\sigma_{\gbp}^2 & \sigma_G^2 & -\sigma_{\gbp}^2\\[0.5em]
0 & \sigma_G^2 & \sigma_G^2 & \sigma_G^2+\sigma_{\grp}^2 & \sigma_{\grp}^2\\[0.5em]
0 & 0 & -\sigma_{\gbp}^2 & \sigma_{\grp}^2 & \sigma_{\gbp}^2+\sigma_{\grp}^2\\[0.5em]
\end{array}
\right) \ .
\end{equation}
To ensure that this matrix is not singular, we use regularisation for the inversion. We found that an offset of the order of 0.1 milimagnitudes in the photometric errors suffices, and has no relevant impact on the results.

\begin{figure}
\includegraphics[width=\hsize]{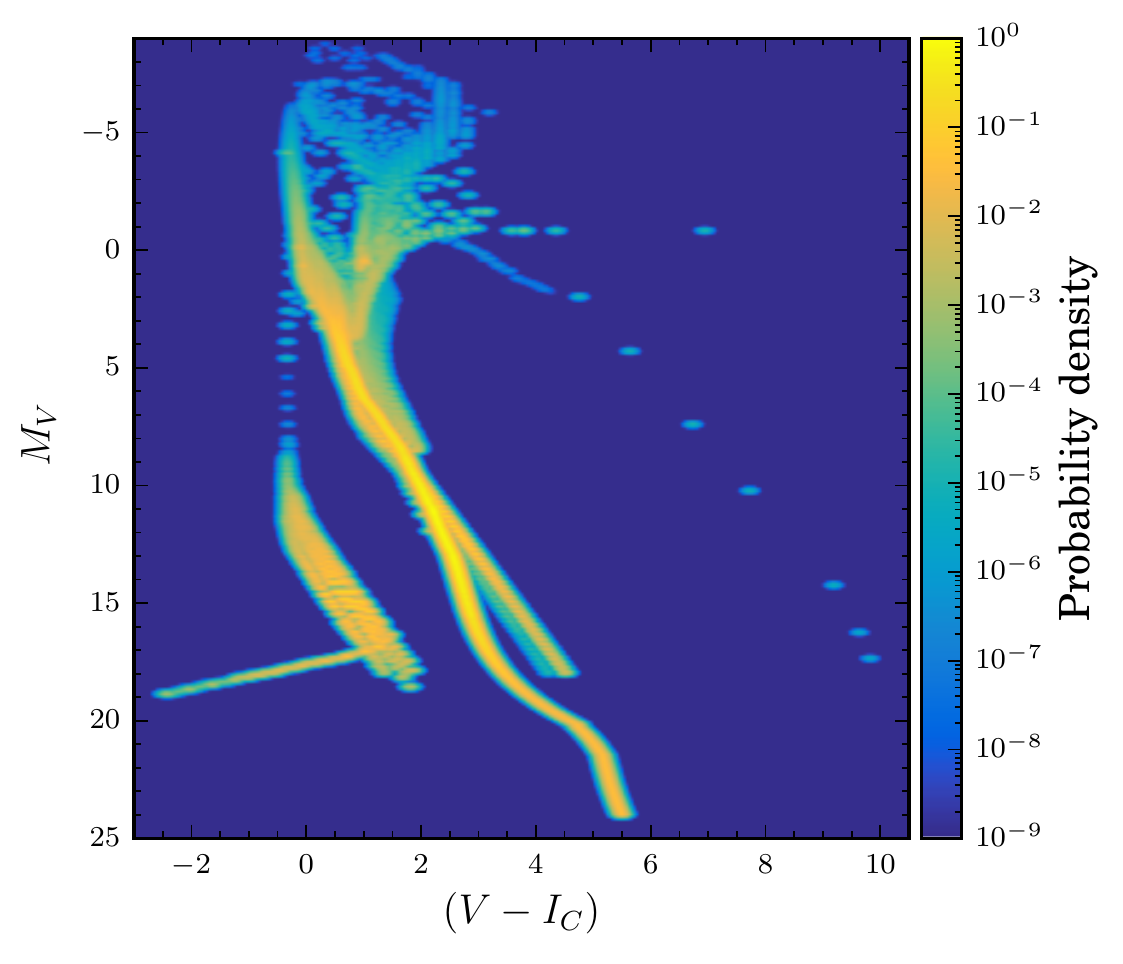}
\caption{The HRD prior $P(\vmini,M_V)$. The colour scale indicates the probability density and is normalised to unity. The smallest value is $1.25\times 10^{-9}$,
which is the value (post-normalization) of the offset added to avoid zero probability.}
\label{fig:hrd}
\end{figure}

\subsection{The priors}
For the prior on distance $P(r)$ we use again the \expp{} prior discussed in Sect.~\ref{subsec:3priors}, using $L=1.35$~kpc, as well as the Milky Way density model $\rho_{\rm MW}(r,l,b)$ discussed in Sect.~\ref{subsec:mw}. The extinction prior $P(A_V)$ is a Gaussian centered on $A_V^{\rm map}(r,l,b)$ with standard deviation $\sigma_{A_V} = 0.1A_V^{\rm map}$, where $A_V^{\rm map}(r,l,b)$ is the extinction calculated from the \cite{dri01} extinction map shown in Fig.~\ref{fig:extMap}.

The HRD prior $P(\vmini,M_V)$ is constructed from the same GUMS data used to construct the luminosity function described in Sect.~\ref{subsec:mw}. Again we combine the CMDs of different components of the Milky Way into a single universal CMD (or HRD), and smooth it using two-dimensional kernel density estimation (a Gaussian kernel with widths of 0.05\,mag in both directions and no covariance is used). The number of stars per volume element is again used as the weighting to calculate the density of points in the HRD on a $2000\times2000$ grid. A small positive constant is added to ensure no region of the HRD has exactly zero probability density. The value of this is set such that its ratio to the total probability (original plus offset density) is $2.87\times 10^{-7}$ (this number corresponds to 5$\sigma$ probability in a Gaussian distribution). 
This total probability is then normalised to unity. The resulting HRD prior is shown in Fig.~\ref{fig:hrd}. It it far from perfect, but it is adequate for demonstration purposes.

\begin{figure}
\includegraphics[width=\hsize]{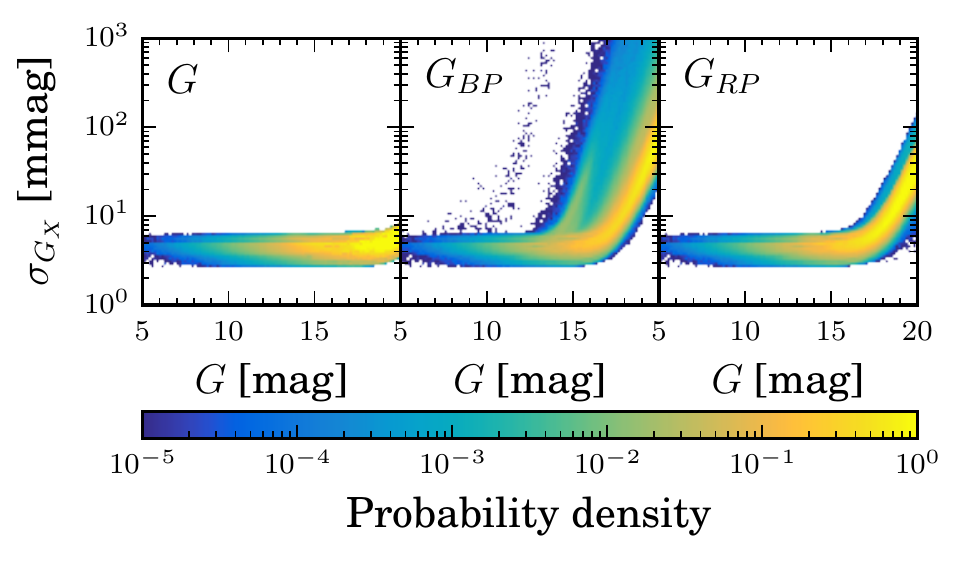}
\caption{The distribution of the end-of-mission photometric errors $\sigma_{G}$ (left), $\sigma_{\gbp}$ (middle), and $\sigma_{\grp}$ (right) as a function of true $G$ magnitude, for all stellar objects in the GUMS catalogue. The colour shows the probability density per unit magnitude per unit log photometric errors.}
\label{fig:photError}
\end{figure}
\subsection{The photometric error model}
The error models for the photometric measurements are the parameterised form\footnote{\url{www.cosmos.esa.int/web/gaia/science-performance}} of the formulation outlined in \cite{jdb022} and \cite{jor10}. This includes the scattered light levels measured post-launch \citep{deb14}. Using the same notation as for our parallax noise model (Eq.~\ref{eq:sigmaVarpi}), the standard errors in the $G$-band magnitude is
\begin{multline}
\sigma_G\;[{\rm mag}] = m\left[\frac{1}{n_{\rm tr}p_{{\rm det},G}(G)}\left(\sigma_{\rm cal}^2 \right.\right.\\ 
\left.\vphantom{\frac{1}{n_{\rm tr}p_{{\rm det},G}(G)}}\left.\vphantom{\sigma_{\rm cal}^2} + 10^{-6}\left[0.0001985 + 1.8633z + 0.04895z^2\right]\right)\right]^{1/2},
\label{eq:sigmaG}
\end{multline}
where
\begin{equation}
z = \max\left(10^{0.4(12 - 15)}, 10^{0.4(G-15)}\right),
\end{equation}
$m = 1.2$ is a contingency margin to account for unknown sources of error, and $\sigma_{\rm cal} = 30$\,mmag \citep{jdb022, jor06, cj047} is the adopted total photocalibration error.

\begin{table}
\caption{The coefficients $c_{{\rm XP},ij}$ used in Eqs.~\ref{eq:sigmaGbprp}--\ref{eq:coeffXp} to calculate the BP/RP photometric errors $\sigma_{\gxp}$.}
\begin{tabular}{clrrrr}
\hline\hline
\multicolumn{1}{c}{XP} & \multicolumn{1}{c}{$c_{{\rm XP},i}$} & \multicolumn{1}{c}{$c_{{\rm XP},i0}$} &\multicolumn{1}{c}{$c_{{\rm XP},i1}$} &\multicolumn{1}{c}{$c_{{\rm XP},i2}$} &\multicolumn{1}{c}{$c_{{\rm XP},i3}$}\\
\hline
\multirow{3}{*}{BP} & $c_{{\rm BP},0}$ & -1.866968 & -0.205807 & 0.060769 & 0.000262\\
& $c_{{\rm BP},1}$ & 1.465592 & 0.195768 & 0.018878 & -0.000400\\
& $c_{{\rm BP},2}$ & 1.043270 & 0.355123 & 0.044390 & -0.000562\\
\hline
\multirow{3}{*}{RP} & $c_{{\rm RP},0}$ & -3.042268 & -0.091569 & 0.027352 & -0.001923\\
& $c_{{\rm RP},1}$ & 1.783906 & -0.318499 & 0.057112 & -0.003803\\
& $c_{{\rm RP},2}$ & 1.615927 & -0.636628 & 0.114126 & -0.007597\\
\hline
\end{tabular}
\label{tab:cxp}
\end{table}
The BP/RP photometric errors are
\begin{multline}
\sigma_{\gxp}\;[{\rm mag}] = \left[\frac{1}{n_{\rm tr}p_{{\rm det},XP}(G)}\right.\times\\
 \left.\vphantom{\frac{1}{n_{\rm tr}p_{{\rm det},XP}(G)}}
 \left(1.44\sigma_{\rm cal}^2 + 10^{-6}\sum_{i=0}^{2}10^{c_{{\rm XP},i}}z^i\right)\right]^{1/2}
\label{eq:sigmaGbprp}
\end{multline}
where
\begin{equation}
z = \max\left(10^{0.4(11 - 15)}, 10^{0.4(G-15)}\right),
\end{equation}
the coefficients $c_{{\rm XP},i}$ are polynomials in $\vmini$:
\begin{equation}
c_{{\rm XP},i} = \sum_{J=0}^{3}c_{{\rm XP},ij}\vmini^j,
\label{eq:coeffXp}
\end{equation}
in which the values of the coefficients $c_{{\rm XP},ij}$ are tabulated in Table~\ref{tab:cxp}, and the BP/RP detection probability $p_{\rm XP}(G)$ is taken from Table~10 in \cite{jor10}. The BP/RP parameterisation in Eqs.~\ref{eq:sigmaGbprp}--\ref{eq:coeffXp} already includes the contingency margin of $m=1.2$.

The distributions of end-of-mission $\sigma_G$, $\sigma_{\gbp}$, and $\sigma_{\grp}$ as a function of the $G$-band magnitude are shown in Fig.~\ref{fig:photError}. BP photometric errors $\sigma_{G_{\gbp}}$ shows larger errors than the other two bands because of the effects of extinction. We can see in Fig.~\ref{fig:photError} that there is an additional lower density locus in the distribution of $\gbp$, which is due to areas with heavy extinction in the GUMS catalogue.

To simulate the photometric observations from GUMS, we draw the $G$ and $\gxp$ magnitudes 
as we did the parallaxes, by drawing at random from their likelihoods (which are independent 1D Gaussians with mean equal to the true magnitude and standard deviation given by the noise model).

\subsection{Computing the posterior}

Since the posterior is now multivariate, we now use the affine-invariant Markov Chain Monte Carlo (MCMC) ensemble sampler \citep{goo10} to sample the multivariate posterior, from which the one-dimensional marginalised posterior $P(r|\mathbf{X})$ is easy to obtain via density estimation over these samples. For each source, samples are taken in a maximum of 30\,000 steps, but for every 10\,000 steps the autocorrelation time is calculated and the chain is stopped when convergence is achieved.

The extinction $A_V^{\rm map}(r,l,b)$ is calculated using a lookup table that tabulates the extinction in distance and \texttt{healpix} cell, with $n_{\rm side} = 64$. For a source with a given $(l,b)$ we find the cell closest to that direction and interpolate $A_V$ in distance. This speeds up the calculation in two ways: We do not have to calculate the extinction using the full set of equations and parameters that describe the extinction model, and we can reuse the extinction curve in a given cell for all stars in the cell (there are far fewer cells than stars).

\begin{figure}
\includegraphics[width=\hsize]{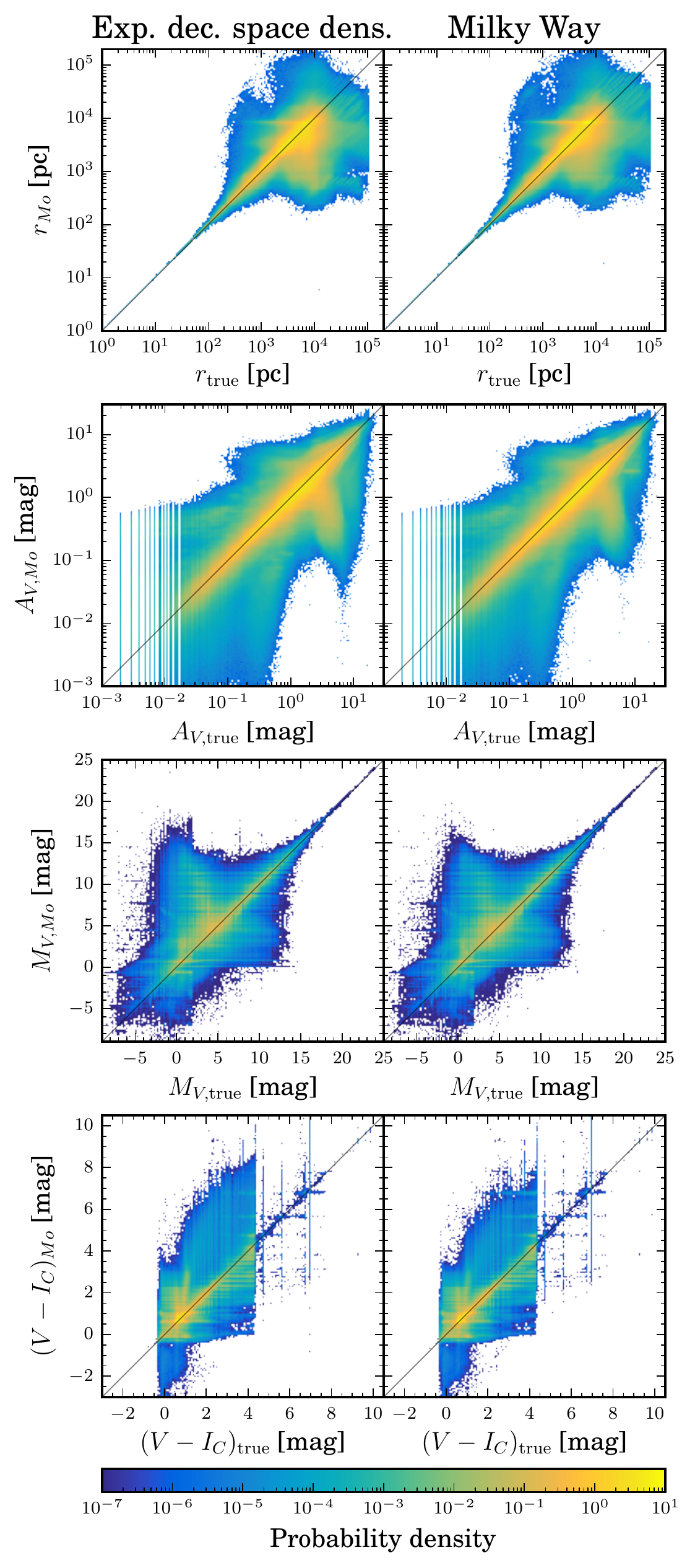}
\caption{Comparison of the mode parameters, from top to bottom: Distance $\rmode$, extinction $A_{V,Mo}$, absolute magnitude $M_{V,Mo}$, and colour $(V-I_C)\mode$ for all stars in the GUMS catalogue, with their corresponding true values, inferred using the \expp{} prior (left column) and the \mw{} (right column) prior as the distance prior. The black lines in each panel indicate a perfect match between the estimated and true value.}
\label{fig:comparisons_mwhrd_mo}
\end{figure}
\begin{figure*}
\centering
\includegraphics{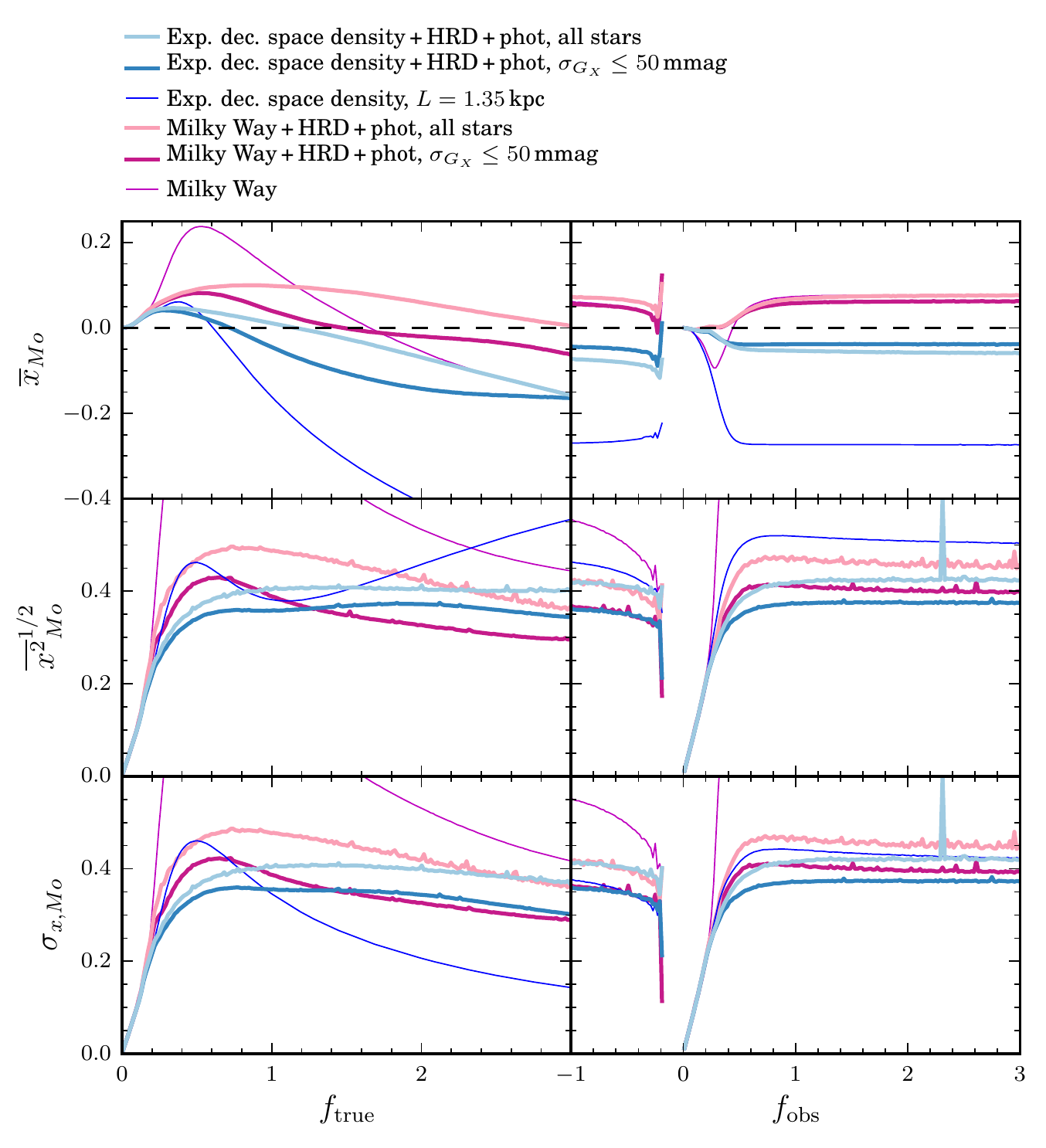}
\caption{Performance of the mode distance $\rmode$ using the \exphrd{} prior (thick teal and cyan lines) and the \mwhrd{} prior (thick violet and pink lines), shown in terms of the bias $\bias\mode$ (top rows), the RMS (middle rows), and the standard deviation $\stddevmod$ (bottom rows), as a function of $\ftrue$ (left columns) and $\fobs$ (right columns). 
For both distance priors, the lines in lighter shades (viz. cyan and pink) are for all stars, whereas the lines in darker shades (teal and violet) are just for stars with photometric errors less than 50\,mmag.
The performance of the corresponding parallax-only distance estimates (thin blue and magenta lines) previously shown in Fig.~\ref{fig:summary_performance_ftrue_mode} and Fig.~\ref{fig:summary_performance_fobs_mode} are included for comparison. Note that the vertical axes have different (smaller) ranges than in Fig.~\ref{fig:summary_performance_ftrue_mode} and Fig.~\ref{fig:summary_performance_fobs_mode}.}
\label{fig:summary_performance_ftrue_mode_mwhrd}
\end{figure*}
\begin{figure*}
\includegraphics[width=\hsize]{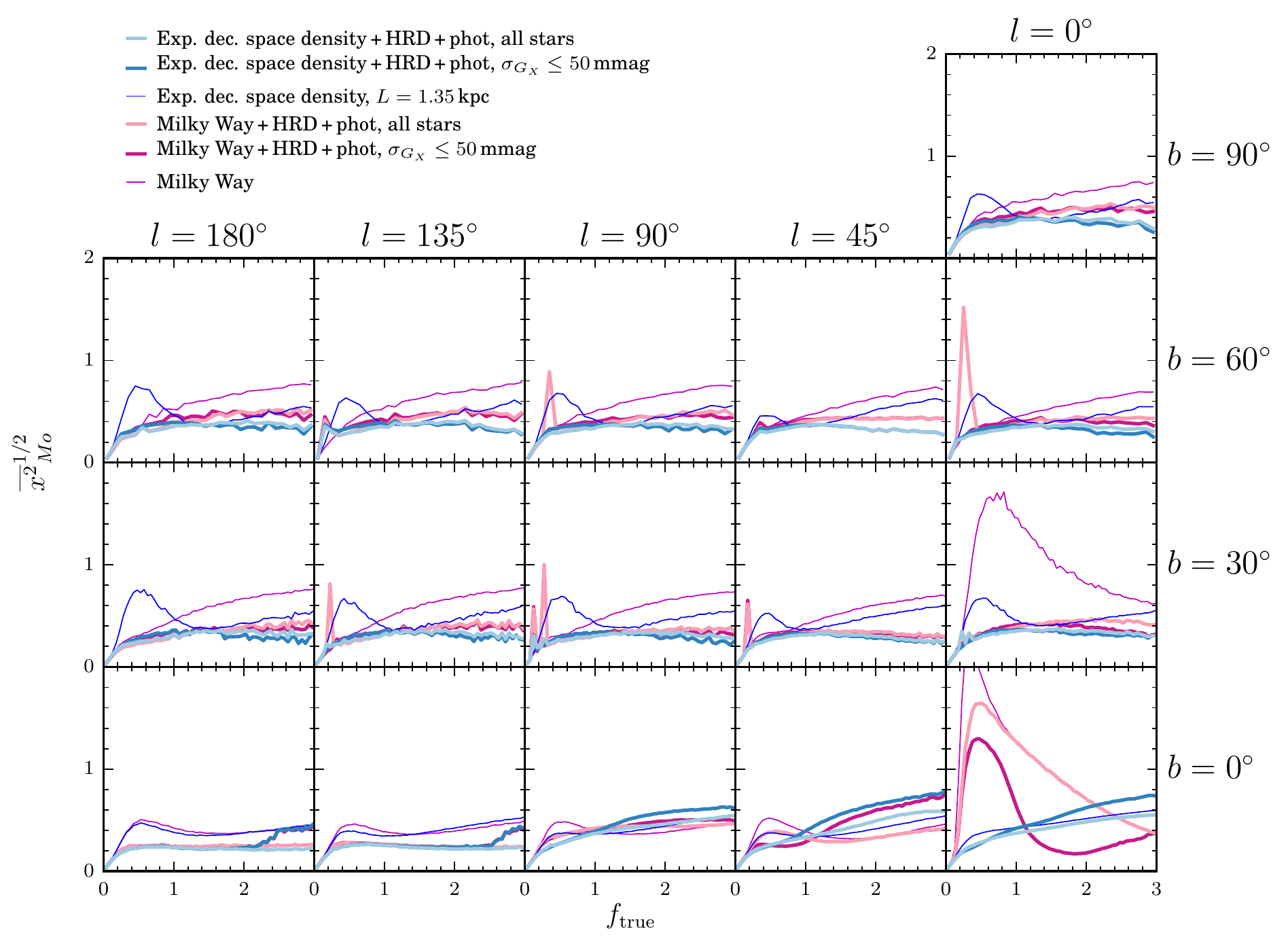}
\caption{As Fig.~\ref{fig:summary_performance_ftrue_mode_dir}, but for the \exphrd{} prior and the \mwhrd{} prior. The results of the corresponding parallax-only distance estimates are also included for comparison. Note that the vertical axes have different (smaller) ranges than in Fig.~\ref{fig:summary_performance_ftrue_mode_dir}.}
\label{fig:summary_performance_ftrue_mode_mwhrd_dir}
\end{figure*}
\begin{figure}
\includegraphics[width=\hsize]{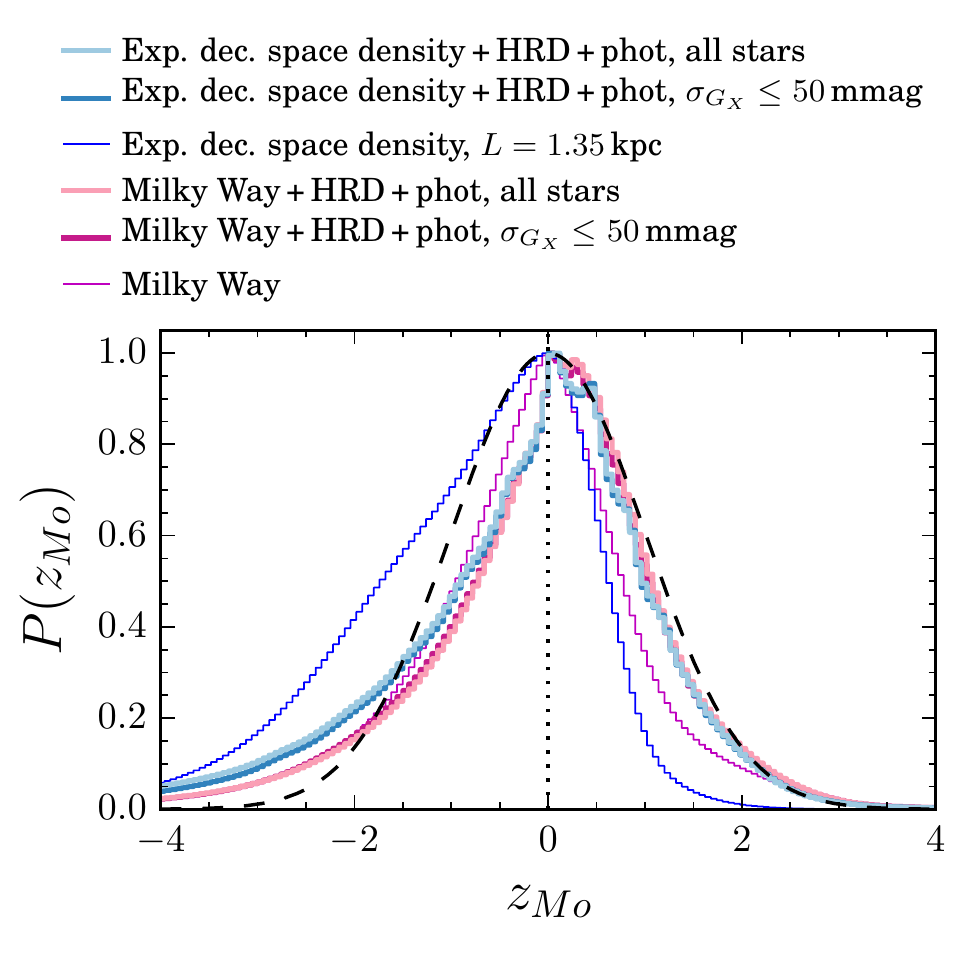}
\caption{As Fig.~\ref{fig:zhist}, but for the \exphrd{} prior and the \mwhrd{} prior, for all stars and with the photometric error cutoff applied. The results for the corresponding parallax-only inferences are shown for comparison. The black dashed line indicates the standard normal distribution. All distributions are scaled to have their maxima at one. Note the different range on the $z$-axis compared to Fig.~\ref{fig:zhist}.}
\label{fig:zhist2}
\end{figure}

\subsection{Results of the joint astrometric and photometric posterior}

Having computed the posterior we find the mode of the parameter vector, $\mathbf{\Theta}\mode$, for each star, by finding the sample in the posterior chain that gives the highest posterior probability.
Fig.~\ref{fig:comparisons_mwhrd_mo} compares this with the true parameters, $\mathbf{\Theta}\true$, for both distance priors. We see that the distance inference---our main goal---is much improved compared to the distance estimate using just the parallax and (the same) distance priors shown in Fig.~\ref{fig:comparisons_mo}. Comparing the two distance priors (left and right column in Fig.~\ref{fig:comparisons_mwhrd_mo}), we see no significant difference, which indicates that any poor distance estimates arising from poorly measured parallaxes can be improved by the inclusion of photometric data.

Despite the improvement, we see that for $\rtrue\gtrsim 2$\,kpc, the inferred distances are underestimated for some stars. This is primarily because we are unable to distinguish dwarfs from giants with these colours. Looking at the inference for the absolute magnitudes, we see that the region with $M_{V,{\rm true}}$ between about -3 and 5 is where the absolute magnitudes are slightly overestimated, i.e.\ they are inferred to be fainter than the true values. The model is therefore placing these stars closer than they really are.

The performance of the distance estimation as a function of $\ftrue$ and $\fobs$ averaged over all directions is shown in Fig.~\ref{fig:summary_performance_ftrue_mode_mwhrd}.
The primary comparison here is between the line labelled ``\exphrd{}'' and line labelled ``\expp{}'', as well as between the line labelled ``\mwhrd{}'' and the line labelled ``\mw{}''. Looking at the plots against $\fobs$ first (right column), we see the significant improvement in performance in both the bias and standard deviation when adding the photometry (and the HRD prior). Indeed, when using the \mw{} model as the distance prior, we see that adding the photometry almost entirely eliminates the bias for small positive $\fobs$ and stabilises it at $\bias\mode\sim 0.08$ for $\fobs\gtrsim 1$. Using the \expp{} prior eliminates the bias even further, with $\bias\mode\sim -0.03$ for the same range of $\fobs$. The negative value indicates that the distances to most stars are still slightly underestimated, due to the constraint imposed by the distance prior.
The inclusion of photometry also significantly reduces the standard deviation in the distance estimates, although interestingly, with the \mw{} model as the distance prior, it does not decrease the standard deviation more than what we achieved using the \expp{} prior with just the parallax data.

Nonetheless, the use of parallaxes and photometry together produces smaller RMS residuals than using the parallax alone (with any of the priors). We can see this in the right column and middle row of Fig.~\ref{fig:summary_performance_ftrue_mode_mwhrd}. Using the \mw{} prior, the RMS reaches a plateau of around $\sim$0.45 with a bias of $\sim$0.08 once $\fobs$ rises above $\sim$1. For negative parallaxes, the bias and the RMS are stable at respectively $\sim$0.08 and $\sim$0.42. In this regime, the inclusion of photometric data makes this prior superior to other parallax-only priors. Restricting the analysis to stars with small photometric uncertainties (less than 50\,mmag in all bands) only reduces the bias by a negligible amount, but improves the RMS errors and standard deviation by $\sim$0.05 for both positive and negative $\fobs$. Further improvements can be achieved when the \expp{} prior is instead used as the distance prior. For positive parallaxes, the bias is only about $\sim$-0.03, and the RMS is about $\sim$0.42. Considering only stars with small photometric uncertainties (the same cut as before) improves these further by about the same amount as using the \mw{} prior. Having said this, it should be recalled that the majority of stars with positive parallaxes and $\fobs\lesssim 3$ are more likely to be disc stars. This is why the \exphrd{} prior performs better than the \mwhrd{} prior. 

In terms of $\ftrue$ (left column of Fig.~\ref{fig:summary_performance_ftrue_mode_mwhrd}), we see that the \mwhrd{} prior is superior to the parallax-only \mw{} prior, although over a range of $\ftrue$ between 0.1 and 1.8 the parallax-only \expp{} prior is actually better in RMS. Using only stars with small photometric uncertainties improves the distance estimates, with the RMS and standard deviation are always below $\sim$0.45 for all $\ftrue$. If we look at the distribution of the photometric errors in Fig.~\ref{fig:photError}, we see that the 50\,mmag cutoff affects mostly red-colored stars (whether this be intrinsic or due to reddening). We see that the \exphrd{} prior improves the estimation further, with the RMS always below $\sim$0.4 even when all stars are included. Taking only stars with small photometric uncertainties reduces the RMS further by $\sim$0.05. For poorly measured stars with $\ftrue\gtrsim 1.2$, however, the \mwhrd{} prior performs better with the RMS stabilising at $\sim$0.3 for large $\ftrue$. We shall see later that this is due to the poor performance of the \exphrd{} prior for directions toward the galactic centre.

The directional performance (in terms of the RMS) of the mode estimator for the \exphrd{} and the \mwhrd{} prior is shown in Fig.~\ref{fig:summary_performance_ftrue_mode_mwhrd_dir}. Here we see that the performance is relatively stable everywhere, both priors performs more-or-less equally well in directions away from the galactic plane, with the \exphrd{} performing $\sim$0.05 better for large $\ftrue$. Both are especially accurate in the galactic plane towards the direction of the anticentre (at $l=0^\circ$ and $b=135^\circ, 180^\circ$), where the RMS stabilise at $\sim$0.2 for large $\ftrue$ (in other directions the RMS stabilise at $\sim$0.25--0.4 for large $\ftrue$). In some directions, however, some ``spikes'' can be seen in the performance curve of the \mwhrd{} prior. These corresponds to some stars which distances are so poorly estimated they disrupt the overall behaviour of the curve. The distances are poorly estimated again because of the inability to distinguish dwarfs from giants. These spikes tend to appear at high latitudes due to the paucity of stars in these directions. At the galactic plane, the stars are more numerous, the overall behaviour is relatively undisturbed by these stars. Making cuts in photometric errors as before can remove most of these stars and improves the overall performance, suggesting that these poor performances are to some degree due to poor photometric measurements. This shows that a combination of bad photometric and parallax measurements will not help in accurately estimating distances. Bad parallax measurements should be complemented by good photometric measurements and vice versa.

These directions aside, the galactic centre (bottom-right panel of Fig.~\ref{fig:summary_performance_ftrue_mode_mwhrd_dir}) remains a problematic direction. For the \mwhrd{} prior, we see here that the RMS error in this direction can reach up to 1.64 even for $\ftrue\lesssim 0.5$. The performance improves as $\ftrue$ increase, but it is hardly better than the parallax-only \mw{} prior. Considering only stars with small photometric uncertainties improves the situation considerably, although it is not better than the \expp{} prior even for $\ftrue\lesssim 1$. For the same range of $\ftrue$, the \exphrd{} prior performs best for this direction, but only up to $\ftrue\sim 1.2$. For larger $\ftrue$, the \mwhrd{} prior (with cuts in photometric uncertainties) performs better. The heavy extinction in this direction seem to be main cause of the poor performance of the \mwhrd{} prior, but for the \exphrd{} prior the poor performance is due to the mismatch between the distance prior and true distribution. Clearly $L=1.35$\,kpc is not a good scale length for directions toward the centre. Since the \mw{} model takes the galactic bulge into account (see Fig~\ref{fig:prior_mw} for the same direction), it provides a better distance constraint. Considering the performance in this direction, we can see that the averaged performance shown in Fig.~\ref{fig:summary_performance_ftrue_mode_mwhrd} is dominated by stars towards the galactic centre. 

Finally, we compare the formal uncertainties (Sect.~\ref{subsec:formalErrors}) in Fig.~\ref{fig:zhist2}. The distributions of the standardised residuals $z$ for the two pairs are similar, an indication that the uncertainties in the distance estimation depends largely on photometric data. The $z$-distribution is asymmetric, although for positive $z$ the distribution is very close to a standard normal distribution (shown with a black dashed line). For the underestimated distances (negative $z$), the width is less than that of a standard normal distribution, suggesting that the uncertainties are more overestimated than the uncertainties for the overestimated distances (positive $z$). Overall, the HWHM of the $z$-distribution is 0.9, which suggests that in general the uncertainties might be overestimated, altough not as much as the parallax-only \mw{} prior (which is 0.84). The HWHM of the HRD + phot priors seems to suggest that the inclusion of photometric data can produce a smaller credible interval on distance (hence a larger value of the HWHM), but the photometric errors also add additional uncertainties thus keeping it smaller than the HWHM of a standard Gaussian.

\section{Summary and conclusions}
We have estimated distances from simulated \gaia{} parallaxes using probability-based inference. We examined four different priors on distances: the three isotropic priors---\ud{}, \usd{}, \expp{}---and the more complex \mw{} prior, which models the Galaxy and its observability in a magnitude-limited survey. We further considered the inclusion of photometric measurements (plus an HRD prior) with the \expp{} prior and the \mw{} prior. The simulated \gaia{} data were taken from the \gaia{} Universe Model Snapshot (GUMS) catalogue. Note that the \mw{} prior was intentionally not taken from this. We investigated the use of both the median and the mode of the posterior as distance estimators. The median was found to be most appropriate for the \usd{} prior, but the mode performed better for the other priors.

For stars with positive parallaxes and observed fractional parallax errors $\fobs$ less than about 0.1, the distance estimates were largely independent of the choice of prior. This will only apply to about $\sim$10\% of the \gaia{} catalogue, however. Taking this approach would not only limit us to a small fraction of hard-won data, it would also lead to truncation biases when using such a sample for astrophysical analyses.

We found that the \usd{} prior with an upper limit of $\rlim=100$\,kpc performs poorly. This is partly because this is an unrealistic prior when applied to a large volume of the Galaxy, since it is obvious that within a radius of $r=100$\,kpc the volume density of the Galaxy is not constant (the mode should not be used with this prior because once $\fobs$ is larger than about 0.35, the mode is at $\rlim$). Yet truncating this prior at smaller distances produces other, no less dramatic, problems.

Viewed as a function of the expected (true) fractional parallax error, $\ftrue$, the \ud{} prior also does very poorly, showing a steep rise in bias and standard deviation above $\ftrue=0.1$ (already over 100\% at $\ftrue=0.3$).
But if we view it as a function of the observed fractional parallax error, $\fobs$, we see much better results. This is because plotting against $\fobs$ automatically ``cleans'' the appearance by assigning poor data (very small or negative parallaxes) to large or negative values of $\fobs$.
Nonetheless, over moderate positive values of $\fobs$ this prior works reasonably well, better than the \mw{} prior and not much worse than the \expp{} prior with $L=1.35$\,kpc. However, a closer inspection shows that most of its good performance comes from a narrow region towards the galactic centre where there are a lot of stars (see Fig.~\ref{fig:summary_performance_fobs_mode_dir}), which therefore dominates the average performance. The performance at other low latitude directions is considerably worse than the \mw{} and \expp{} priors, and at high latitudes it is no better than the \expp{} (and significantly worse once $\fobs\gtrsim 1$).

The \expp{} prior performs well, altough fine-tuning is required to find the appropriate scale length $L$ (the only parameter in the prior). The RMS of the scaled residual in distance (Eq.~\ref{eq:x}), averaged over all directions in the galaxy, is roughly equal to $\ftrue$ for $\ftrue\lesssim 0.4$, but does not increase further as $\ftrue$ grows to 1. In terms of these RMS errors, we found that $L=1.35$\,kpc gives the best performance. The mode of this prior is $2L = 2.7$\,kpc, and this is on the order of the scale length of the stellar disc of the Milky Way, where most stars are located. Thus it can accurately infer the distance of a large majority of stars in the catalogue. We see, however, that the RMS increases with $\ftrue$, because poor parallaxes ``force'' the mode of the posterior to be close to the mode of the prior at $2L$. This is far from the true distances for many of these stars, as they are more likely to be distant stars.

The \mw{} prior performs well in directions away from the galactic centre. The poor performance toward the galactic centre is due a prior mismatch problem, specifically due to a different adopted bulge-to-disc mass ratio for our prior. This can be seen from a comparison between its distance dependence and that in the ``true'' Galaxy: We see in Fig.~\ref{fig:gums_rdist_directional} significant discrepancies at directions close to the galactic centre. Again, as in the case of the \ud{} prior, the performance in this direction dominates the direction-averaged performance. The prior could be improved by adopting a different mass density model and by using a different luminosity function $\phi(M_G)$ for each component of the Milky Way, rather than a single universal luminosity function. However, such discrepancies between the prior and the underlying true distribution we are trying to infer will inevitably occur in any real application. Our results show that although constructing a detailed model of the Milky Way is a good idea in principle, in practice it can go wrong due to the strong assumptions involved and subsequent likelihood of a mismatch.

Rather than creating a detailed prior, a better way to improve distance estimation is to include other distance-sensitive measurements, where possible. We have shown that photometric data can help for stars with poor parallax measurements. This requires us to make further assumptions, namely about the intrinsic nature of the sources. To make use of photometry we (must) adopt a Hertzsprung--Russell diagram and an extinction map. We find that this improves the averaged performance significantly, with an RMS error which is always less than about 0.5 for $\ftrue\lesssim 3$. It is, however, still difficult to estimate distances for stars toward the galactic centre, due to the heavy extinction in this direction. At higher latitudes the performance is more or less stable.

While adding photometry helps, \gaia{} broad band photometry on its own does not allow us to distinguish giants from dwarfs, so this remains a limitation in the distance accuracy even when including photometric data. One could of course use more sophisticated priors or more data such as spectroscopy (e.g. \citet{sch14, san16, liu12}) to help estimate absolute luminosity, but one must be aware that distance estimates which do not use just the parallax necessarily require a different set of assumptions, such as the physical properties source. The stronger the prior, the greater the risk of mismatch problems. Increased use of priors may reduce the variance in the distance estimations, but they can also increase the bias.

\acknowledgements
We thank Jos de Bruijne for the discussion on \gaia{} astrometric errors, Berry Holl for providing us with the transit data in Fig.~\ref{fig:ntr}, the DPAC/CU2 developers of \texttt{GaiaSimu} for the code to calculate the extinction map in Fig.~\ref{fig:extMap}, Joe Hennawi and the Max Planck Computing and Data Facility (MPCDF) in Garching for the computational resources, and the IT group of the Max Planck Institute for Astronomy for their technical assistance. We thank Anthony G.A. Brown, Morgan Fouesneau, and Jan Rybizki for useful comments and discussions. We also thank the anonymous referee for the comments and suggestions that improve the analysis of this paper.

This research made use of NASA's Astrophysics Data System; \texttt{Astropy}, a community-developed core Python package for astronomy \citep{2013A&A...558A..33A}; the \texttt{IPython} package \citep{PER-GRA:2007}; \texttt{matplotlib}, a Python library for publication quality graphics \citep{Hunter:2007}; and \texttt{SciPy} \citep{jones_scipy_2001}. 

\appendix
\section{The Milky Way: the density model and the extinction map}
\label{app:mwmodel}
\begin{figure*}
\begin{center}
\includegraphics[height=0.415\textwidth]{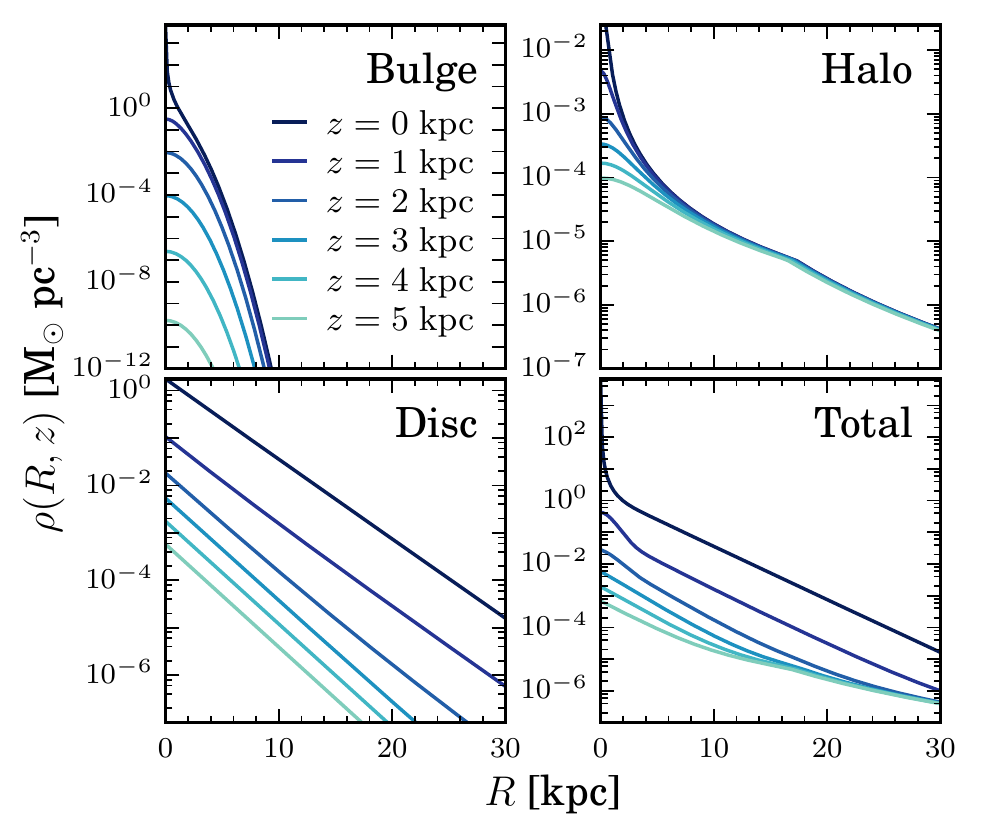}
\includegraphics[height=0.42\textwidth]{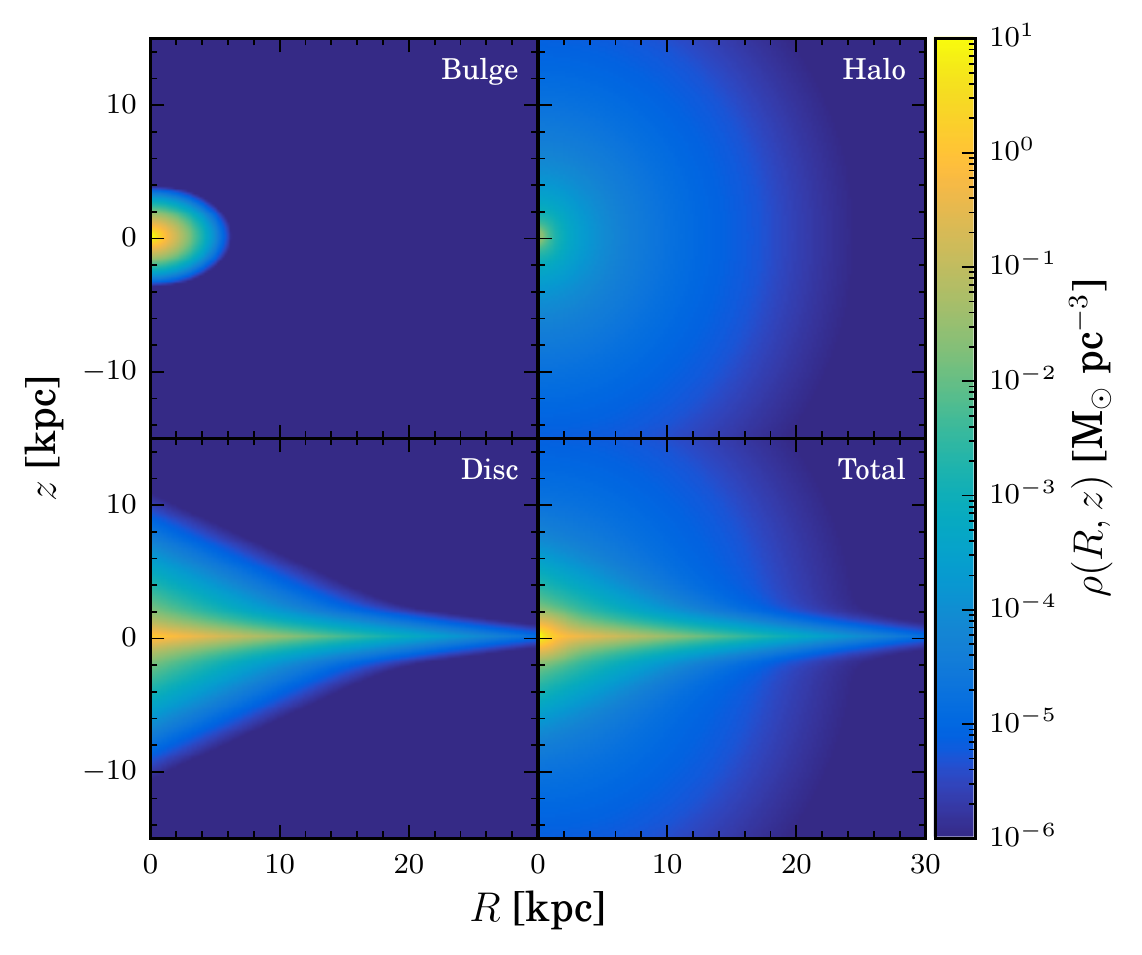}
\end{center}
\caption{The stellar mass density used in our Milky Way prior, shown in galactocentric cylindrical coordinates. The set of plots on the left show the variation as a function of galactic radius $R$ for various values of the galactic height, $z$. The four panels depict the profile for each of the three components, plus the sum of all three (lower right). Note the different scales on the vertical axes. The right plots show the density as a function of $(R,z)$ on the colour scale given in the legend.}
\label{fig:mwdens}
\end{figure*}
\begin{figure*}
\begin{center}
\includegraphics[width=0.48\hsize]{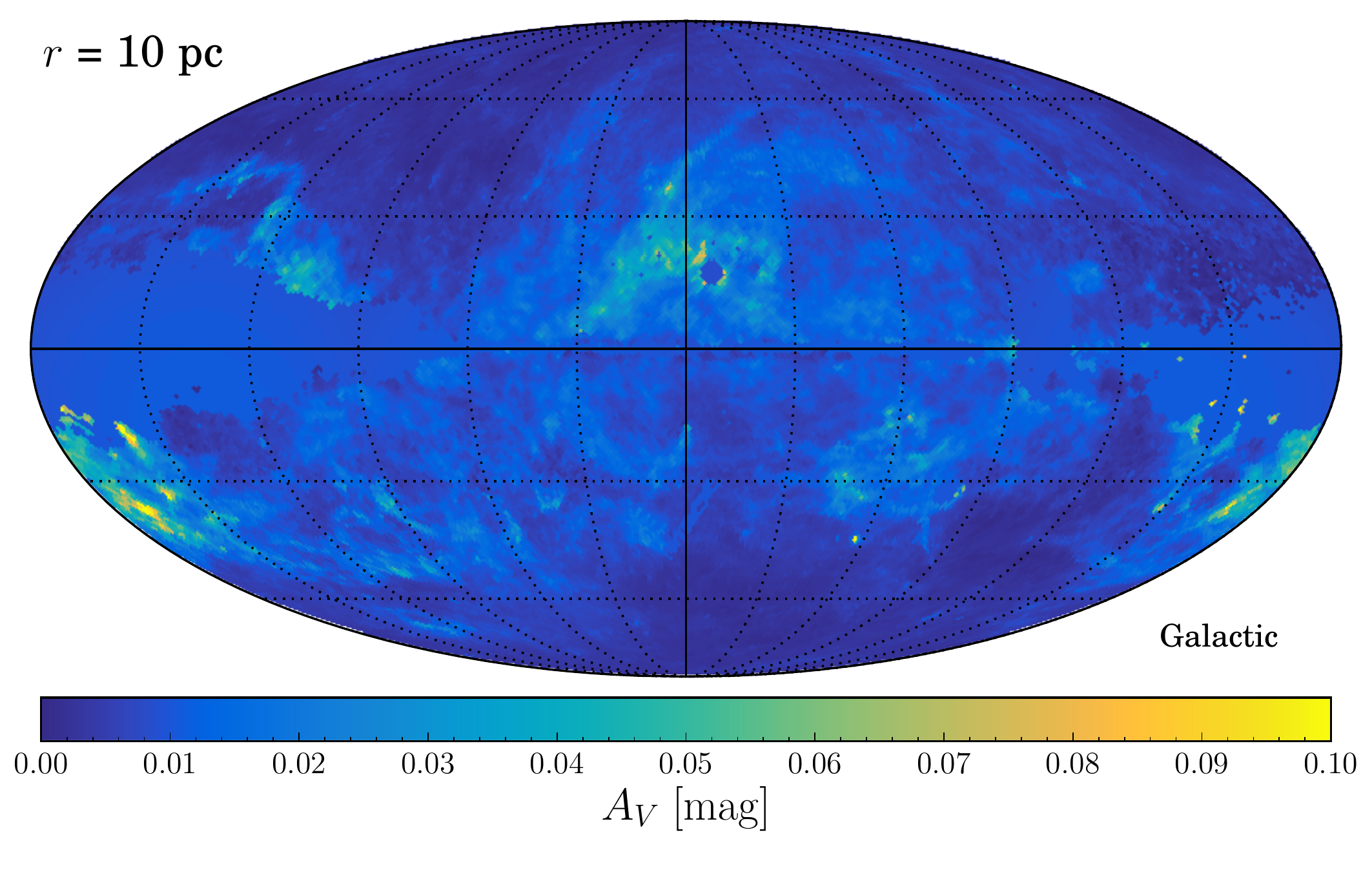}
\includegraphics[width=0.48\hsize]{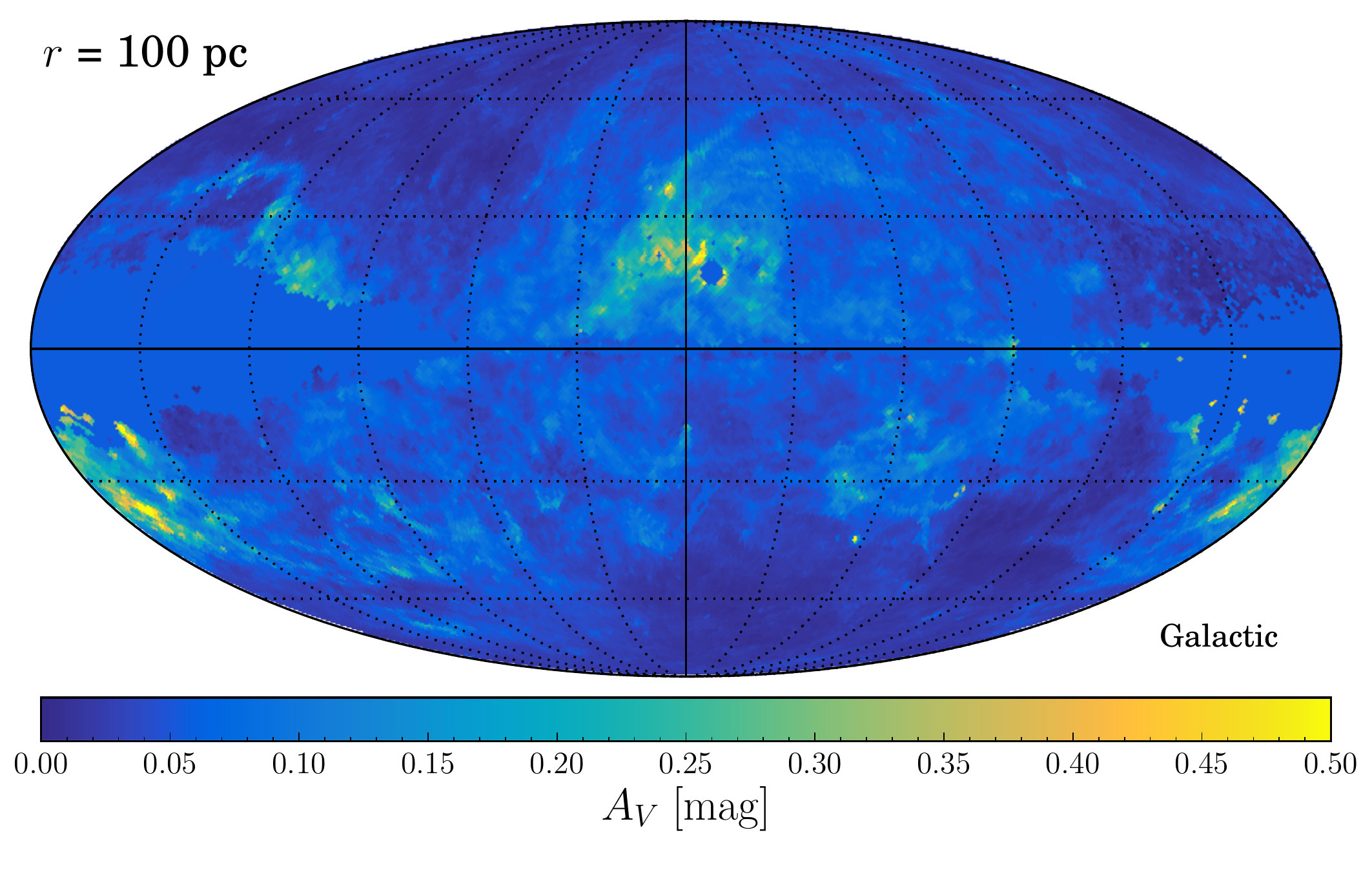}\\
\includegraphics[width=0.48\hsize]{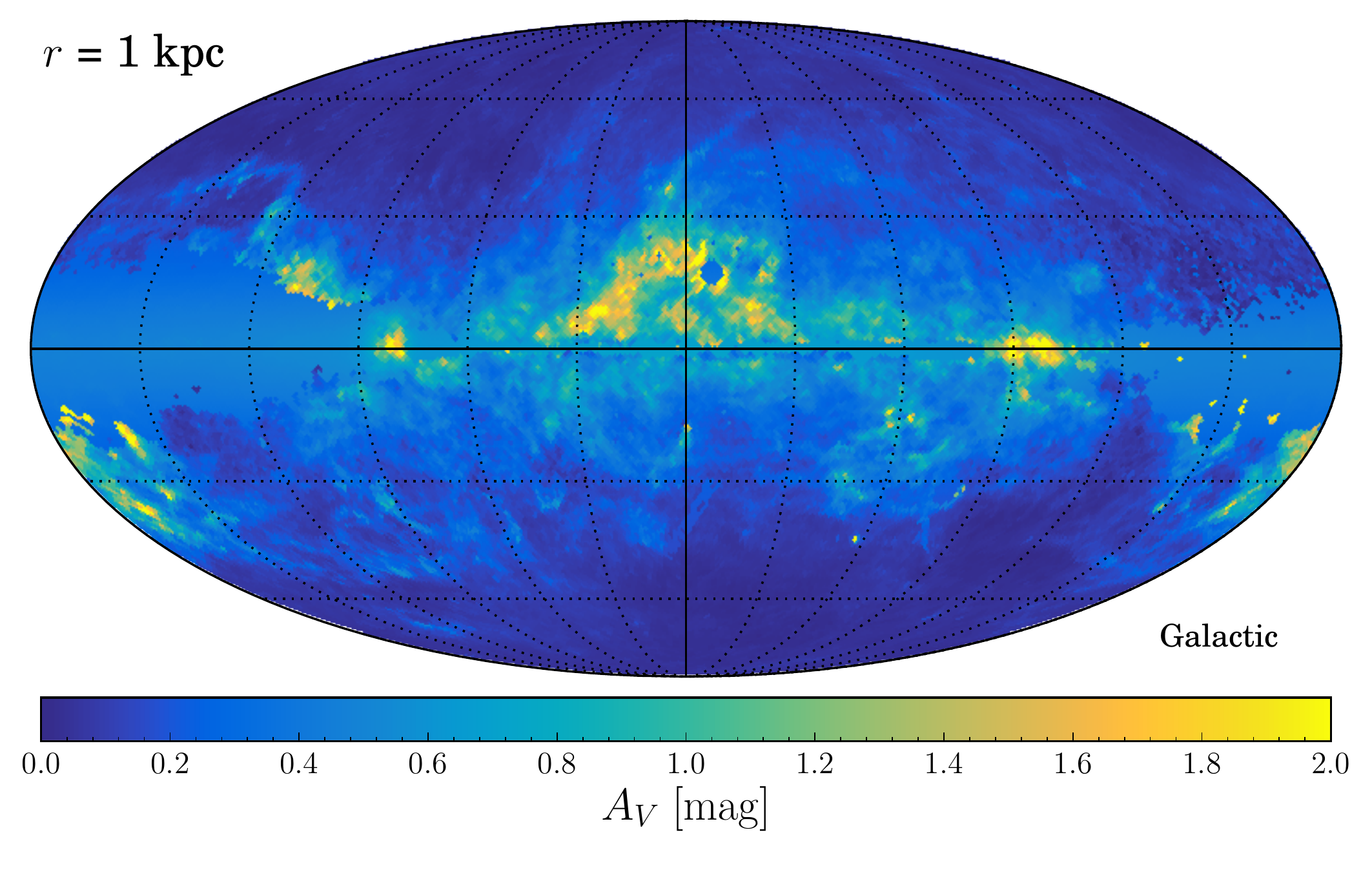}
\includegraphics[width=0.48\hsize]{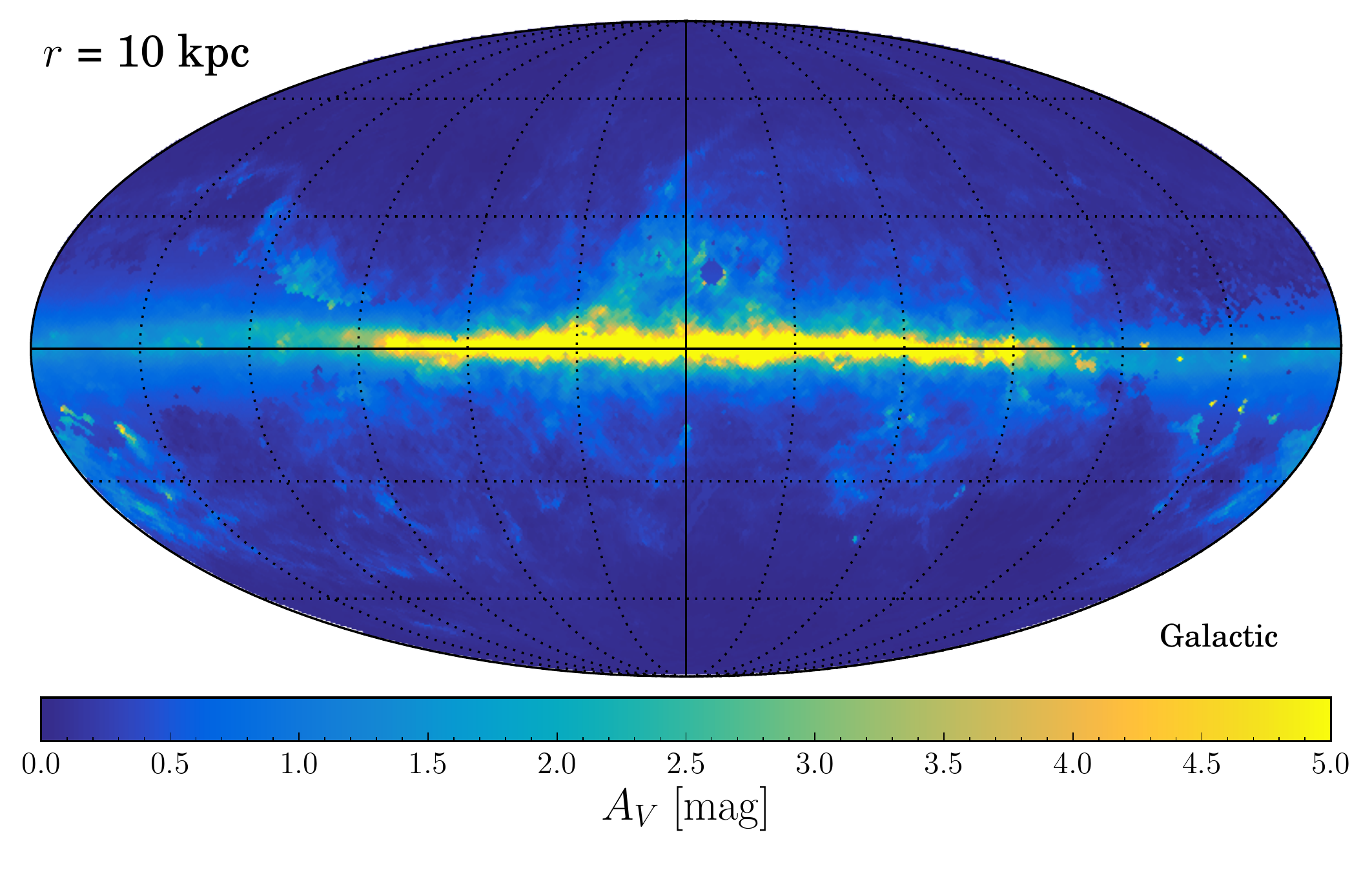}
\end{center}
\caption{Mollweide projections of the extinction map constructed by \cite{dri03}, for distances up to 10\,pc, 100\,pc, 1\,kpc, and 10\,kpc. Note the different scale of the color map, shown in the color bar under each map.}
\label{fig:extMap}
\end{figure*}

The stellar number density, $\rho_{\rm MW}(r,l,b)$, is expressed as the sum of three components: the bulge $\rho_b$, the disc $\rho_d$, and the halo $\rho_h$:
\begin{equation}
\rho_{\rm MW}(r,l,b) = \rho_b(r,l,b) + \rho_d(r,l,b) + \rho_h(r,l,b).
\end{equation}

To model the bulge, we use the spheroidal power-law model described in \cite{BT2}, with a modification in the power law to prevent infinite density at the galactic centre:
\begin{equation}
\rho_b(R,z) = \rho_{b,0}\times
\begin{dcases*}
\left(\frac{q}{a_b}\right)^{-\alpha_b}\exp\left(-\frac{q^2}{r^2_b}\right) & for $q > q_{\rm min}$,\\
\left(\frac{q_{\rm min}}{a_b}\right)^{-\alpha_b}\exp\left(-\frac{q^2}{r^2_b}\right) & otherwise,
\end{dcases*}
\end{equation}
where $(R,z)$ is the galactocentric distance and height in cylindrical coordinates, and 
\begin{equation}
q = \left(R^2 + \frac{z^2}{q^2_b}\right)^{1/2}.
\end{equation}
The model for the disc is a two-component exponential disc, which represents the thin and thick disc \citep{BT2}:
\begin{equation}
\rho_d(R,z) = \frac{\Sigma_t}{2z_t}\exp\left(-\frac{R}{R_t} - \frac{|z|}{z_t}\right) + \frac{\Sigma_T}{2z_T}\exp\left(-\frac{R}{R_T} - \frac{|z|}{z_T}\right),
\end{equation}
and the mass density of the stellar halo is modelled as an isotropic double-power law function with an exponential decline beyond a certain distance \citep{kaf14}:
\begin{equation}
\rho_h(r_G) = \rho_{h,0}\times
\begin{dcases*}
\left(\frac{r_{h,{\rm min}}}{r_{h,b}}\right)^{-\alpha_{h,1}} & for $r_G < r_{h,{\rm min}}$,\\
\left(\frac{r_G}{r_{h,b}}\right)^{-\alpha_{h,1}} & for $r_{h,{\rm min}}\leq r_G < r_{h,b}$,\\
\left(\frac{r_G}{r_{h,b}}\right)^{-\alpha_{h,2}} & for $r_{h,b}\leq r_G < r_{h,t}$,\\
\left(\frac{r_{h,t}}{r_{h,b}}\right)^{-\alpha_{h,2}}\times\\
\left(\frac{r_G}{r_{h,t}}\right)^{\epsilon_h}\times\\
\exp\left(-\frac{r_G-r_{h,t}}{\Delta_h}\right) & for $r_G\geq r_{h,t}$,
\end{dcases*}
\end{equation}
where $r_G$ is the galactocentric radius in spherical coordinates. As in the bulge model, the first part of the halo model for the condition $r_G < r_{h,{\rm min}}$ is a modification from \cite{kaf14} to prevent infinite density at the galactic centre.

\begin{table}[t]
\caption{The parameters of the three components of the Milky Way's mass density. Except for the central mass densities $\rho_{b,0}$, $\rho_{h,0}$, and the surface densities $\Sigma_t$, $\Sigma_T$---which are calculated by integrating the mass density---the values are taken from \cite{BT2} for the bulge component, \cite{bla16} for the disc component, and \cite{kaf14} for the halo component.}
\begin{center}
\begin{tabular}{ll}
\hline\hline
\multicolumn{2}{c}{Bulge}\\
\hline
$\rho_{b,0}$ & $1.722\;M_{\odot}\;{\rm pc}^{-3}$\\
$\alpha_b$ & $1.8$\\
$a_b$ & $1.0\;{\rm kpc}$\\
$r_b$ & $1.9\;{\rm kpc}$\\
$q_b$ & $0.6$\\
$q_{\rm min}$& $10^{-2}\;{\rm kpc}$\\
\hline\hline
\multicolumn{2}{c}{Disc}\\
\hline
$\Sigma_t$ & $970.294\;M_{\odot}\;{\rm pc}^{-2}$\\
$\Sigma_T$ & $268.648\;M_{\odot}\;{\rm pc}^{-2}$\\
$R_t$ & $2.6\;{\rm kpc}$\\
$R_T$ & $2\;{\rm kpc}$\\
$z_t$ & $0.3\;{\rm kpc}$\\
$z_T$ & $0.9\;{\rm kpc}$\\
\hline\hline
\multicolumn{2}{c}{Halo}\\
\hline
$\rho_{h,0}$ & $5.075\times 10^{-6}\;M_{\odot}\;{\rm pc}^{-3}$\\
$\alpha_{h,1}$ & $2.4$\\
$\alpha_{h,2}$ & $4.5$\\
$r_{h,b}$ & $17.2\;{\rm kpc}$\\
$r_{h,t}$ & $97.7\;{\rm kpc}$\\
$\Delta_h$ & $7.1\;{\rm kpc}$\\
$\epsilon_h$ & $\tfrac{r_{h,t}}{\Delta_h} - 4.5$\\
$r_{h,{\rm min}}$ & $0.5\;{\rm kpc}$\\
\hline
\end{tabular}
\end{center}
\label{tab:mwpars}
\end{table}
The adopted values for model parameters are given in Table~\ref{tab:mwpars}. The bulge and halo central mass densities $\rho_{b,0}$ and $\rho_{h,0}$ are calculated assuming that the total masses of the bulge and the halo are respectively 30\% and 1\% of the total mass of the Galaxy, with the mass of the disc is set to be $M_d = 4.8\times 10^{10}\;M_{\odot}$ \citep{bla16}. The surface densities of the disc, $\Sigma_T$ and $\Sigma_t$, are derived from this value of $M_d$ and assuming that the ratio of these surface densities is 0.11 at the solar distance $R_\odot$ \citep{bla16}.

Fig.~\ref{fig:mwdens} shows the resulting profile density $\rho_{\rm MW}$ in one and two dimensions in galactocentric coordinates. To use this density model in our prior we transform a position in heliocentric coordinates $(r,l,b)$ into galactocentric cylindrical coordinates $(R,\theta,z)$ assuming that the Sun is $R_\odot = 8$\,kpc from the galactic centre, and lies in the galactic plane.

To estimate the $V$-band extinction $A_V$ used throughout this work, we use the extinction map of \cite{dri03}, which is based on the dust distribution model of the Milky Way built by \cite{dri01} and fitted to the far-infrared and near-infrared observations of the COBE/DIRBE instrument. The map allows us to calculate $A_V$ at any $(r,l,b)$ position in the Galaxy. Examples of the extinction map is shown in Fig.~\ref{fig:extMap}, which shows $A_V$ as a function of $(l,b)$ direction for distances at 1\,pc, 100\,pc, 1\,kpc, and 10\,kpc.

\section{The median: comparison with the true values and statistical performance}
\label{app:comparisons_md}
\begin{figure*}
\includegraphics[width=\hsize]{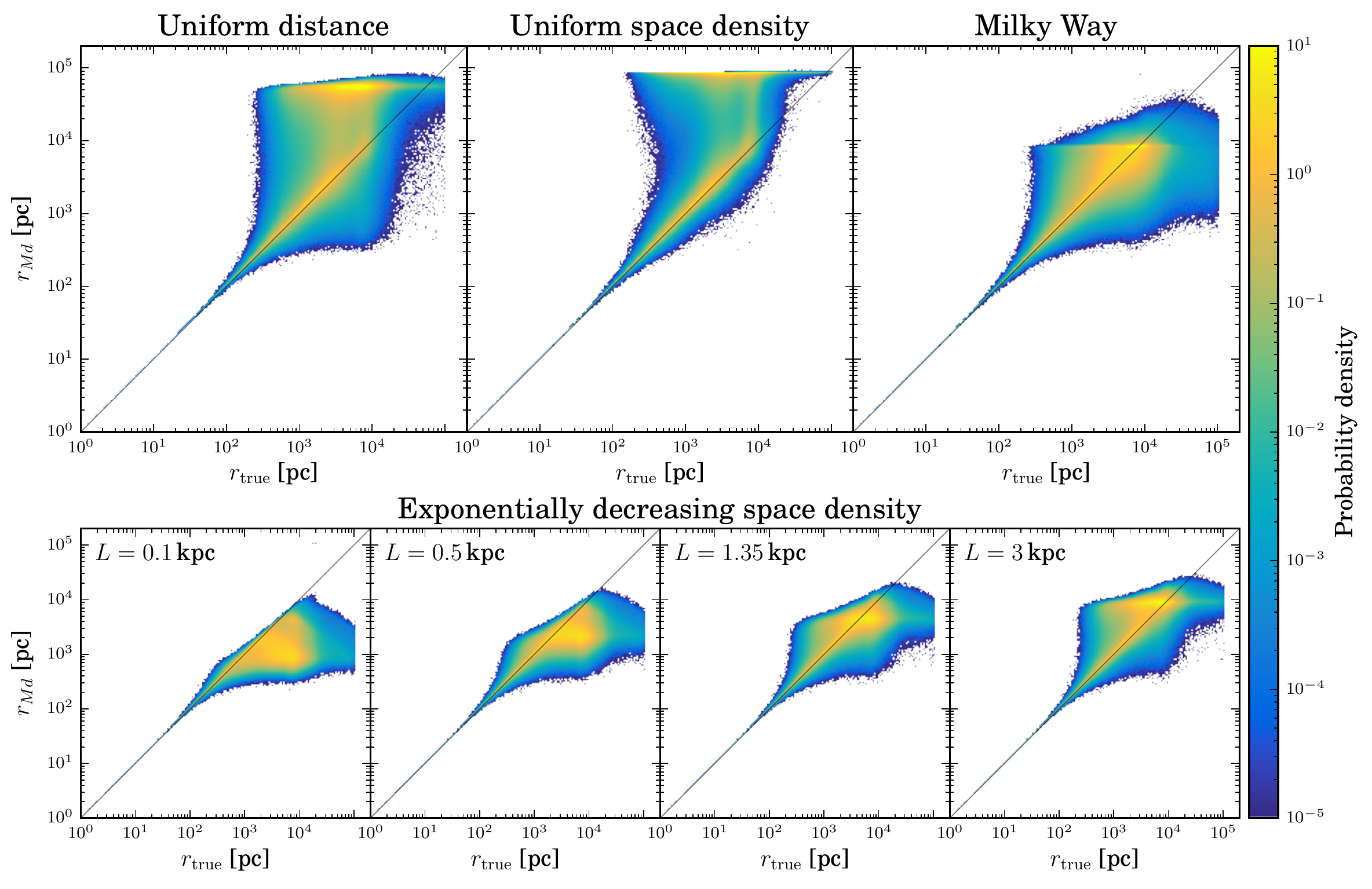}
\caption{As Fig.~\ref{fig:comparisons_mo}, but for the median distance estimator $\rmed$.}
\label{fig:comparisons_md}
\end{figure*}
Fig.~\ref{fig:comparisons_md} compares the median distance $\rmed$ with the true distances, using each of the four priors (cf. Fig.~\ref{fig:comparisons_mo}).
We see some clustering around a narrow range of median distance. These are stars with very large $\fobs$, so the posterior is practically the same as the prior, and the inferred distance is close to the median of the prior. The median of the \ud{} and the \usd{} priors are $\rmed = \tfrac{1}{2}\rlim$ and $\rmed = 2^{-1/3}\rlim$ respectively. For $\rlim=100$\,kpc, these are $\rmed = 50$\,kpc and $\rmed=79.4$\,kpc respectively, which is indeed where we see the clustering.
The median of the \expp{} prior is the (numerical) solution of
\begin{equation}
(2L^2 + 2Lr + r^2)\exp\left(-\frac{r}{L}\right) - L^2 = 0,
\end{equation}
which gives $\rmed\approx 2.674L$. This likewise explains the clustering in the bottom row of plots.
Concerning the \mw{} prior, we saw again the clustering at around 8\,kpc; the distance to the galactic center in this prior. For the \usd{} prior, we find that the median of the posterior moves rapidly to large distances as $\fobs$ increases beyond about 0.3 (see Fig.~\ref{fig:posteriors}), which is why we see fewer stars in the area between $\rmed=\rtrue$ and $\rmed=2^{-1/3}\rlim$ (79\,kpc).

\begin{figure}
\centering
\includegraphics[width=\hsize]{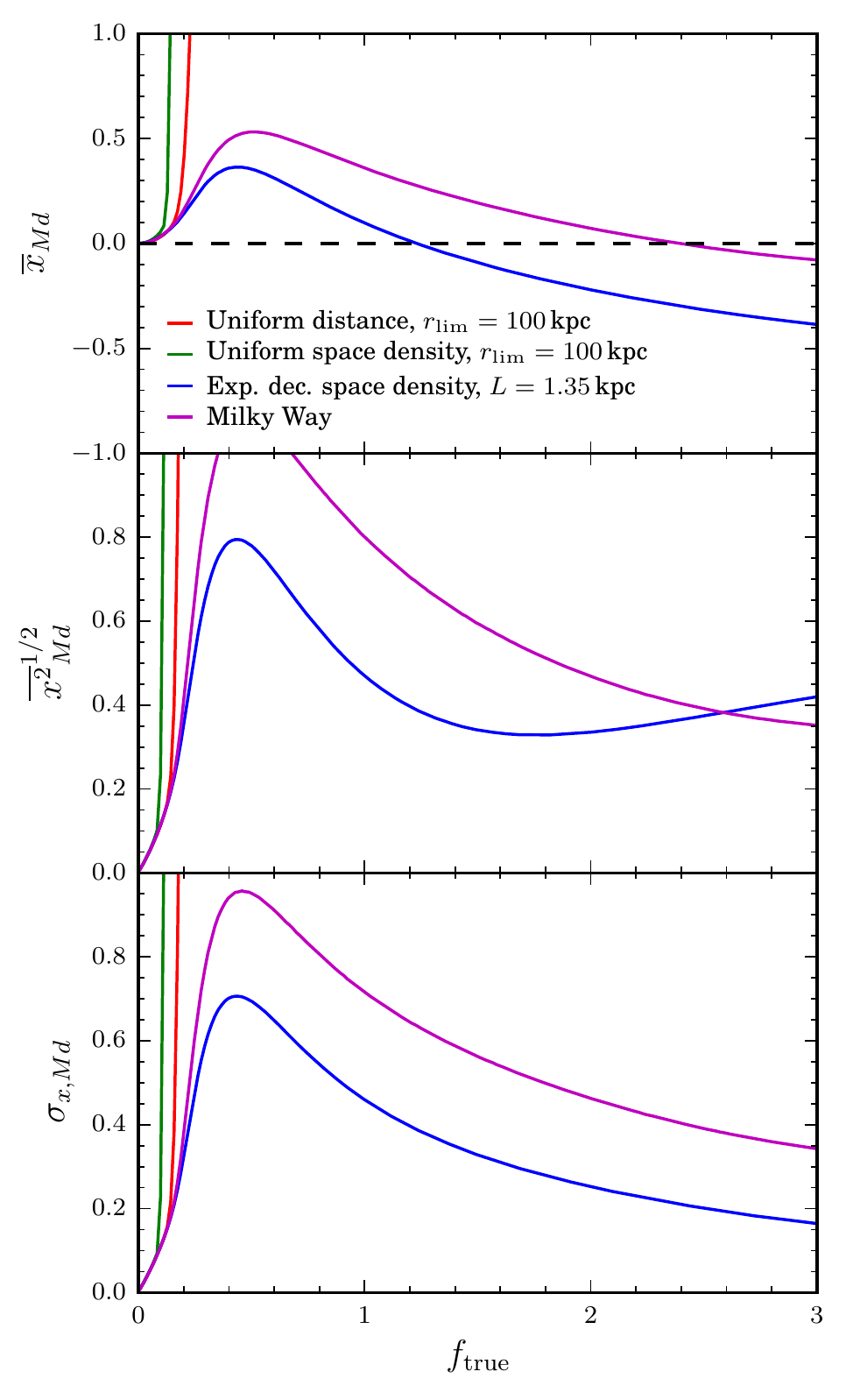}
\caption{As Fig.~\ref{fig:summary_performance_ftrue_mode}, but for the median distance $\rmed$ of all priors.}
\label{fig:summary_performance_ftrue_median}
\end{figure}
Comparing Fig.~\ref{fig:comparisons_md} with Fig.~\ref{fig:comparisons_mo}, it is not obvious which is the better-performing distance estimator. For this purpose we show the performance of the median in Fig.~\ref{fig:summary_performance_ftrue_median}, in the same way as shown for the mode in Fig.~\ref{fig:summary_performance_ftrue_mode}. A comparison of these two figures shows that the median is a worse estimator in terms of smaller bias, RMS, and standard deviation for values of $\ftrue$ less than about one. Only for worse expected data does the median perform slightly better. The \ud{} and \usd{} priors remain poor once $\ftrue$ is greater than about 0.1. The worse performance of the median over the mode is also in part because of the bimodality of the posterior, which as discussed in section \ref{subsec:3priors} occurs at intermediate parallax errors. The median is a compromise between the two modes and so explains neither very well.

\bibliographystyle{aa}
\bibliography{bibliography}

\begin{thebibliography}{33}
\expandafter\ifx\csname natexlab\endcsname\relax\def\natexlab#1{#1}\fi
\expandafter\ifx\csname url\endcsname\relax
  \def\url#1{{\tt #1}}\fi
\expandafter\ifx\csname urlprefix\endcsname\relax\def\urlprefix{URL }\fi

\bibitem[{{Arenou} \& {Luri}(1999)}]{are99}
{Arenou} F., {Luri} X., 1999, In: {Egret} D., {Heck} A. (eds.) Harmonizing
  Cosmic Distance Scales in a Post-HIPPARCOS Era, vol. 167 of Astronomical
  Society of the Pacific Conference Series, 13--32

\bibitem[{{Astropy Collaboration} et~al.(2013){Astropy Collaboration},
  {Robitaille}, {Tollerud} et~al.}]{2013A&A...558A..33A}
{Astropy Collaboration}, {Robitaille} T.P., {Tollerud} E.J., et~al., Oct. 2013,
  \aap, 558, A33

\bibitem[{{Bailer-Jones}(2015)}]{cbj15}
{Bailer-Jones} C.A.L., Nov. 2015, \pasp, 127, 994, paper I

\bibitem[{{Bailer-Jones} et~al.(2013){Bailer-Jones}, {Andrae}, {Arcay}
  et~al.}]{cbj13}
{Bailer-Jones} C.A.L., {Andrae} R., {Arcay} B., et~al., Nov. 2013, \aap, 559,
  A74

\bibitem[{{Binney} \& {Tremaine}(2008)}]{BT2}
{Binney} J., {Tremaine} S., 2008, {Galactic Dynamics: Second Edition},
  Princeton University Press

\bibitem[{{Bland-Hawthorn} \& {Gerhard}(2016)}]{bla16}
{Bland-Hawthorn} J., {Gerhard} O., 2016, \araa, 54, 529

\bibitem[{{Bovy} et~al.(2014){Bovy}, {Nidever}, {Rix} et~al.}]{bov14}
{Bovy} J., {Nidever} D.L., {Rix} H.W., et~al., Aug. 2014, \apj, 790, 127

\bibitem[{{de~Bruijne} et~al.(2005){de~Bruijne}, Perryman, Lindegren
  et~al.}]{jdb022}
{de~Bruijne} J., Perryman M., Lindegren L., et~al., June 2005, {G}aia
  astrometric, photometric, and radial-velocity performance assessment
  methodologies to be used by the industrial system-level teams, Tech. Rep.
  GAIA-JDB-022, European Space Research and Technology Centre,
  \urlprefix\url{http://www.rssd.esa.int/cs/livelink/open/448635}

\bibitem[{{de Bruijne} et~al.(2014){de Bruijne}, {Rygl}, \& {Antoja}}]{deb14}
{de Bruijne} J.H.J., {Rygl} K.L.J., {Antoja} T., Jul. 2014, In: EAS
  Publications Series, vol.~67 of EAS Publications Series, 23--29

\bibitem[{{Drimmel} \& {Spergel}(2001)}]{dri01}
{Drimmel} R., {Spergel} D.N., Jul. 2001, \apj, 556, 181

\bibitem[{{Drimmel} et~al.(2003){Drimmel}, {Cabrera-Lavers}, \&
  {L{\'o}pez-Corredoira}}]{dri03}
{Drimmel} R., {Cabrera-Lavers} A., {L{\'o}pez-Corredoira} M., Oct. 2003, \aap,
  409, 205

\bibitem[{{Fitzpatrick}(1999)}]{fit99}
{Fitzpatrick} E.L., Jan. 1999, \pasp, 111, 63

\bibitem[{{Goodman} \& {Weare}(2010)}]{goo10}
{Goodman} J., {Weare} J., Jan. 2010, Comm. App. Math. And Comp. Sci, 5, 65

\bibitem[{{G{\'o}rski} et~al.(2005){G{\'o}rski}, {Hivon}, {Banday}
  et~al.}]{gor05}
{G{\'o}rski} K.M., {Hivon} E., {Banday} A.J., et~al., Apr. 2005, \apj, 622, 759

\bibitem[{Hunter(2007)}]{Hunter:2007}
Hunter J.D., 2007, Computing In Science \& Engineering, 9, 90

\bibitem[{{Ivezi\'{c}} et~al.(2008){Ivezi\'{c}}, {Tyson}, {Abel}
  et~al.}]{ive08}
{Ivezi\'{c}} Z., {Tyson} J.A., {Abel} B., et~al., May 2008, ArXiv e-prints

\bibitem[{Jones et~al.(2001--)Jones, Oliphant, Peterson
  et~al.}]{jones_scipy_2001}
Jones E., Oliphant T., Peterson P., et~al., 2001--, {SciPy}: Open source
  scientific tools for {Python}, \urlprefix\url{http://www.scipy.org/}

\bibitem[{{Jordi} et~al.(2006){Jordi}, {H{\o}g}, {Brown} et~al.}]{jor06}
{Jordi} C., {H{\o}g} E., {Brown} A.G.A., et~al., Mar. 2006, \mnras, 367, 290

\bibitem[{Jordi et~al.(2009)Jordi, Fabricius, Figueras et~al.}]{cj047}
Jordi C., Fabricius C., Figueras F., et~al., February 2009, {E}rror model for
  photometry and spectrophotometry, Tech. Rep. GAIA-C5-TN-UB-CJ-047, University
  of Barcelona (Dept. Astronomia i Meteorologia),
  \urlprefix\url{http://www.rssd.esa.int/cs/livelink/open/2871236}

\bibitem[{{Jordi} et~al.(2010){Jordi}, {Gebran}, {Carrasco} et~al.}]{jor10}
{Jordi} C., {Gebran} M., {Carrasco} J.M., et~al., Nov. 2010, \aap, 523, A48

\bibitem[{{Jung}(1971)}]{jun71}
{Jung} J., Apr. 1971, \aap, 11, 351

\bibitem[{{Kafle} et~al.(2014){Kafle}, {Sharma}, {Lewis}, \&
  {Bland-Hawthorn}}]{kaf14}
{Kafle} P.R., {Sharma} S., {Lewis} G.F., {Bland-Hawthorn} J., Oct. 2014, \apj,
  794, 59

\bibitem[{{Lindegren} et~al.(2012){Lindegren}, {Lammers}, {Hobbs}
  et~al.}]{lin12}
{Lindegren} L., {Lammers} U., {Hobbs} D., et~al., Feb. 2012, \aap, 538, A78

\bibitem[{{Liu} et~al.(2012){Liu}, {Bailer-Jones}, {Sordo} et~al.}]{liu12}
{Liu} C., {Bailer-Jones} C.A.L., {Sordo} R., et~al., Nov. 2012, \mnras, 426,
  2463

\bibitem[{P\'erez \& Granger(2007)}]{PER-GRA:2007}
P\'erez F., Granger B.E., May 2007, Computing in Science and Engineering, 9,
  21, \urlprefix\url{http://ipython.org}

\bibitem[{{Perryman} et~al.(1997){Perryman}, {Lindegren}, {Kovalevsky}
  et~al.}]{per97}
{Perryman} M.A.C., {Lindegren} L., {Kovalevsky} J., et~al., Jul. 1997, \aap,
  323

\bibitem[{{Robin} et~al.(2003){Robin}, {Reyl{\'e}}, {Derri{\`e}re}, \&
  {Picaud}}]{rob03}
{Robin} A.C., {Reyl{\'e}} C., {Derri{\`e}re} S., {Picaud} S., Oct. 2003, \aap,
  409, 523

\bibitem[{{Robin} et~al.(2012){Robin}, {Luri}, {Reyl{\'e}} et~al.}]{rob12}
{Robin} A.C., {Luri} X., {Reyl{\'e}} C., et~al., Jul. 2012, \aap, 543, A100

\bibitem[{{Roman}(1952)}]{rom52}
{Roman} N.G., Jul. 1952, \apj, 116, 122

\bibitem[{{Santiago} et~al.(2016){Santiago}, {Brauer}, {Anders} et~al.}]{san16}
{Santiago} B.X., {Brauer} D.E., {Anders} F., et~al., Jan. 2016, \aap, 585, A42

\bibitem[{{Sch{\"o}nrich} \& {Bergemann}(2014)}]{sch14}
{Sch{\"o}nrich} R., {Bergemann} M., Sep. 2014, \mnras, 443, 698

\bibitem[{{Smith}(2006)}]{smi06}
{Smith} H., Jan. 2006, \mnras, 365, 469

\bibitem[{{van Leeuwen} \& {Evans}(1998)}]{van98}
{van Leeuwen} F., {Evans} D.W., May 1998, \aaps, 130, 157

\end{thebibliography}

\end{document}